\documentclass[fleqn,usenatbib]{mnras}

\usepackage{newtxtext,newtxmath}

\usepackage[T1]{fontenc}

\usepackage{amsmath}

\DeclareRobustCommand{\VAN}[3]{#2}
\let\VANthebibliography\thebibliography
\def\thebibliography{\DeclareRobustCommand{\VAN}[3]{##3}\VANthebibliography}

\usepackage{graphicx}	
\usepackage{amsmath}	

\usepackage{hyperref}
\usepackage{academicons}
\usepackage{xcolor}

\newcommand{\orcid}[1]{\href{https://orcid.org/#1}{\textcolor[HTML]{A6CE39}{\aiOrcid}}}

\title[The JWST Weather Report from WISE1049AB]{The JWST Weather Report from the Nearest Brown Dwarfs  I: multi-period JWST NIRSpec + MIRI monitoring of the benchmark binary brown dwarf WISE 1049AB}

\author[Beth Biller et al.]{Beth A. Biller$^{1,2}$\thanks{E-mail: bb@roe.ac.uk}, Johanna M. Vos$^{3}$, Yifan Zhou$^{4}$, Allison M. McCarthy$^{5}$, Xianyu Tan$^{6, 7}$, 
Ian J.M. Crossfield$^{8}$, \newauthor 
Niall Whiteford$^{9}$, Genaro Suarez$^{9}$, Jacqueline Faherty$^{9}$, Elena Manjavacas$^{10, 11}$, Xueqing Chen$^{1,2}$, \newauthor Pengyu Liu$^{1,2,12}$,  Ben J. Sutlieff$^{1,2}$, Mary Anne Limbach$^{13}$, Paul Molliere$^{14}$, Trent J. Dupuy$^{1,2}$, \newauthor Natalia Oliveros-Gomez$^{11}$, Philip S. Muirhead$^{5}$, Thomas Henning$^{14}$, Gregory Mace$^{15}$, Nicolas Crouzet$^{12}$, \newauthor Theodora Karalidi$^{16}$, Caroline V. Morley$^{15}$, Pascal Tremblin$^{17}$, Tiffany Kataria$^{18}$
\\
$^{1}$ Institute for Astronomy, University of Edinburgh, Royal Observatory, Edinburgh EH9 3HJ,UK \\
$^{2}$ Centre for Exoplanet Science, University of Edinburgh, Edinburgh, UK \\
$^{3}$ School of Physics, Trinity College Dublin, The University of Dublin, Dublin 2, Ireland \\
$^{4}$ Department of Astronomy, University of Virginia, 530 McCormick Rd., Charlottesville, VA 22904, USA \\
$^{5}$ Department of Astronomy \& The Institute for Astrophysical Research, Boston University, 725 Commonwealth Ave., Boston, MA 02215, USA \\
$^{6}$ Tsung-Dao Lee Institute, Shanghai Jiao Tong University, 520 Shengrong Road, Shanghai, 200127, People's Republic of China\\
$^{7}$ School of Physics and Astronomy, Shanghai Jiao Tong University, 800 Dongchuan Road, Shanghai,200240, People's Republic of China\\
$^{8}$ Department of Physics and Astronomy, University of Kansas, Lawrence, KS, USA \\
$^{9}$ Department of Astrophysics, American Museum of Natural History, Central Park West at 79th Street, NY 10024, USA \\
$^{10}$ AURA for the European Space Agency (ESA), ESA Office, Space Telescope Science Institute, 3700 San Martin Drive, Baltimore, MD, 21218 USA \\
$^{11}$ William H. Miller III Department of Physics and Astronomy, Johns Hopkins University, Baltimore, MD 21218, USA \\
$^{12}$ Leiden Observatory, Leiden University, PO Box 9513, 2300 RA Leiden, The Netherlands \\
$^{13}$ Department of Astronomy, University of Michigan, Ann Arbor, MI 48109, USA \\
$^{14}$ Max-Planck-Institut f\"ur Astronomie, K\"onigstuhl 17, D-69117 Heidelberg Germany \\ 
$^{15}$ Department of Astronomy, The University of Texas at Austin, Austin, TX 78712, USA \\
$^{16}$ Department of Physics, University of Central Florida, 4111 Libra Dr., Orlando, FL 32816 \\
$^{17}$ Maison de la Simulation, CEA, CNRS, Univ. Paris-Sud, UVSQ, Universit\'e Paris-Saclay, 91191 Gif-sur-Yvette, France \\
$^{18}$ Jet Propulsion Laboratory, California Institute of Technology, Pasadena, CA 91109 USA \\
}

\date{Accepted XXX. Received YYY; in original form ZZZ}

\pubyear{2024}

\begin{document}
\label{firstpage}
\pagerange{\pageref{firstpage}--\pageref{lastpage}}
\maketitle

\begin{abstract}
We report results from 8 hours of JWST/MIRI LRS spectroscopic monitoring directly followed by 7 hours of JWST/NIRSpec prism spectroscopic monitoring of the benchmark binary brown dwarf WISE 1049AB, the closest, brightest brown dwarfs known.  We find water, methane, and CO absorption features in both components, including the 3.3 $\mu$m methane absorption feature and a tentative detection of small grain ($<$ 1$\mu$m) silicate absorption at $>$8.5 $\mu$m in WISE 1049A.  Both components vary significantly ($>$1$\%$), with WISE 1049B displaying larger variations than WISE 1049A.  Using K-means clustering, we find three main transition points in wavelength for both components of the binary:
1) change in behavior at $\sim$2.3 $\mu$m coincident with a CO absorption bandhead, 2) change in behavior at 4.2 $\mu$m, close to the CO fundamental band at $\lambda >$ 4.4 $\mu$m, and 3) change in behavior at 8.3-8.5 $\mu$m,
potentially corresponding to silicate absorption.  We interpret the lightcurves observed with both NIRSpec and MIRI as likely stemming from 1) a deep pressure level driving the double-peaked variability seen in WISE 1049B at wavelengths $<$2.3 $\mu$m
and $>$8.5 $\mu$m, 2) an intermediate pressure level shaping the lightcurve morphology between 2.3 and 4.2 $\mu$m, and 3) a higher-altitude pressure level producing single-peaked and plateaued lightcurve behavior between 4.2 and 8.5 $\mu$m.
\end{abstract}




\section{Introduction}

The vast majority of spectroscopy of directly imaged exoplanet and brown dwarf atmospheres consists of time-averaged snapshots of a single hemisphere.
However, brown dwarfs, and likely young directly imaged exoplanets as well, are known rapid rotators with rotational periods of $\sim$1-20 hours \citep{ZapateroOsorio2006, Snellen2014, Bryan2020, Tannock2021, Vos2022} and have evidence of inhomogeneous top-of-atmosphere (TOA) structure driving significant photometric variability \citep[c.f.~among~others][]{Artigau2009, Radigan2012, Radigan2014, Metchev2015a, Biller2015, Vos2019, Vos2022}.
To truly understand the physics of brown dwarf atmospheres, and by extension, their lower mass gas giant planet cousins, requires time-resolved observations that characterize emission from these objects as a function of rotational phase.  Obtaining longitudinally-resolved observations of the TOA structure of these objects as a function of wavelength will provide key tests for existing 1-D atmospheric models \citep[see][for~one~example~of~such~a~test]{Luna2021} as well as for 3-D models currently in development that take into account the rapid rotation of these objects \citep[c.f.][among others]{Tan2022, Tan2021a, Tan2021b, Lee2024}.

The James Webb Space Telescope \citep[henceforth~JWST,][]{Gardner2006, Rigby2023} uniquely enables time-resolved studies of brown dwarfs over a broad wavelength range and also opens up investigations of variability in the silicate absorption features at $>$8.5 $\mu$m \citep{Cushing2006, Suarez2022} for the first time, directly tracing the impact of small-grain ($\lesssim$1 $\mu$m), high-altitude clouds on the variability of these objects.  Previous generation surveys with Spitzer (photometric monitoring at 3.6 $\mu$m and 4.5 $\mu$m) found that variability is very common at mid-IR wavelengths across all spectral types \citep{Metchev2015a, Vos2022}.  Variability is very common at near-IR wavelengths as well \citep{Radigan2012, Radigan2014, Radigan2014a, Wilson2014, Eriksson2019, Vos2019, Liu2024}, although most near-IR variability monitoring to date is from ground-based facilities, which limits the measurable variability to variations $>$1$\%$ at wavelengths $<$3 $\mu$m and the length of continuous observations to $\sim$8-10 hours.  
In a $J$-band survey of 57 L4 to T9 brown dwarfs, \citet{Radigan2014} found
4 strong variables (measured variation $>$2$\%$) among L/T transition brown dwarfs (L9 to T3.5 spectral types) and 5 weak variables (measured variation 0.6-1.6$\%$) at other spectral types.  \citet{Radigan2014} interpret the higher variability amplitudes measured for their L/T transition targets as likely driven by the breakup of silicate clouds over the transition between L spectral types, which are dominated by silicate clouds, and T spectral types, which have clear atmospheres.  This interpretation is supported by atmospheric modeling 
\citep{Marley2012} and time-resolved spectroscopic observations \citep{Apai2013}. Other authors have suggested hot spots \citep{Robinson2014}, potentially driven by non-equilibrium chemistry \citep{Tremblin2016}, or aurorae \citep{Hallinan2015} as mechanisms contributing to the observed variability.  

WISE J104915.57$-$531906.1AB (henceforth WISE 1049AB, also known as Luhman 16AB)
are the two closest brown dwarfs to the Earth and form a benchmark binary brown dwarf system \citep{Luhman2013, Kniazev2013, Burgasser2013, Faherty2014}, with measured masses \citep{Garcia2017, Bedin2017, Lazorenko2018} and ages \citep{Gagne2023}.  WISE 1049A has a spectral type of L7.5 and WISE 1049B has a spectral type of T0.5 \citep{Burgasser2013}, bracketing the L/T spectral type transition.  Along the L/T transition, variability appears to be both more common and with higher amplitudes compared to other spectral types \citep{Radigan2014}.  This L/T transition brown dwarf binary pair are the closest and brightest brown dwarfs \citep{Luhman2013}.  WISE 1049AB share similar compositions as they formed from the same natal cloud.  Both have dynamical mass measurements (34.2$\pm$1.2\,M$_\mathrm{Jup}$ for A, 27.9$\pm$1.0\,M$_\mathrm{Jup}$ for B, \citealt{Garcia2017, Bedin2017, Lazorenko2018}), with a mass ratio of $\sim$0.82.  The system has a well-measured age of 510$\pm$95\.Myr via membership in the newly discovered Oceanus moving group \citep{Gagne2023}.  Along the L/T transition, effective temperature remains nearly constant; both binary components have effective temperatures $\sim$1200--1300 K \citep{Filippazzo2015}. \cite{Apai2021} find that both are observed with a nearly equator-on viewing angle.  

WISE 1049AB are both variable at optical and IR wavelengths.  Given their proximity and brightness, extensive studies have sought to characterize this variability \citep{Gillon2013, Biller2013a, Burgasser2014, Mancini2015, Street2015, Buenzli2015, Buenzli2015a, Kellogg2017, Heinze2021, Apai2021}. 
These studies have found that, despite their similar masses, compositions, temperatures, and viewing angles, WISE 1049B is highly variable (5-15$\%$) with a well-measured period of $\sim$5 hours \citep{Gillon2013,Apai2021} and with differences in measured rotational phase found between the $i$, $z$, $J$, $H$, and $K$ band lightcurves \citep{Biller2013}, while
WISE 1049A is somewhat less variable \citep[$<$3$\%$~in~the~near-IR,][]{Buenzli2015, Biller2013} with a less-constrained period estimate of $\sim$7 hours \citep{Apai2021}. Thus, it is predominantly the differences in their TOA structure that cause the difference in variability properties.  JWST observations of this system present a unique opportunity to isolate two points along this critical spectral type transition, removing confounding variations in properties that would exist for any other brown dwarfs, such as differences in age, composition, and viewing angle.  WISE 1049AB is the pivotal first system to test the unique capabilities of JWST to probe the atmospheres of similar objects, as the wide wavelength coverage of JWST opens wavelengths inaccessible from the ground or with any other telescope and enables tests of specific variability mechanisms.   

\section{Observations}

Light curves for brown dwarfs and exoplanet analogues can change dramatically over the course of a few rotation periods \citep{Artigau2009, Gillon2013, Apai2017} and followup scheduled even days later may probe significantly different variability patterns. Thus, as part of JWST GO 2965 (PI Biller), we observed at least one full period of both the A and B components sequentially with MIRI followed by NIRSpec in an uninterruptible sequence in order to capture similar variability properties with both instruments.  MIRI LRS timeseries observations were acquired from UT 12:24:18 to UT 21:05:33 on 8 July 2023, followed by a brief MIRI background observation from UT 21:05:38 to UT 21:37:17 on 8 July 2023.  These observations were directly followed by NIRSpec Bright Object Time Series (BOTS) observations from UT 21:37:21 on 8 July 2023 to UT 05:12:53 on 9 July 2023.

We obtained MIRI time-series observations of WISE 1049AB with the LRS slitless mode with the P750L disperser to prevent slit losses, disabling dithers to ensure photometric stability, and utilizing the FAST readout mode to provide many samples up the ramp.  This yields spectral resolutions of R=40--160 from 5 to 10 $\mu$m, matching the resolution of archival Spitzer IRS spectra of field brown dwarfs.  \cite{Bell2023} recommend a 1 hour "burn-in" period to mitigate ramp effects and allow the detector to reach the stability required for time series observations, thus we obtained 7 hours + 1.0 hour = 8.0 hours on-sky time with MIRI.  We performed target acquisition using the LRS SLITLESSPRISM subarray with the FAST readout pattern with 4 groups, followed by 2193 integrations ($\sim$8 hours) using the LRS SLITLESSPRISM subarray with the FASTR1 readout pattern with 80 groups per integration, providing a base exposure time of 12.72 s during the time series observations.  The timeseries observations were followed up by a brief background observation of an empty field offset from the target by $-$10 arcsec respectively in Right Ascension and Declination.  We obtained 10 integrations with the same FASTR1 readout pattern and 80 groups per integration at the background position.

After completing the MIRI observations, we observed WISE 1049AB using the NIRSpec BOTS mode in the low-resolution PRISM/CLEAR mode, providing simultaneous wavelength coverage from 0.6 $\mu$m to 5.3 $\mu$m at a resolution ranging from R=30-300.  This resolution matches that of earlier HST WFC3 0.98 to 1.7 $\mu$m studies of this object \citep{Buenzli2015, Buenzli2015a} and is sufficient to resolve important water, CO, and CH$_4$ features.  As both components fall within the S1600A1 aperture during a WATA target acquisition and given their similar magnitudes, WATA target acquisition was performed using the mid-point between components as the target coordinates, and thus letting the target acquisition algorithm centre on whichever component of the binary lay closest to these coordinates during the epoch of observations.  This happened to be the B component, which was then acquired at the nominal pointing position for NIRSpec BOTS observations, with the trace for the A component appearing north of that of the B component. 
WISE 1049AB are the two brightest brown dwarfs known, so short exposure times were necessary to avoid saturation.  Adopting the SUB512S subarray, NRSRAPID readout, and 2 groups per integration, yielded a cadence of 0.45 s and avoided saturation of either component.  

WISE 1049AB was observed at a V3 PA (position angle of the observatory V3 reference axis) of 130.6$^{\circ}$, corresponding to a MIRI aperture position angle (APA) of 135.5$^{\circ}$ and a NIRSpec APA of 269.8$^{\circ}$.  During the epoch of the JWST observation, WISE 1049B was at a separation of $\sim$0.65" and a position angle of 113$^{\circ}$ from WISE 1049A, as measured from the NIRSpec acquisition and MIRI target verification images.  As a result, the NIRSpec traces are well-resolved ($>$5 pixel separation between traces), however, the MIRI traces are separated by only $\sim$2 pixels and must be disentangled via careful point spread function (PSF) modelling and extraction.

\section{Data Reduction and Spectral Extraction}

\subsection{NIRSpec reduction and spectral extraction}

For our NIRSpec time-series data, we performed a standard BOTS reduction using JWST STScI pipeline version 1.11.3, CRDS version 11.17.2, and CRDS context file \texttt{jwst\_1236.pmap}. We used STRUN version 0.5.0 to process all data through pipeline stages 1 and 2 (\texttt{calwebb\_detector1} and \texttt{calwebb\_spec2}), with default settings.  We then processed all data through JWST pipeline stage 3 (\texttt{calwebb\_tso3}) using the standard BOTS extraction region.  

\begin{figure*}
	\includegraphics[width=\textwidth]{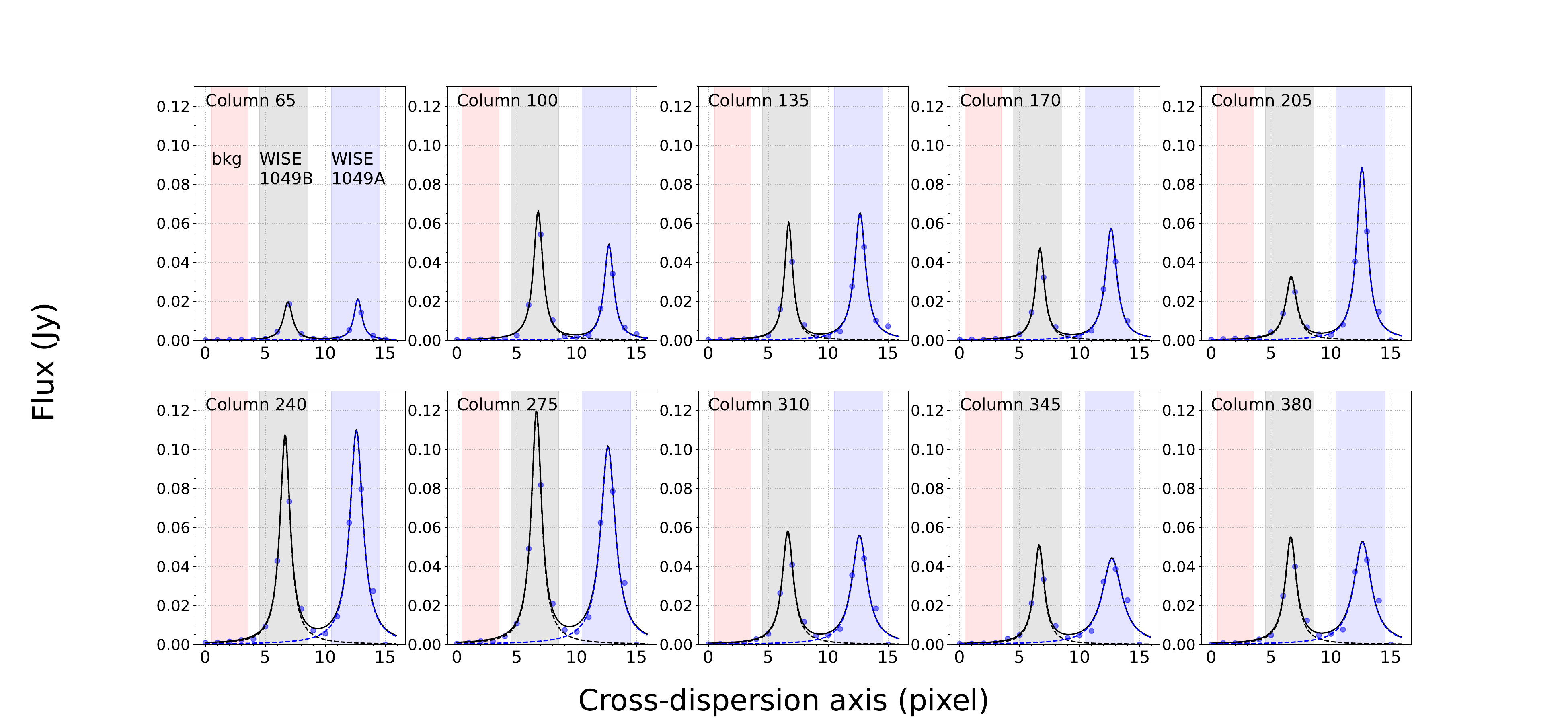}
    \caption{Spectral profile and fit of two Lorentzian PSF profiles to different data columns for 1 2-d NIRSpec spectral image.  The best fit to WISE 1049A is shown as a blue dashed line, the best fit to WISE 1049B is shown as a black dashed line, and the best combined fit is shown as a solid black line.  The column number for each profile is labeled next to its respective profile, with lower numbers corresponding to shorter wavelengths and higher numbers corresponding to longer wavelengths on the detector.  Extraction regions for the background, WISE 1049A, and WISE 1049B are denoted by the red, blue, and black shaded regions respectively.}
    \label{fig:NIRSpec_psffit}
\end{figure*}

The B component of the binary was acquired at the nominal pointing position, thus wavelengths from the default 1d extraction from the pipeline are correct.  However, the A component is offset in both the vertical and horizontal direction from the B component.  The horizontal offset will lead to a shift in the correct wavelengths for the A component, as NIRSpec disperses along the x-axis and thus the trace on the detector is dispersed in a slightly offset position in the x direction compared to the B component.  We measured a positional offset between A and B in the NIRSpec BOTS acquisition images of -2.5 pixels (-0.25") in x and by 5.7 pixels (0.57") in y using \texttt{astropy.photutils.DaoStarfinder} and then reran the stage 2 pipeline to account for the position offset of A by treating the offset as an additional dither position.  
To do so, we first set \texttt{TSOVISIT=False} in the file headers, as the time series observation (TSO) version of the \texttt{spec2} pipeline skips the \texttt{wavecorr} step, which adjusts the wavelengths in the case of a dithered observation.  We set the xoffset and yoffset keywords in the file headers in arcsec, with positive values, e.g. \texttt{XOFFSET = 0.25} and \texttt{YOFFSET = 0.57}, before rerunning stage 2 of the pipeline for the offset A component.  
We also override the default extraction region using an updated json extraction region file to use a rectangular extraction region centered on the A component (rows 11 to 14), with a background subtraction region placed in a source-free region at the bottom of the detector (rows 1 to 3).  This re-reduction produces a wavelength vector for A that aligns major absorption features (in particular the 1.45 $\mu$m water absorption feature) in the spectra of both WISE 1049A and WISE 1049B.  We also reran the stage 2 pipeline for the B component, with the original wavelength scaling but with a custom json extraction region file with an extraction region centered on the B component (rows 5 to 8) and the same background subtraction region used for the A component. The spectral profiles and a 2-Lorentzian fits (using \texttt{astropy.modeling})  to each component for representative columns from a single 2-d NIRSpec spectral image are shown in Fig.~\ref{fig:NIRSpec_psffit}.   
The two components of the binary are generally well-resolved, but each component is slightly contaminated by the other component.  From the Lorentzian fits to both components, we estimate 1--5$\%$ contamination of component B by component A (with longer wavelengths experiencing more contamination), and 1--3$\%$ contamination of component A by component B.  This level of contamination will not affect our relative lightcurve measurements or qualitative analysis of the spectra, but a future dedicated spectroscopic analysis will use full PSF-fitting techniques to fully de-blend the two components on NIRSpec.  The final spectra generally looked clean, however, 3-4 bad pixels per spectral trace had to be manually removed from the final spectral products.

\subsection{MIRI reduction and PSF-fitting spectral extraction}

We performed basic data reduction steps on the LRS time-series using JWST STScI pipeline version 1.11.3, CRDS version 11.17.2 and CRDS context file \texttt{'jwst\_1100.pmap'}.  The uncalibrated files for the LRS time-series observation were divided into 11 segments due to the length of the observation.  We ran the JWST pipeline on the command line using STRUN version 0.5.0 to process all segments of data through pipeline stages 1 and 2 (\texttt{calwebb\_detector1} and \texttt{calwebb\_spec2}), with default settings. We then used custom scripts to extract spectra for both components of the binary from the resulting calibrated 2-d spectral image (\texttt{\_calints}) products.

In the LRS slitless spectroscopy mode, the spectral traces for the A and B components of the binary are dispersed in the y-direction starting from different y-positions, corresponding to the offset in y position between each component and the nominal pointing position used by the pipeline.  Thus, the appropriate wavelength as a function of detector row will vary depending on the position of each component.  Due to the offset in the y-direction between the two binary components, we first had to correct the wavelength calibration for both components to account for their offsets from the nominal pointing position.  Before starting the MIRI time-series observation, we obtained a MIRI LRS target verification image in the F560W filter, with the FASTR1 readout, 5 groups, and 1 integration. We measured the positions of the A and B components in the MIRI target verification image.
We found a best WISE1049A position of 37.33, 300.08 in x,y pixel coordinates on the detector and a best position for WISE1049B of 39.69, 305.70 using \texttt{astropy.photutils.DaoStarfinder}.  The nominal pointing position for the MIRI LRS slitless mode is <x> = 37.66 $\pm$ 0.05, <y> = 300.43 $\pm$ 0.09 from JWST commissioning data (private communication, G. Sloan, STScI).  Thus, the A component was placed at the nominal pointing position and wavelengths from 1d extraction from the STScI JWST pipeline are correct for this component.  The B component is offset by $\sim$5 pixels from the nominal pointing position.  To produce the correct wavelength scale for B, we interpolated from the wavelength scale for A, offsetting by the difference between the B y-position and the nominal pointing y-position.  Because of this offset, the first wavelength bin in MIRI spectrum for WISE 1049B is at 5.17 $\mu$m, while the spectrum for WISE 1049A is usable down to 5 $\mu$m.


\subsubsection{PSF-fitting using the MIRI LRS commissioning PSF}

To extract and deblend both spectral traces, we used a MIRI LRS PSF built from commissioning data and provided by STScI (private communication, S. Kendrew and G. Sloan, STScI).   The provided PSF has dimensions of 416 rows, with 2001 subsampled pixels per row, with the observed source placed in column 1000.  In contrast, each 2-d spectral image has a dimension of 416 rows and 72 columns, with units of MJy / steradian per pixel.  The PSF has been super-sampled by a factor of 50, thus, the PSF covers the inner 40 data pixels around the source on the detector.  The area under the source profile for each row of the PSF has been normalised to 1 MJy / steradian.    

For each 2-d spectral image in the sequence we then fit the 1-d spectral profiles on a row-by-row basis.  For each row, we followed the following procedure:

\begin{enumerate}
    \item We estimated and subtracted out the background as the median of the last 11 pixels for each 1-d spectral profile.
    \item For the inner 40 data pixels around each binary component, we interpolated appropriate super-sampled PSF spectral profiles for both components from the provided super-sampled PSF using \texttt{scipy.interpolate.interp1d}, accounting for the wavelength shift due to the offset of the y-positions of each component on the detector from the nominal MIRI LRS pointing position.
    \item We then interpolated the super-sampled PSFs for each component at the actual measured sampling positions (40 measured data points), fixing the position of each component at its measured x-position from our measurement of the positions of both components on the MIRI target verification image.  The full two-component model is then the interpolated PSF profile for each component, multiplied by a flux-scaling factor for each component.  Thus, our final model has only two parameters.
    \item Finally, we found the best fit flux-scaling factor for each component by subtracting the interpolated two-component model from the data for the row, and then minimizing $\chi^2$ using \texttt{scipy.optimize.minimize}.
\end{enumerate}

\begin{figure*}
	\includegraphics[width=\textwidth]{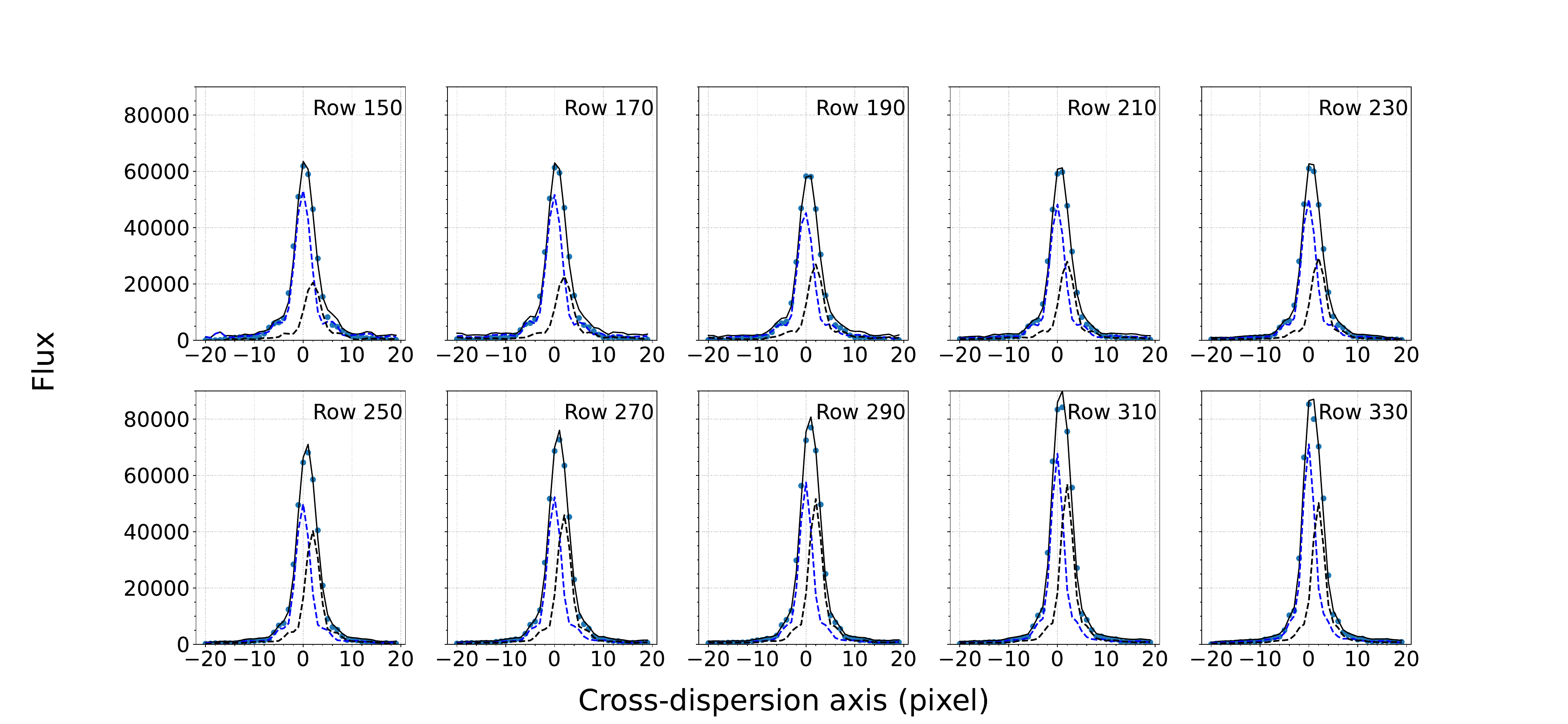}
    \caption{Sample 2-component PSF fit to different data rows for 1 2-d MIRI spectral image.  Flux is given in units of the native counts on the detector.  The best fit to WISE 1049A is shown as a blue dashed line, the best fit to WISE 1049B is shown as a black dashed line, and the best combined fit is shown as a solid black line.  The row number for each profile is labeled next to its respective profile, with lower numbers corresponding to longer wavelengths and higher numbers corresponding to shorter wavelengths on the detector.}
    \label{fig:MIRI_psfsub_1}
\end{figure*}

Fig.~\ref{fig:MIRI_psfsub_1} shows the result of this two-component fitting procedure for a number of sample rows for a single 2-d MIRI spectral image.

As the input PSF was normalized to 1 MJy / steradian for each row and the input 2-d spectral data image is in units of MJy / steradian per pixel, the flux-scaling factors fit for each component should also be in units in MJy / steradian.  However, because we only interpolate at 40 data points out of the 2000 datapoints used for the normalization from the super-sampled PSF, the area under the PSF profile for each component before fitting the scaling factor is 0.02, instead of 1.0.  Thus, to correct our retrieved fluxes into units of MJy, we must first divide by a factor of 50, to account for the different sampling between the data and super-sampled PSF, then multiply by the solid angle in steradians of one 0.11"$\times$0.11" MIRI pixel (2.23$\times$10$^{-13}$ steradians).

We adopt the flux-scaling factors (converted to mJy) as the flux at the appropriate wavelength for each component, thus reconstructing the spectra of each component row-by-row.  For a number of test 2-d spectral images, the fits were performed using both \texttt{scipy.optimize.minimize} and a Markov-Chain Monte Carlo fit using the \texttt{emcee} package \citep{Foreman-Mackey2019}.  Both methods yielded similar results, but we elected to use the \texttt{scipy.optimize.minimize} method for the full lightcurve / spectral extraction, as it was computationally much quicker.

\subsubsection{Lorentzian and Moffat Profile PSF-fitting}

We adopt the full PSF-fitting using the commissioning PSF for the spectra and lightcurves presented in this paper.  However, to validate the results from the full empirical PSF-fitting, we also extracted the spectral traces using two parameterized model fits, utilising a Lorentzian and Moffat function PSF profile respectively.

For the Lorentzian method, we developed a model of the MIRI LRS PSF as a function of wavelength using a time-series dataset from GO 3548 (PI Vos).  This is a MIRI LRS dataset of a similar variable brown dwarf, reduced using the same pipeline version and CRDS files.  We used the binned time series to fit a Lorentzian profile with \texttt{astropy.modeling} with parameters: profile centre (pixels in x direction), profile amplitude (detector counts), and profile full-width half max (FWHM, in pixels) to each detector row.  We retained the best fit FWHM as a function of wavelength / detector row as a model of the PSF.  We then fit each spectral image in the WISE 1049AB dataset row-by-row with a double Lorentzian model, after estimating and subtracting out the background for each row as the median of the last 11 pixels in that row. The centres of both Lorentzians were fixed to the measured x-axis positions for each component from the target verification image, while the FWHM was drawn from the FWHM model built from the GO 3548 data.  As noted above, because of the y-position offset between the targets and the nominal pointing position, we adjusted the wavelength scale for each component separately, interpolating the wavelength vs. FWHM model on a component-by-component basis to choose the correct FWHM for the differing wavelengths of each component due to their positional shifts.  We then used \texttt{astropy.modeling} to fit the appropriate amplitudes for each component as a function of wavelength.  For a Lorentzian profile, the amplitude of a profile normalized to an area of 1 is given by amplitude $a = 2 / (\pi FWHM)$.
As the FWHM at any given wavelength is set by the FWHM model built from the GO 3548 data, we can calculate the amplitude expected for a profile with that given FWHM and a total area under the curve of 1 MJy / steradian (the units per pixel of the MIRI spectral images). For each row, the flux in MJy for each component at the appropriate wavelength for the row will be the fitted amplitude for that component, divided by the amplitude of the normalized profile, and multiplied by the solid angle in steradians of one 0.11"$\times$0.11" MIRI pixel (2.23$\times$10$^{-13}$ steradians). Thus, after looping through all rows of each spectral 2-D image, we retrieve the spectra of both components as a function of wavelength.   

The Moffat profile realistically reproduces stellar PSFs so we use its 1-D form to disentangle the two spectral traces in the MIRI images
\begin{equation}
    P(x) = A\Bigl[\frac{(x-\mu)^2}{\sigma^2} + 1\Bigr]^{-\beta},
\end{equation}
in which $A$ is the peak amplitude, $\mu$ is the peak location, $\sigma$ and $\beta$ set the width and wings of the peak. In fitting the Moffat model to the W1049AB data, we assume the two peaks share the same shape ($\sigma$ and $\beta$) while differing in amplitude and position. To enhance the precision in extracting the time series spectra, we further assume 1. the two spectral traces are strictly aligned with the $y$-axis and their peak positions ($\mu_A$, $\mu_B$) do not change with time; 2. $\sigma$ and $\beta$ are also temporally invariant thanks to the stability of JWST in time series observations; and 3. $\sigma$ varies continuously and slowly with wavelength ($\lambda$), following a second-order polynomial function; As a result, the combined spectral cut of W1049 A and B can be written as

\begin{multline}
\label{eq:moffat}
    M(x, t, \lambda) = A_A(t, \lambda)\Bigl[\frac{(x-\mu_A)^2}{\sigma(\lambda)^2} + 1\Bigr]^{-\beta(\lambda)} +  \\ A_B(t, \lambda)\Bigl[\frac{(x-\mu_B)^2}{\sigma(\lambda)^2} + 1\Bigr]^{-\beta(\lambda)} + C(t, \lambda),
\end{multline}    
in which the new parameter $C(t, \lambda)$ is the time- and wavelength-dependent background.

Fitting starts with determining $\mu_A$, $mu_B$, $\sigma(\lambda)$ and $\beta(\lambda)$ using the median-combined frame. We first select a subregion of the median frame where the two traces are visibly separated ($y\in[387, 392]$), collapse the subregion in the $y$-direction, fit the profile with Equation~\ref{eq:moffat}, and find the best-fitting values of $\mu_A, \mu_B, A_A, A_B, \sigma, \beta$, and $C$. We then fit each spectral slice in the median frame with fixed $\mu_A, \mu_B$ and obtain wavelength-dependent $A_A, A_B, \sigma, \beta$, and $C$. The resulting series of $\sigma$ is smoothed by replacing them with the best-fitting, second-order polynomial between $\sigma(\lambda)$ and $\lambda$. This allows us to re-do the median frame fit with fixed $\sigma$ and determine the common $\beta(\lambda)$ shared by individual frames. Finally, we extract $A_A(t, \lambda)$, $A_B(t, \lambda)$, and $C(t, \lambda)$ by fitting Equation~\ref{eq:moffat} to individual frames and determine the flux density using the integral of the Moffat profile
\begin{equation}
\label{eq:integral}
    F = \int_{-\infty}^{\infty} A\Bigl[\frac{(x-\mu)^2}{\sigma^2} + 1\Bigr]^{-\beta} \mathrm{d}x = A\sigma\sqrt{\pi} \cdot\frac{\Gamma(\beta-0.5)}{\Gamma{(\beta)}},
\end{equation}
in which $\Gamma$ is the Gamma function.

\begin{figure*}
	\includegraphics[width=0.48\textwidth]{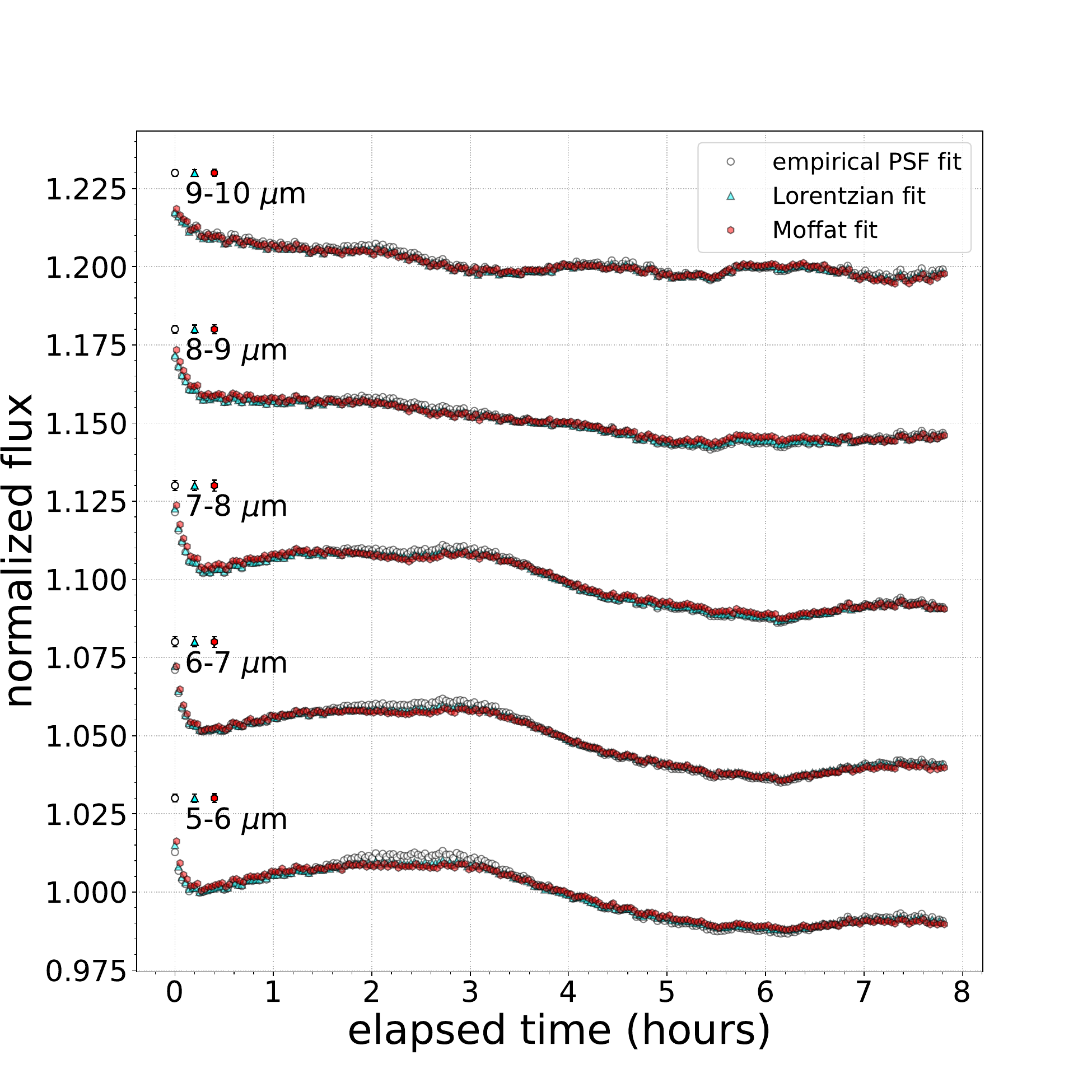}
     \includegraphics[width=0.48\textwidth]{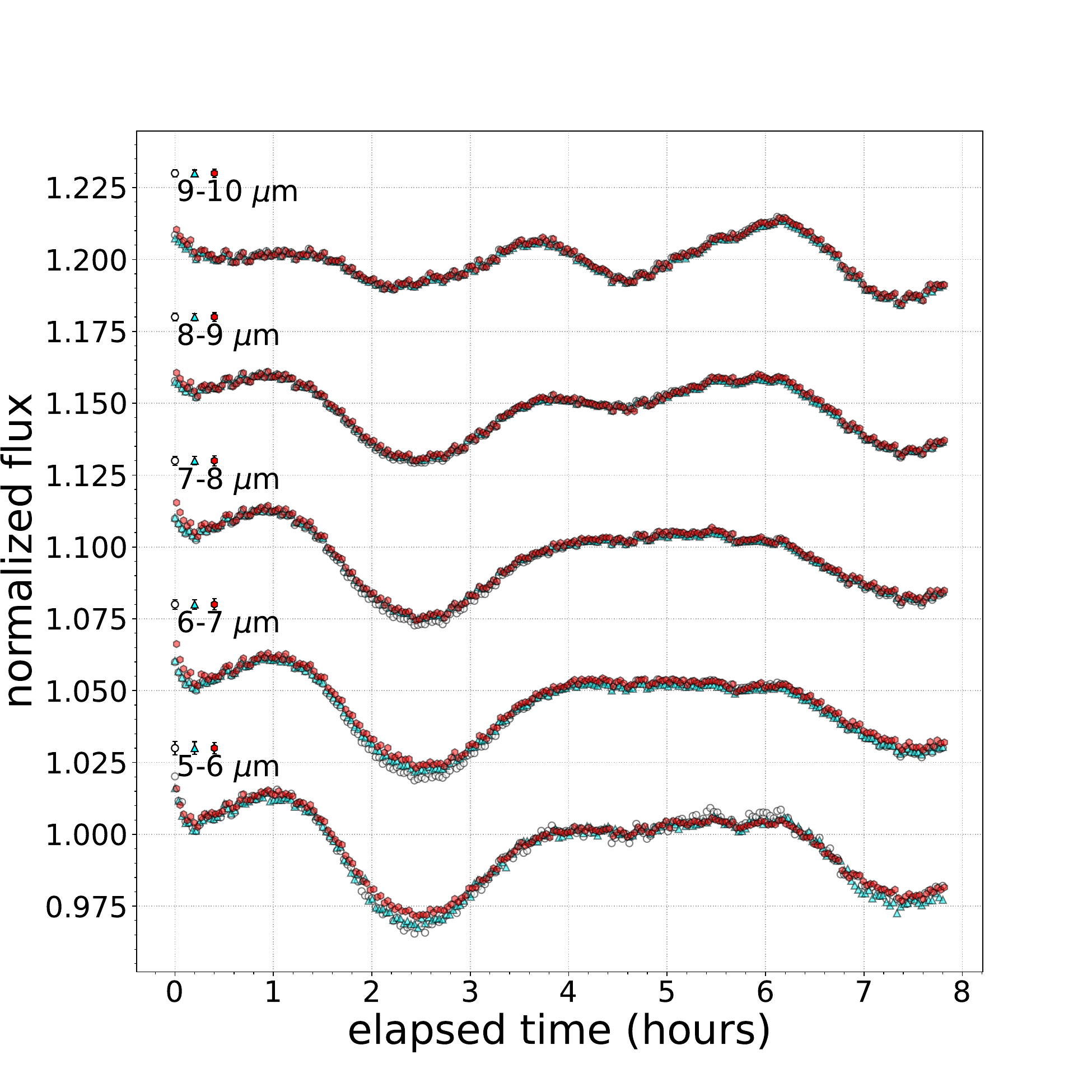}
    \caption{Sample lightcurves for WISE 1049AB extracted via the full PSF-fitting, Lorentzian, and Moffat-fitting methods for WISE 1049A (left) and WISE 1049B (right).  Lightcurves have been normalized to their median values, to highlight fractional variations.  Uncertainties for each lightcurve have been estimated using $\sigma_{pt}$, as defined in \citet{Radigan2014a}, specifically the standard deviation of the lightcurve subtracted from a version of itself shifted by one bin in time, divided by $\sqrt{2}$. The uncertainty for each lightcurve is shown as a plot symbol with an errorbar on the left side of the plot.}
    \label{fig:MIRI_lightcurve_method_comparison}
\end{figure*}

Fig.~\ref{fig:MIRI_lightcurve_method_comparison} shows sample lightcurves extracted using all three extraction methods for both components of the binary. To construct the lightcurves, we summed the flux across a given spectral bandwidth, taking into account the varying widths of each spectral bin. Lightcurves were binned by a factor of 10 in time, to yield a cadence of 129 s. Lightcurves have been normalized to their median values, to highlight fractional variations.  All extraction methods yield similar lightcurves within uncertainties, thus lending confidence in the robustness of the extractions.  For the rest of this work, we show results using the full PSF-fitting extraction method, as the Lorentzian method provides a robust fit to the PSF peak but tends to miss some flux in the wings of the PSF and the Moffat method is complicated to flux-calibrate.

\section{Results}

\subsection{Spectra}

\begin{figure*}
	\includegraphics[width=\textwidth]{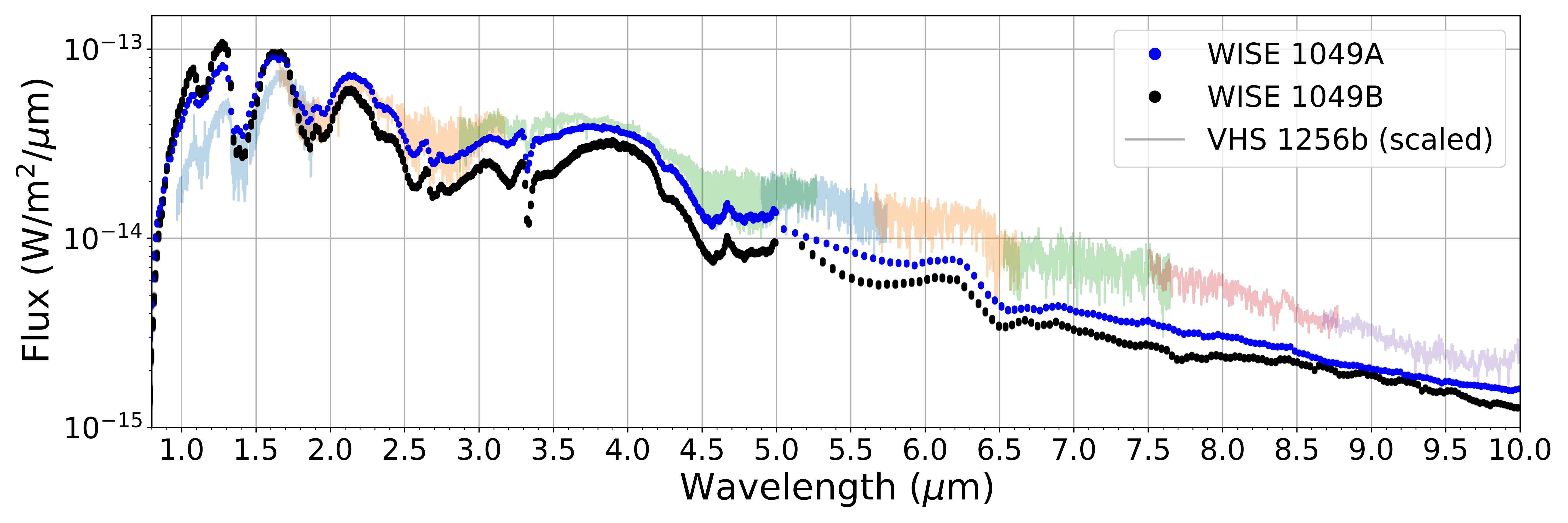}
    \caption{Full NIRSpec+MIRI spectra in $F_\lambda$ units.  WISE 1049A is plotted in blue while WISE 1049B is plotted in black.  The NIRSpec spectra have been binned to a cadence of 45 s for each spectral point shown while the MIRI observations have been binned to a cadence of 129 s for each spectral point shown;  the spread of points illustrate the full spectral variability of each component.  The NIRSpec IFU and MIRI MRS spectra for the young, L6/L7 planetary mass object VHS 1256-1257ABb \citep{Miles2023} is shown for comparison, multiplied by a factor of 100.}
    \label{fig:spectra}
\end{figure*}

Full NIRSpec + MIRI 0.8--10 $\mu$m spectra for WISE 1049AB are shown in Fig.~\ref{fig:spectra}, across the whole time-series observations. NIRSpec observations have been binned by a factor of 100 in time, for a resulting exposure time of 45 s for each spectral point shown, while MIRI observations have been binned by a factor of 10, for a resulting exposure time of 129 s.  We cut off the NIRSpec spectra at 5 $\mu$m, as at longer wavelengths, the spectra were severely affected by 1/f noise.  Due to the increased PSF FWHM at longer wavelengths, we were not successful at sufficiently deblending the MIRI spectral traces to extract absolute photometry beyond 10 $\mu$m, although we were able to extract reasonable relative photometry out to 11 $\mu$m.  Hence, for MIRI, we only show the spectra out to 10 $\mu$m.  We overplot as well NIRSpec IFU and MIRI MRS spectra for VHS 1256$-$1257ABb, a young L6/L7 planetary mass object observed as part of the JWST Early Release Science (ERS) observations \citep{Miles2023}.

\begin{figure*}     
     \includegraphics[width=\textwidth]{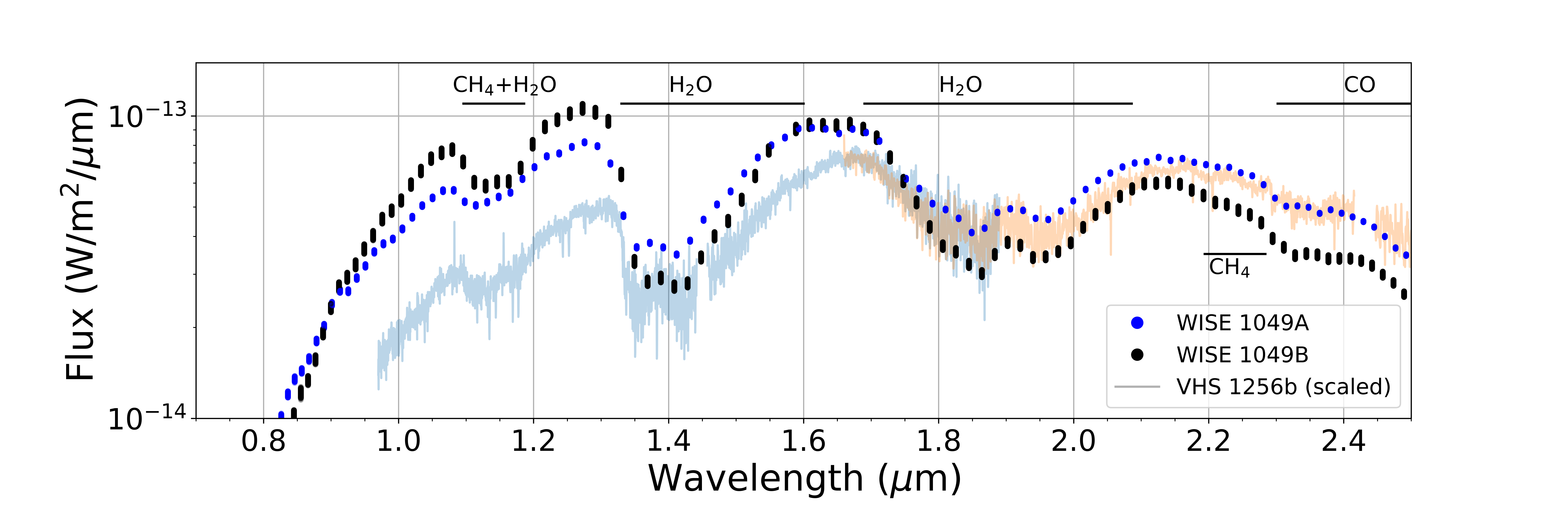}
     \includegraphics[width=\textwidth]{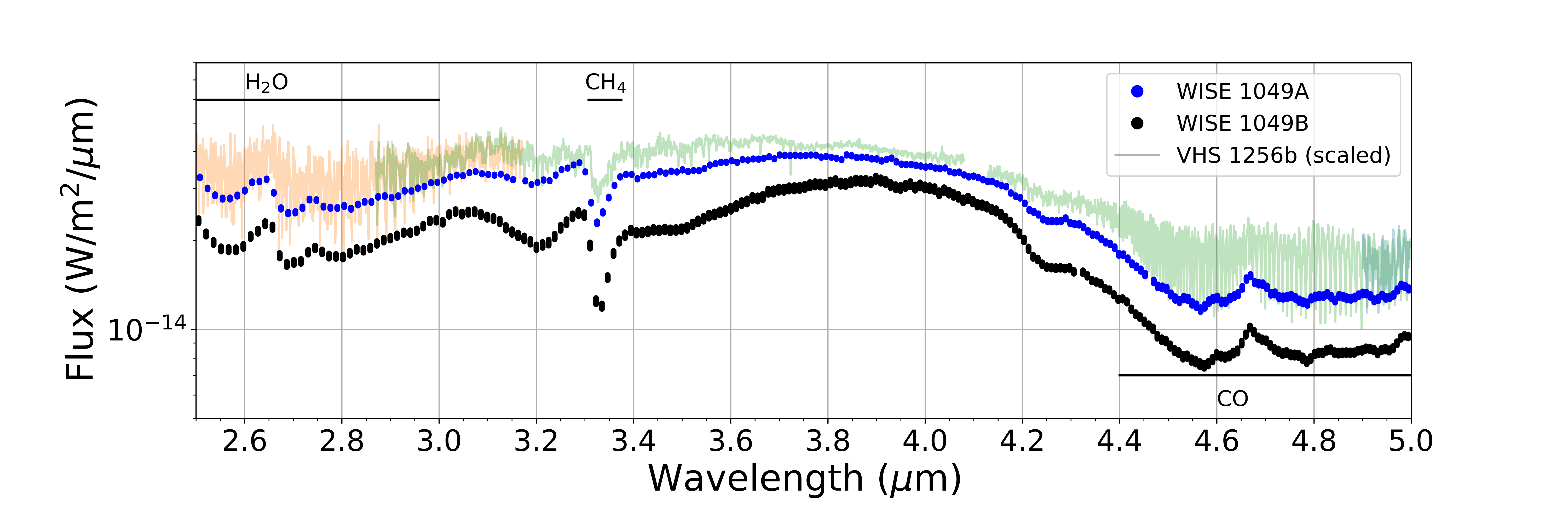}
    \caption{Zoomed in view of the NIRSpec spectra. WISE 1049A is plotted in blue while WISE 1049B is plotted in black.  The NIRSpec spectra have been binned to a cadence of 45 s for each spectral point;  the spread of points illustrate the full spectral variability of each component.  The NIRSpec IFU spectrum for the young, L6/L7 planetary mass object VHS 1256-1257ABb \citep{Miles2023} is shown for comparison, multiplied by a factor of 100.  Prominent absorption features from water, methane, and CO are evident for both components, as well as the fact that WISE 1049AB is a flux-reversal binary, with WISE 1049A only becoming clearly brighter than B at wavelengths $>$ 1.8 $\mu$m, despite the fact that WISE 1049A has an earlier spectral type and higher mass and luminosity than WISE 1049B.}
    \label{fig:NIRSpecspectra}
\end{figure*}

In Fig.~\ref{fig:NIRSpecspectra}, we provide a zoomed-in view of the NIRSpec spectra, again with VHS 1256-1257ABb provided as a comparison, with key absorption features labeled \citep{Miles2023, Faherty2014}.  Prominent water absorption features are evident in both components at 1.45 $\mu$m, 1.8 $\mu$m, and 2.6 $\mu$m.  Strong CO bandheads are also found in each component at 2.3 $\mu$m and from 4.4 $\mu$m to 5.0 $\mu$m (although the many lines found in each bandhead for VHS 1256-1257ABb using the NIRSpec IFU are blurred in the lower-resolution PRISM spectra of WISE 1049AB).  Both components of WISE 1049AB show methane absorption at 3.3 $\mu$m, but only WISE 1049B shows methane absorption at 2.2 $\mu$m.  At 1.1 $\mu$m, both components display an absorption feature in a region with degenerate methane and water absorption opacity.  WISE 1049B is clearly the more variable of the two, with a wider spread in flux of spectral points, especially at $\lambda <$ 1.8 $\mu$m and at $\lambda$ between 3.4 and 4.2 $\mu$m.  WISE 1049AB is also a well-known flux-reversal binary \citep{Faherty2014}, where WISE 1049B is brighter than A in the $J$ and $H$ bands, despite its later spectral type and lower mass.  This flux-reversal is quite evident in the JWST spectra, with WISE 1049B brighter than A at $\lambda <$ 1.3 $\mu$m, but with WISE 1049A clearly brighter than B at all wavelengths $>$ 1.8 $\mu$m.

\begin{figure*}        
     \includegraphics[width=\textwidth]{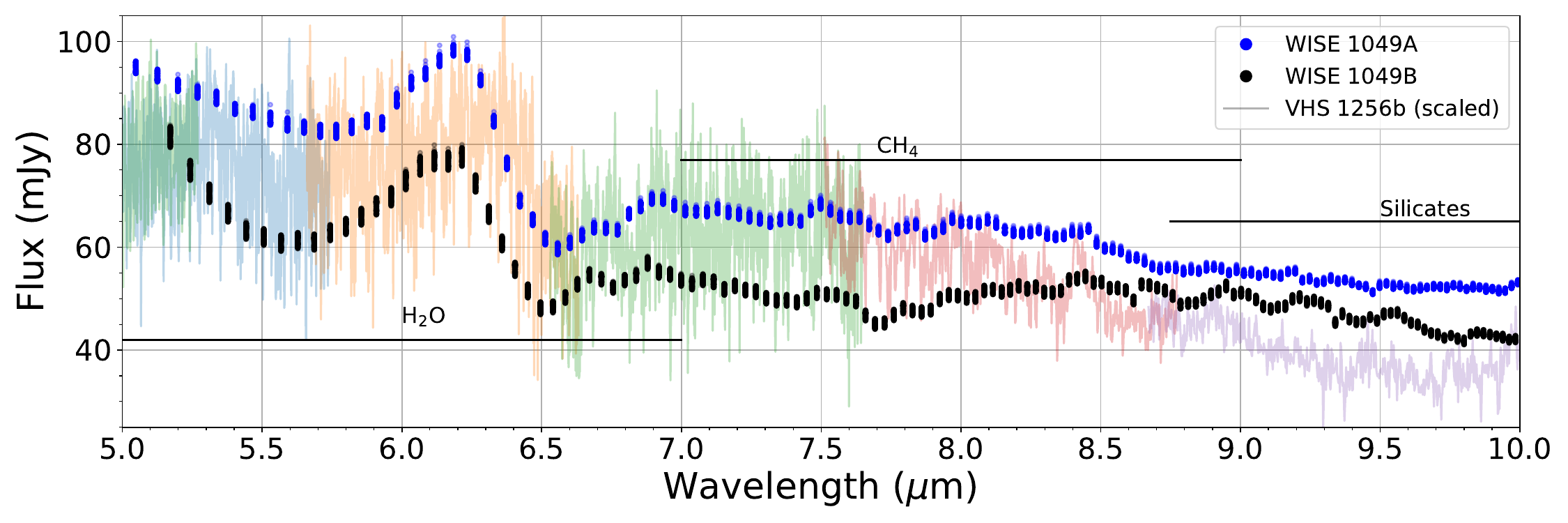}
    \caption{Zoomed in view of the MIRI spectra, in $F_{\nu}$ units (which better display the structure of the spectra at longer wavelengths). WISE 1049A is plotted in blue while WISE 1049B is plotted in black.  The MIRI observations have been binned to a cadence of 129 s for each spectral point shown;  the spread of points illustrates the full spectral variability of each component. The MIRI MRS spectrum for the young, L6/L7 planetary mass object VHS 1256-1257ABb \citep{Miles2023} is shown for comparison, multiplied by a factor of 50. Both components have prominent water absorption features around 6 $\mu$m and possible methane absorption at $\sim$7.7 $\mu$m; WISE 1049A also displays a tentative plateau feature in its spectra at $\lambda >$ 8.7 $\mu$m, attributed to small grain silicates for other mid-L dwarfs \citep{Cushing2006, Suarez2022, Miles2023}.}
    \label{fig:MIRIspectra}
\end{figure*}

In Fig.~\ref{fig:MIRIspectra}, we provide a zoomed-in view of the MIRI spectra, with key absorption features labeled.  We choose to display the MIRI spectra in $F_{\nu}$ units as this better displays the intrinsic structure in the spectra at longer wavelengths.  Prominent water absorption features are seen for both components around 6 $\mu$m, with some evidence for methane absorption at $\sim$7.7 $\mu$m.  WISE 1049A displays the flat plateau feature at wavelengths $>$8.5 $\mu$m seen in mid-L dwarfs and attributed to small grain silicates \citep{Cushing2006, Suarez2022, Miles2023}, while WISE 1049B has a somewhat steeper spectral slope at these wavelengths.  The wider spread of spectral fluxes for WISE 1049B confirms its higher variability compared to WISE 1049A. 
     
\subsection{NIRSpec lightcurves\label{sec:NSlc}}

\begin{figure*}
	\includegraphics[width=0.4\textwidth]{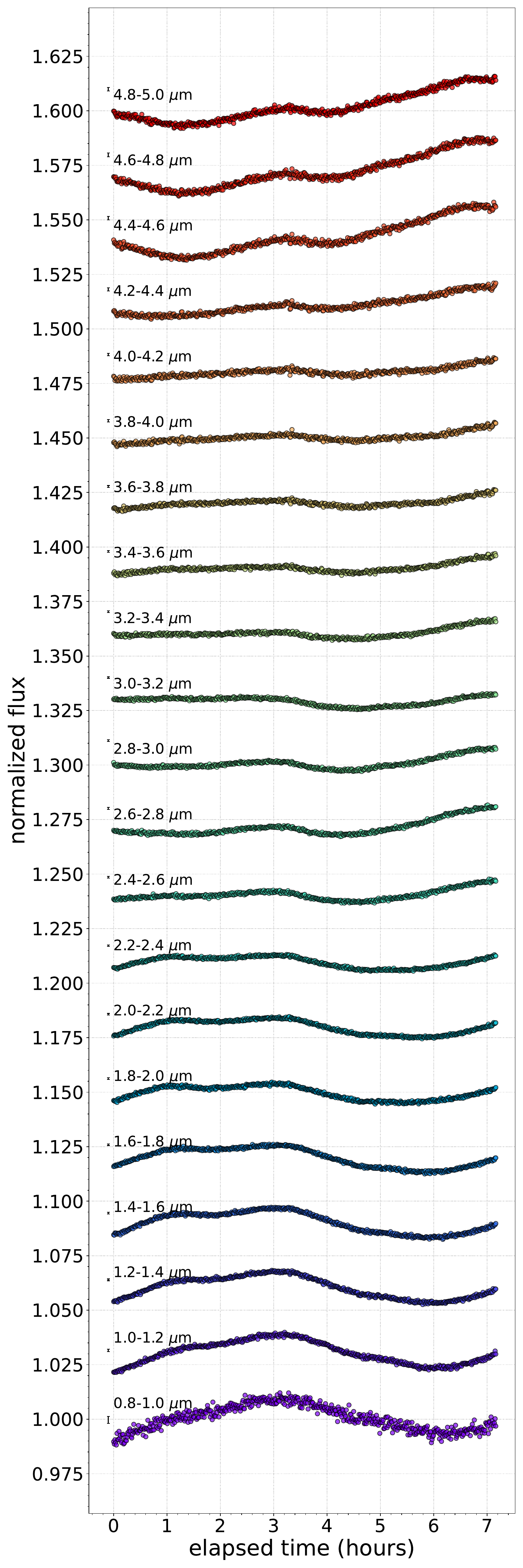}
     \includegraphics[width=0.4\textwidth]{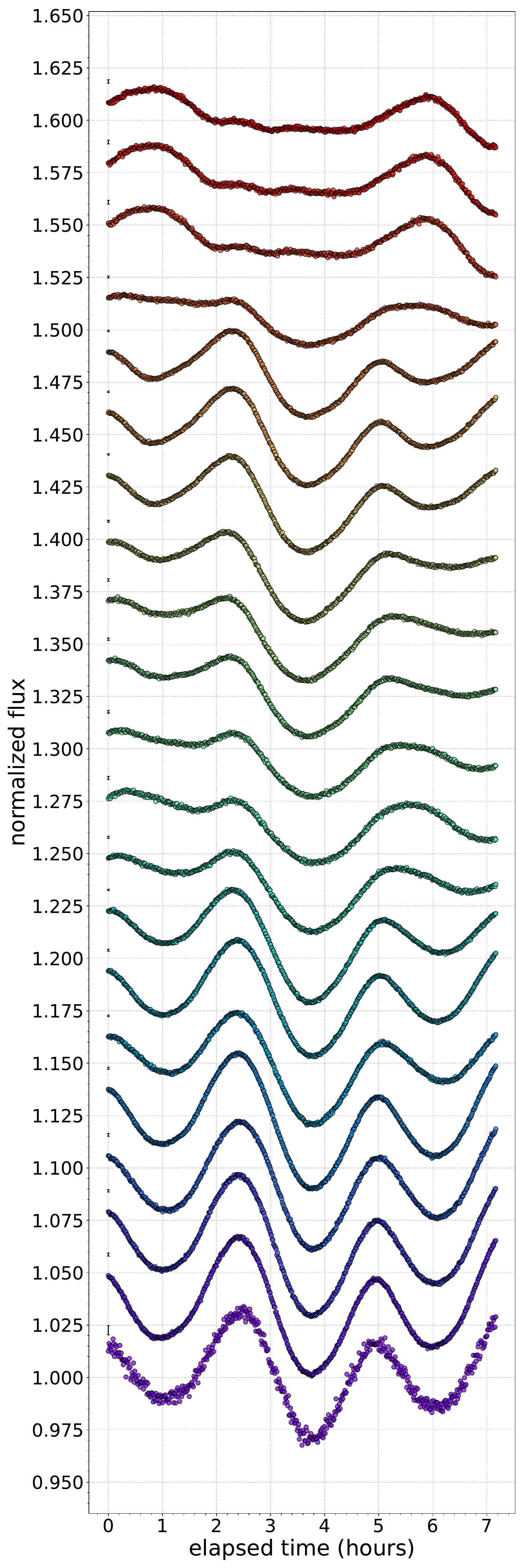}
    \caption{NIRSpec lightcurves for WISE 1049A (left) and WISE 1049B (right), with 0.2 $\mu$m bins in wavelength and a cadence of 45 s. Lightcurves have been normalized to their median values, to highlight fractional variations. A constant has been added to each lightcurve to visually separate them on the plot.  The uncertainty for each lightcurve is shown as an errorbar on the left side of the plot. }
    \label{fig:NIRSpec_lightcurves_spaced}
\end{figure*}

From the series of NIRSpec spectra, we produced  lightcurves for both binary components, binned in wavelength by 0.2 $\mu$m and combining every 100 integrations, for a cadence of 45 s.  We constructed the  lightcurves by summing the flux across a given spectral bandwidth, taking into account the varying widths of each raw wavelength spectral element. To highlight fractional variations, the lightcurves for each wavelength bin were normalized to their median values.   Lightcurves for WISE 1049AB are presented in Fig.~\ref{fig:NIRSpec_lightcurves_spaced}.  Uncertainties for each lightcurve have been estimated using $\sigma_{pt}$, as defined in \citet{Radigan2014a}, specifically the standard deviation of the lightcurve subtracted from a version of itself shifted by one bin in time, divided by $\sqrt{2}$.  From 1--5 $\mu$m, uncertainties range from 0.04$\%$ to 0.09$\%$ per wavelength bin for WISE 1049A and from 0.04$\%$ to 0.1$\%$ per wavelength bin for WISE 1049B. We thus estimate that we are generally sensitive to variations down to the 0.1$\%$ level.    

Both components of the binary show significant variability, although, as in previous studies, WISE 1049B presents a higher amplitude of variability than WISE 1049A, as well as more variation in variability properties as a function of wavelength.  In particular, WISE 1049B displays a striking shift in variability properties around 4.2 $\mu$m.  Wavelengths shorter than 4.2 $\mu$m exhibit double-peaked, quasi-sinusoidal lightcurves, with successive maxima / minima spaced by $\sim$2.5 hours, half of the known $\sim$5 hour period of this object.  These lightcurves are qualitatively similar to stellar "double-dipping" lightcurves \citep{Basri2018}.  In contrast, the NIRSpec lightcurves for wavelengths longer than 4.2 $\mu$m transition back to single-peaked behavior, with two maxima separated by approximately the known rotation period of WISE 1049B and with a shallow trough between the maxima.

\begin{figure*}
	\includegraphics[width=0.49\textwidth]{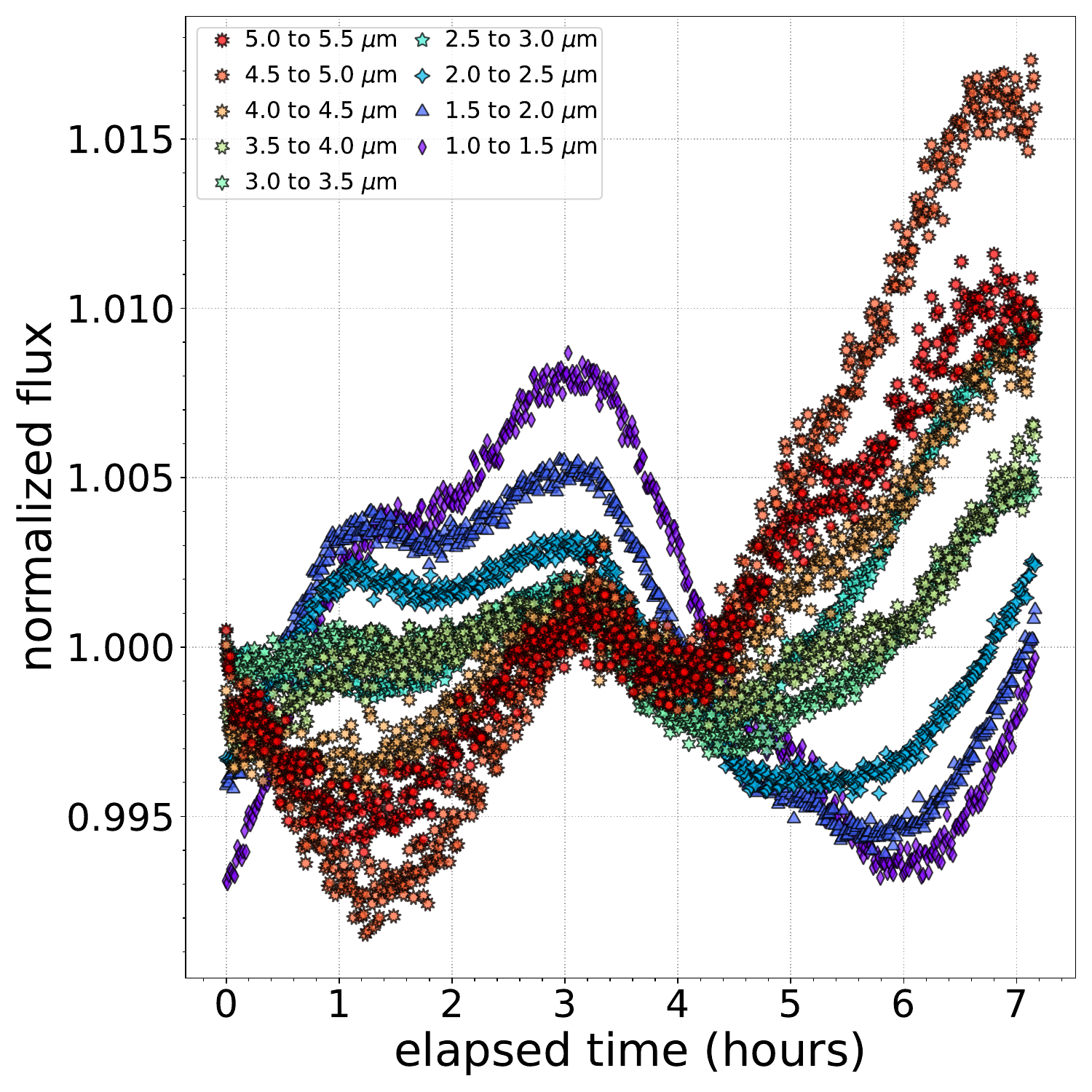}
     \includegraphics[width=0.49\textwidth]{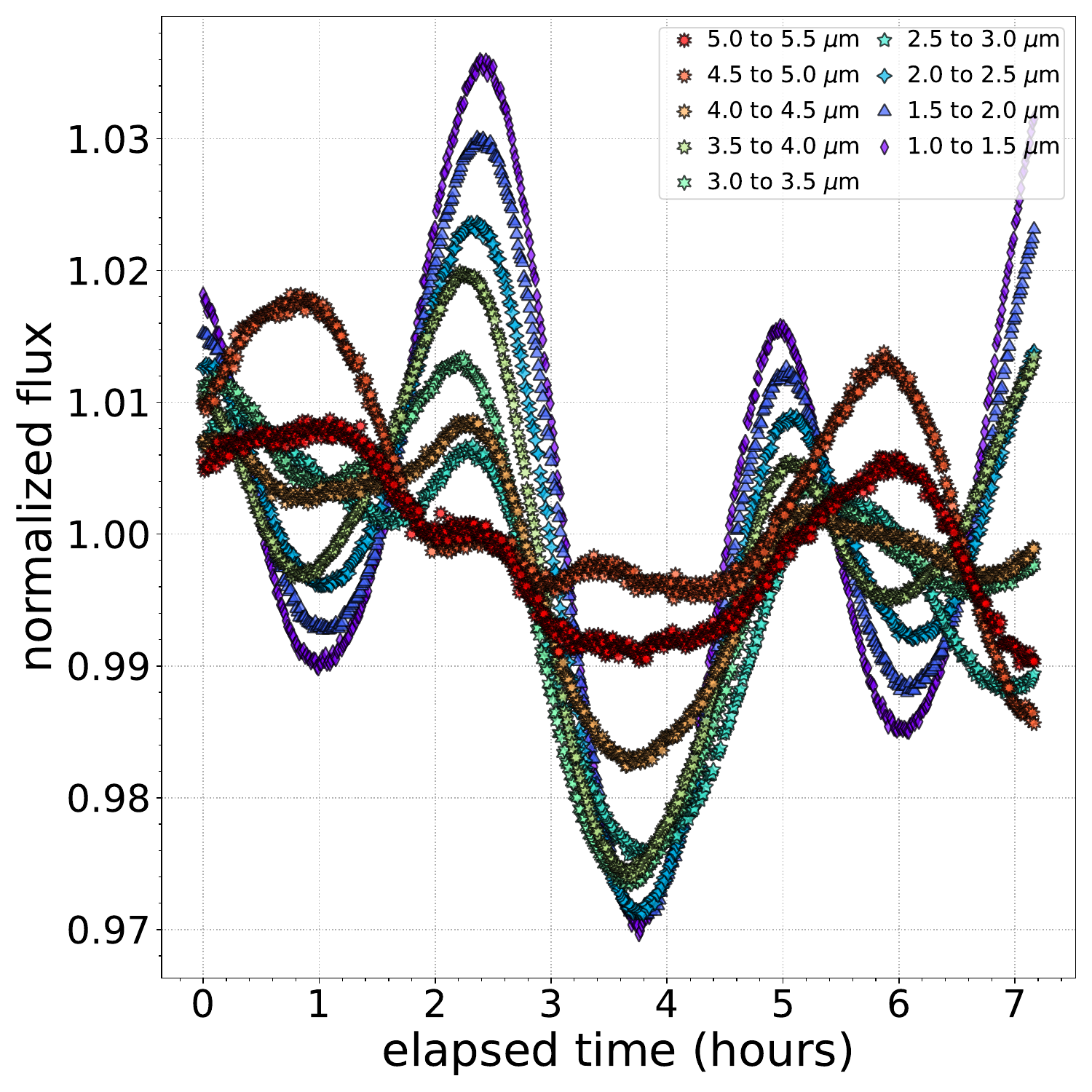}
    \caption{NIRSpec lightcurves for WISE 1049A (left) and WISE 1049B (right), with 0.5 $\mu$m bins in wavelength and cadence of 45 s. Lightcurves have been normalized to their median values, to highlight fractional variations, and are overplotted to highlight phase shifts / correlations / anti-correlations between wavelengths.}
    \label{fig:NIRSpec_lightcurves_overplotted}
\end{figure*}

In Fig.~\ref{fig:NIRSpec_lightcurves_spaced}, we have plotted the lightcurves with spaces between each curve, to highlight the amplitude and apparent period variation as a function of wavelength.  However, this makes it difficult to identify correlations / anti-correlations / phase shifts between different wavelengths.  To highlight differences in timing between maxima / minima as a function of wavelength, in Fig.~\ref{fig:NIRSpec_lightcurves_overplotted} we present  lightcurves in 0.5 $\mu$m bins overplotted as a function of time.  While previous variability studies have used the term "phase shift" to refer to apparent differences in rotational phase between wavelengths, this implies the presence of a base lightcurve behavior which does not evolve with time (in many cases, assumed to be sinusoidal).  The significant evolution of lightcurve morphology as a function of wavelength for WISE 1049AB demonstrates that these observations are not simply capturing a phase-shifted version of some base light curve, but rather, different intrinsic lightcurve shapes at different wavelength ranges, potentially corresponding to different shaping mechanisms as a function of depth / pressure in these atmospheres.  While our observations cover only a single period for WISE 1049A, we cover $\sim$1.4 periods for WISE 1049B, and it is already visually clear that the lightcurve structure changes significantly from period-to-period.  The lightcurves for WISE 1049B display a global minimum at $\sim$3.8 hours, but otherwise, the times of maxima / minima vary significantly as a function of wavelength, with maxima / minima at wavelengths $>$4 $\mu$m, often appearing anti-correlated or nearly anti-correlated ($\sim$5--10 degrees phase shift) from maxima / minima at shorter wavelengths. WISE 1049A also displays a distinct change in behavior at wavelengths $>$4 $\mu$m, although with $\leq$1 period of coverage, it is difficult to interpret these behavioral changes using the phase shift model.

\begin{figure*}
	\includegraphics[width=\textwidth]{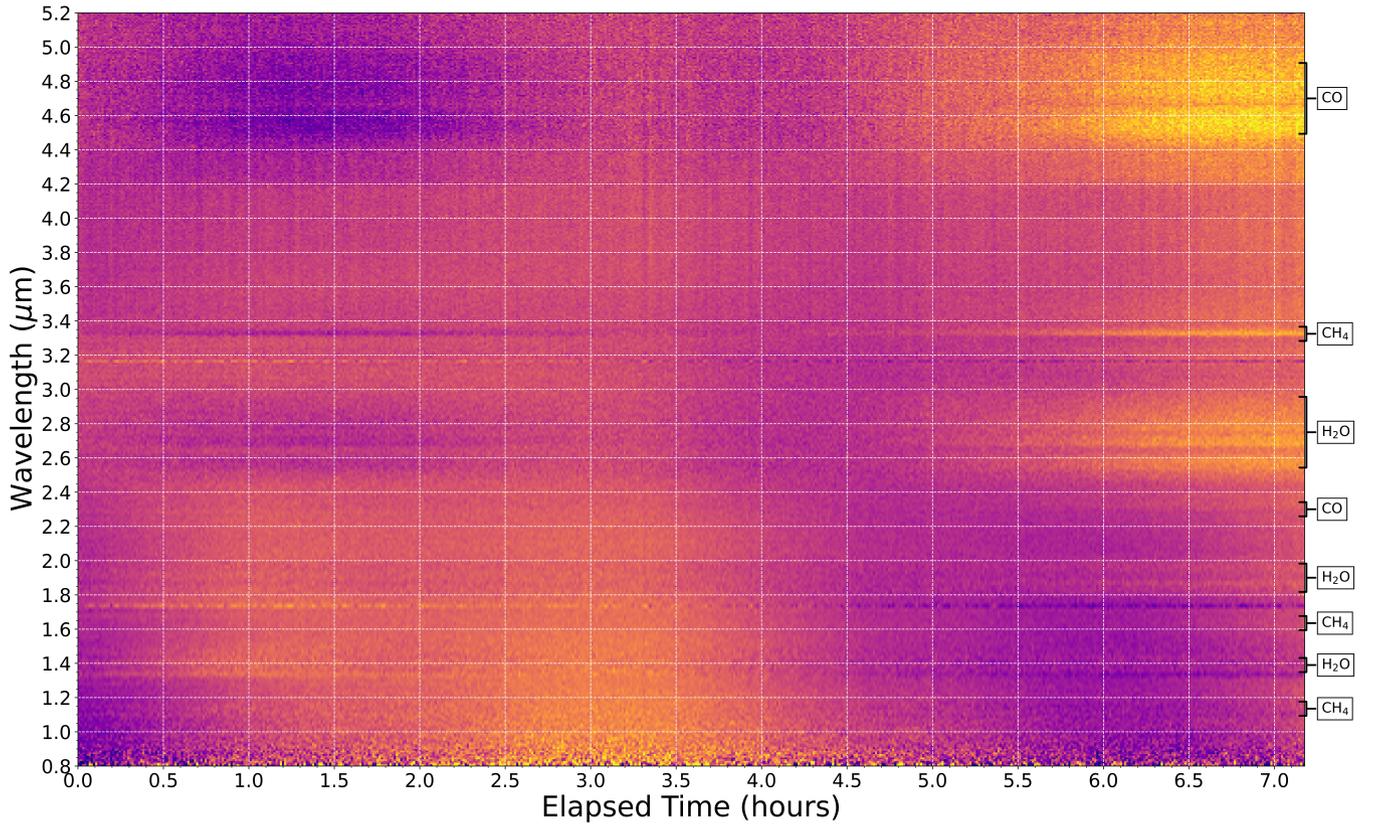}
    \caption{NIRSpec variability map for WISE 1049A, generated by producing median-normalized lightcurves for each wavelength bin in the NIRSpec spectra, then plotting as a function of time and wavelength.  The colour-bar ranges from fractional fluxes of 0.98 to 1.02, in order to highlight wavelength-dependent changes in amplitude, with darker colour representing lower fractional fluxes, and brighter colour representing higher fractional fluxes. We label the wavelength ranges where water, methane, and CO absorption features are present in the spectrum. }
    \label{fig:NIRSpec_WISE1049A_map}
\end{figure*}

\begin{figure*}
	\includegraphics[width=\textwidth]{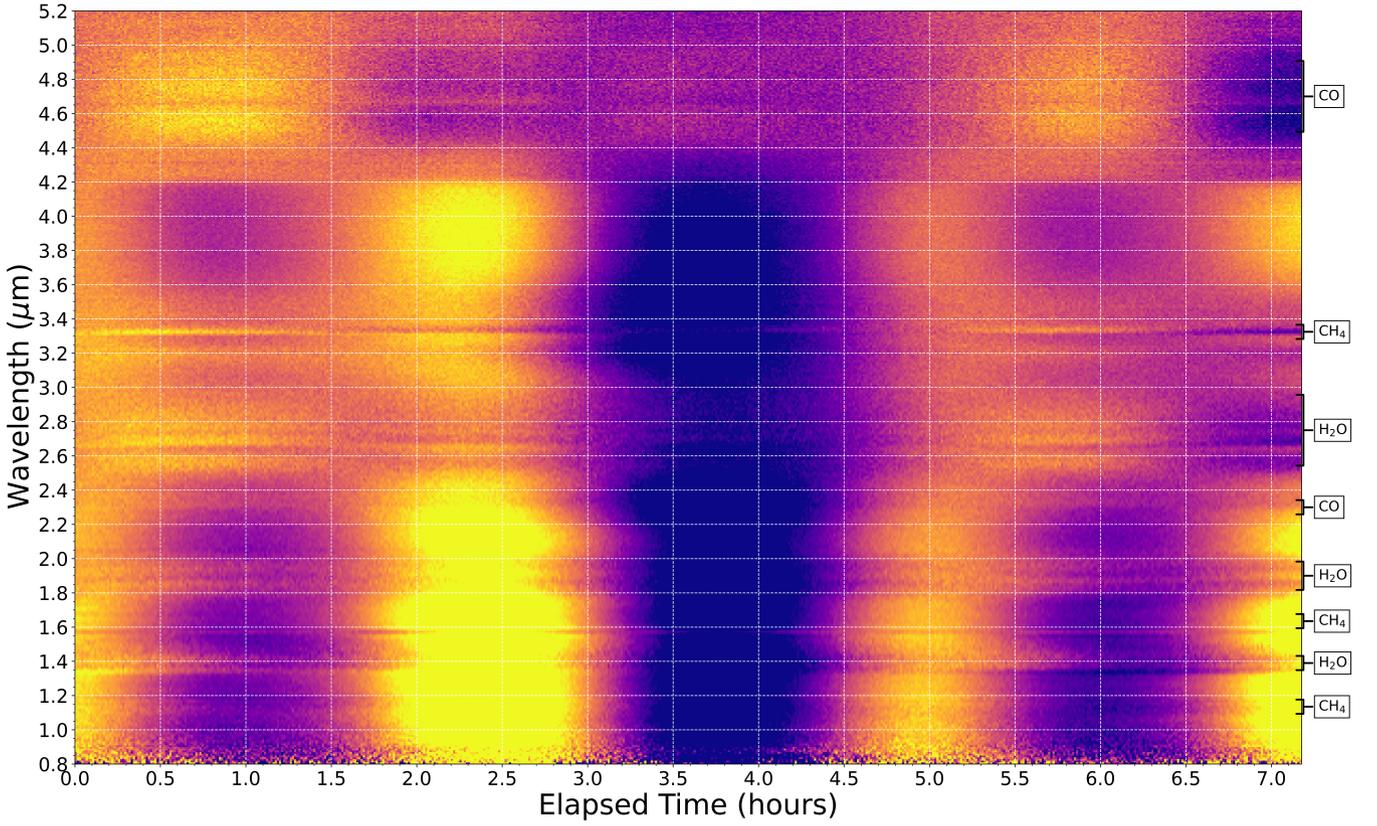}
    \caption{NIRSpec variability map for WISE 1049B, generated by producing median-normalized lightcurves for each wavelength bin in the NIRSpec spectra, then plotting as a function of time and wavelength.  The colour-bar ranges from fractional fluxes of 0.98 to 1.02, in order to highlight wavelength-dependent changes in amplitude, with darker colour representing lower fractional fluxes, and brighter colour representing higher fractional fluxes. We label the wavelength ranges where water, methane, and CO absorption features are present in the spectrum.}
    \label{fig:NIRSpec_WISE1049B_map}
\end{figure*}

From Fig.~\ref{fig:NIRSpec_lightcurves_spaced}, it is also clear that there is often significant variation in lightcurve shape over small changes of wavelength.  To highlight lightcurve evolution as a function of wavelength, we present variability maps for both components in Fig.~\ref{fig:NIRSpec_WISE1049A_map} for WISE 1049A and Fig.~\ref{fig:NIRSpec_WISE1049B_map} for WISE 1049B.  These maps were generated by producing an array of median-normalized lightcurves for every intrinsic wavelength bin in the NIRSpec spectra (resolution$\sim$30 at 1 $\mu$m to resolution$\sim$300 at 5 $\mu$m), with the same 45 s cadence as before.  To account for the changing spectral resolution across the NIRSpec bandpass, we rebin the lightcurve array onto a uniformly spaced wavelength grid of 0.01 $\mu$m per bin. 
We then plotted the lightcurve array as an image in time and wavelength, with color-bar ranging from fractional fluxes of 0.98 to 1.02, in order to highlight changes in amplitude.  While the variability map for WISE 1049A generally presents smooth variation as a function of wavelength, the map for WISE 1049B shows some regions of very abrupt change with wavelength around wavelengths where molecular absorption is present in the spectrum.  For instance, between 2.5 and 3 $\mu$m (water absorption) and around 3.35 $\mu$m (methane absorption), lightcurves in adjacent wavelength bins appear to be anti-correlated, with one lightcurve reaching a maximum when the other reaches a minimum.  
Maps for both A and B show distinct regions of different lightcurve behavior as a function of wavelength, with clear breaks occurring around 2.4 $\mu$m and 4.2 $\mu$m for both binary components.  In Section~\ref{sec:pressure}, we discuss whether these changes in behavior correspond to different pressure levels (and hence depths) in the atmospheres of these objects.  It is clear, however, that the TOA structure for both objects would look distinctly different depending on the wavelength observed, likely with a double-spotted appearance at wavelengths with double-peaked lightcurves, but only a single dominant spot at wavelengths with single-peaked lightcurves.

\subsection{MIRI lightcurves\label{sec:mirilc}}

\begin{figure*}
	\includegraphics[width=0.4\textwidth]{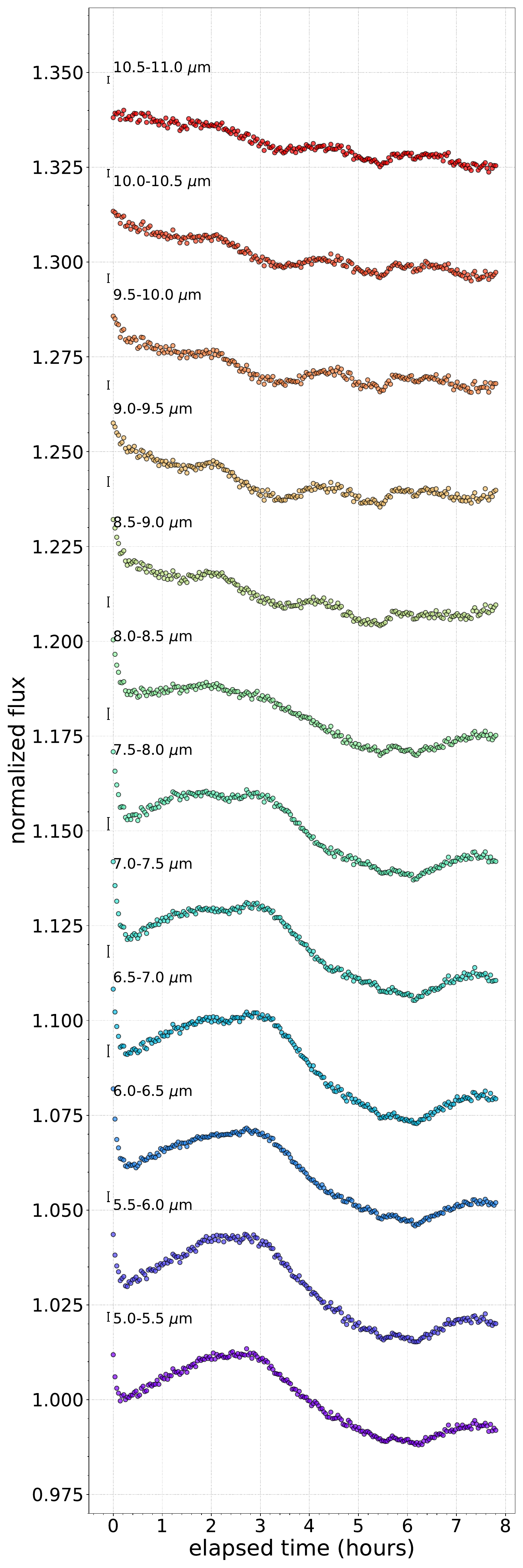}
     \includegraphics[width=0.4\textwidth]{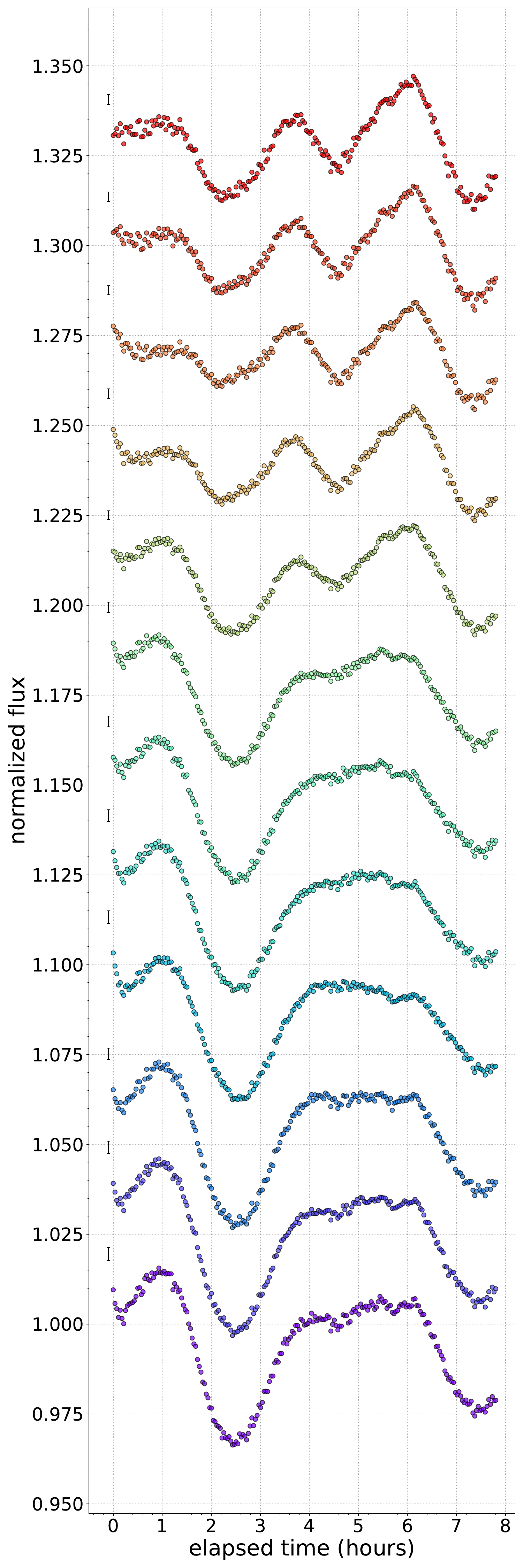}
    \caption{MIRI lightcurves for WISE 1049A (left) and WISE 1049B (right), with 0.5 $\mu$m bins in wavelength and a cadence of 129 s. Lightcurves have been normalized to their median values, to highlight fractional variations. A constant has been added to each lightcurve to visually separate them on the plot.  The uncertainty for each lightcurve is shown as an errorbar on the left side of the plot.  Lightcurves at some wavelengths show the downward ramp effect artifact \citep{Bell2023} in the first 20 minutes of the observation.}
    \label{fig:MIRI_lightcurves_spaced}
\end{figure*}

From the series of MIRI spectra, we produced lightcurves for both binary components, binned in wavelength by 0.5 $\mu$m and combining every 10 integrations, for a cadence of 129 s.  To highlight fractional variations, the lightcurves for each wavelength bin were normalized to their median values. Lightcurves for WISE 1049AB are presented in Fig.~\ref{fig:MIRI_lightcurves_spaced}, with uncertainties estimated like with NIRSpec as $\sigma_{pt}$.  From 5--11 $\mu$m, uncertainties range from 0.09$\%$ to 0.16$\%$ per wavelength bin for WISE 1049A and from 0.12$\%$ to 0.18$\%$ per wavelength bin for WISE 1049B. We thus estimate that we are generally sensitive to variations down to the 0.2$\%$ level.     

Lightcurves at some wavelengths display a steep downward trend in the first 15-20 minutes of the observation; this is the downward ramp effect reported by \citet{Bell2023}, but this effect appears to be negligible after the first half-hour of the observation.  Like in the NIRSpec lightcurves, the MIRI lightcurves for WISE 1049B transition as a function of wavelength between single peaked and double peaked behavior, with double-peaked behavior re-emerging at wavelengths $\geq$8.5 $\mu$m.  The lightcurves at wavelengths $>$8.5 $\mu$m appear qualitatively similar to the NIRSpec double-peaked lightcurves, with successive maxima / minima separated by about half of the rotation period of the object.  However, the MIRI peaks appear more saw-toothed than the peaks in the comparable short wavelength NIRSpec lightcurves and the degree of varation between successive peaks and troughs in the $>$8.5 $\mu$m MIRI lightcurves is also less than that in the comparable short wavelength NIRSpec lightcurves.  The WISE 1049A lightcurves also exhibit a notable change in shape around 8.5 $\mu$m. 

\begin{figure*}
	\includegraphics[width=0.49\textwidth]{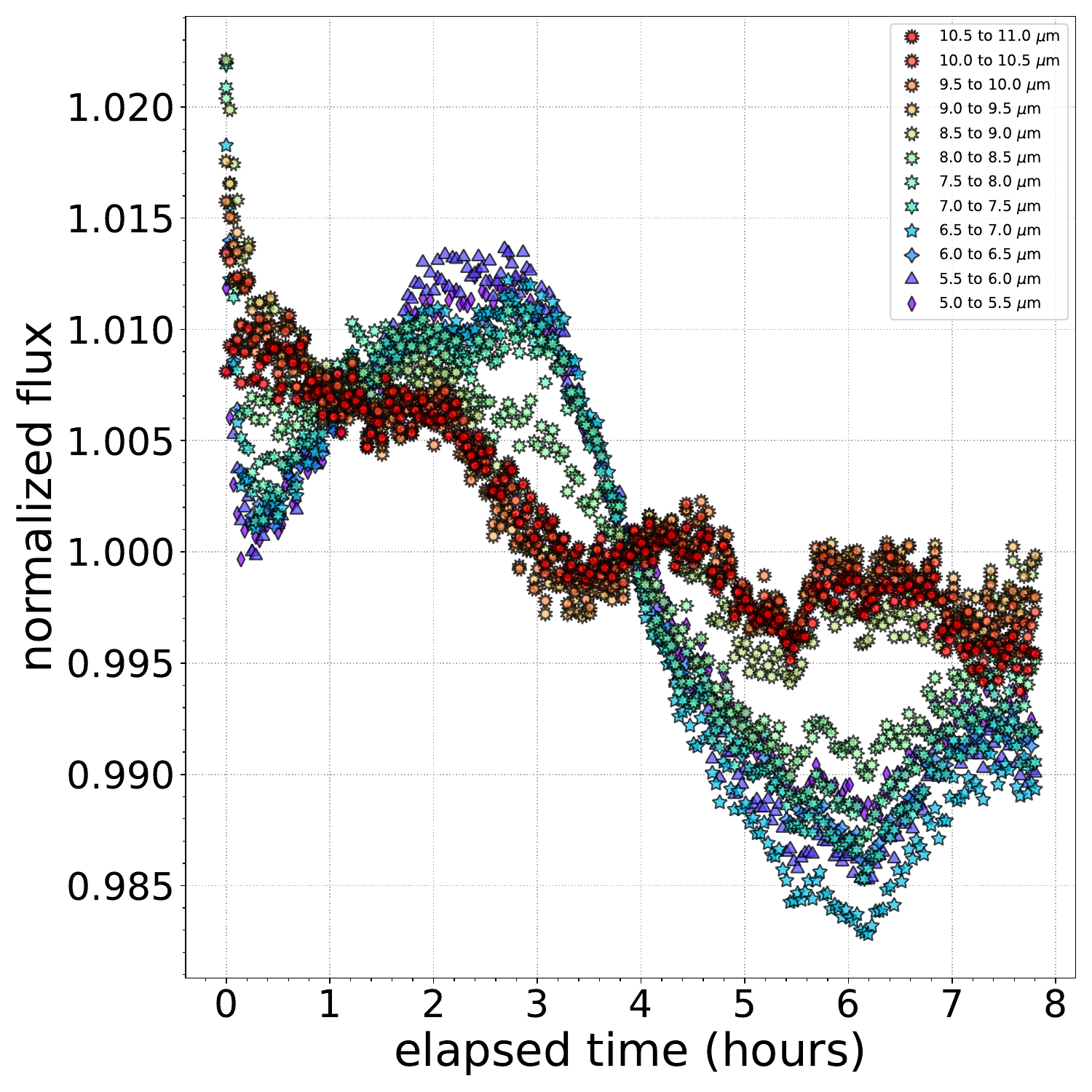}
     \includegraphics[width=0.49\textwidth]{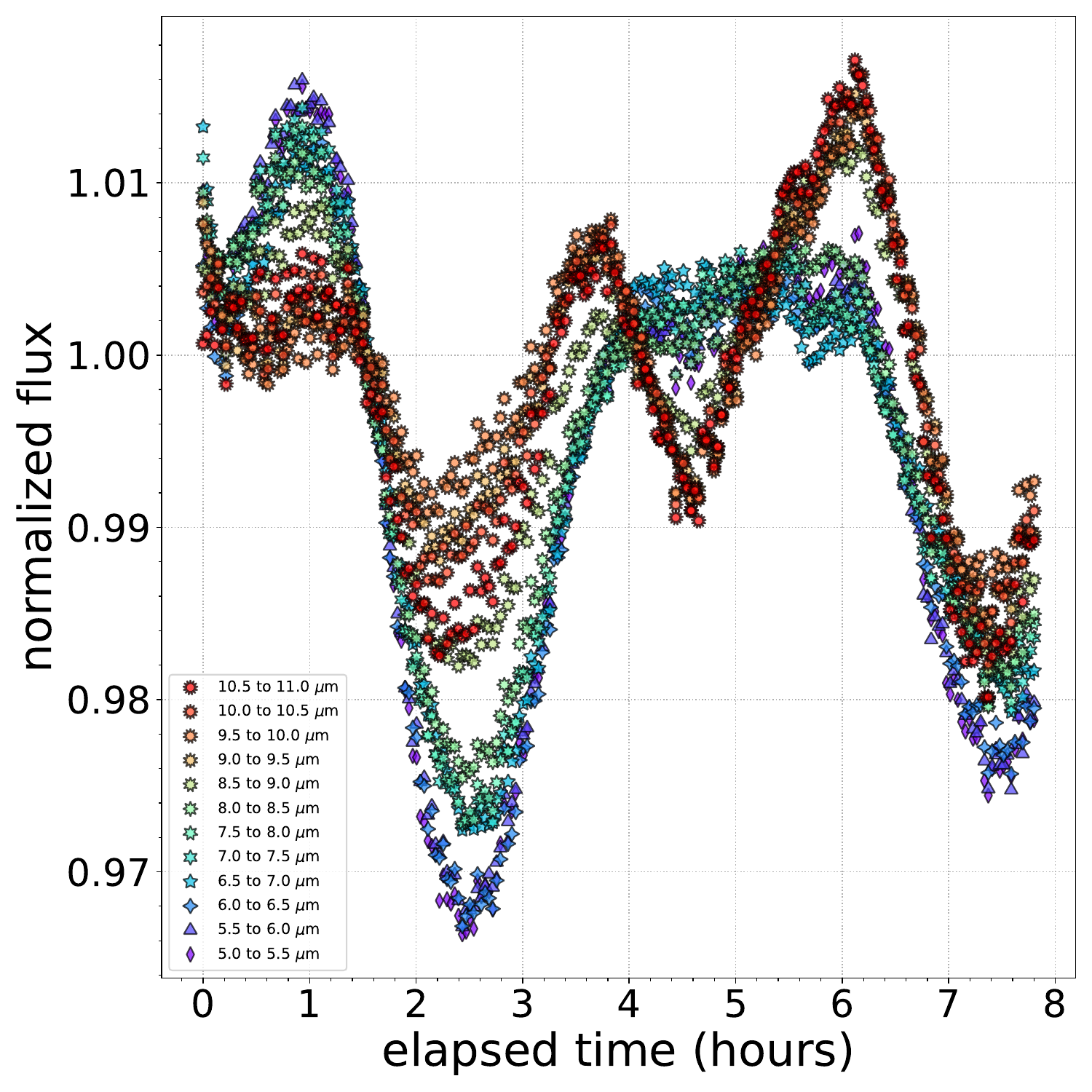}
    \caption{MIRI  lightcurves for WISE 1049A (left) and WISE 1049B (right), with 0.5 $\mu$m bins in wavelength and a cadence of 129 s. Lightcurves have been normalized to their median values, to highlight fractional variations, and are overplotted to highlight phase shifts / correlations / anti-correlations between wavelengths. Lightcurves at some wavelengths show the downward ramp effect artifact \citep{Bell2023} in the first 20 minutes of the observation.} 
    \label{fig:MIRI_lightcurves_overplotted}
\end{figure*}

To highlight phase shifts, in Fig.~\ref{fig:MIRI_lightcurves_overplotted} we present lightcurves in 0.5 $\mu$m bins overplotted as a function of time.  Lightcurves for both WISE 1049A and B exhibit striking changes in shape at wavelengths $\geq$8.5 $\mu$m. The transition in A is not well-described as a "phase shift", and appears as a general shift in lightcurve shape from a quasi-sinusoidal structure at shorter wavelengths, with a single peak and trough during the observation, to a downward slope at longer wavelengths.  WISE 1049B also displays two different lightcurve "behaviors", with a transition to double-peaked structure at 8.5 $\mu$m.  Despite the transition between single and double-peaked lightcurve shapes, the minima for B seem to be fairly well aligned in phase, while the two maxima at wavelengths $\geq$8.5 $\mu$m coincide with a plateau feature at shorter wavelengths.  

\begin{figure*}
	\includegraphics[width=\textwidth]{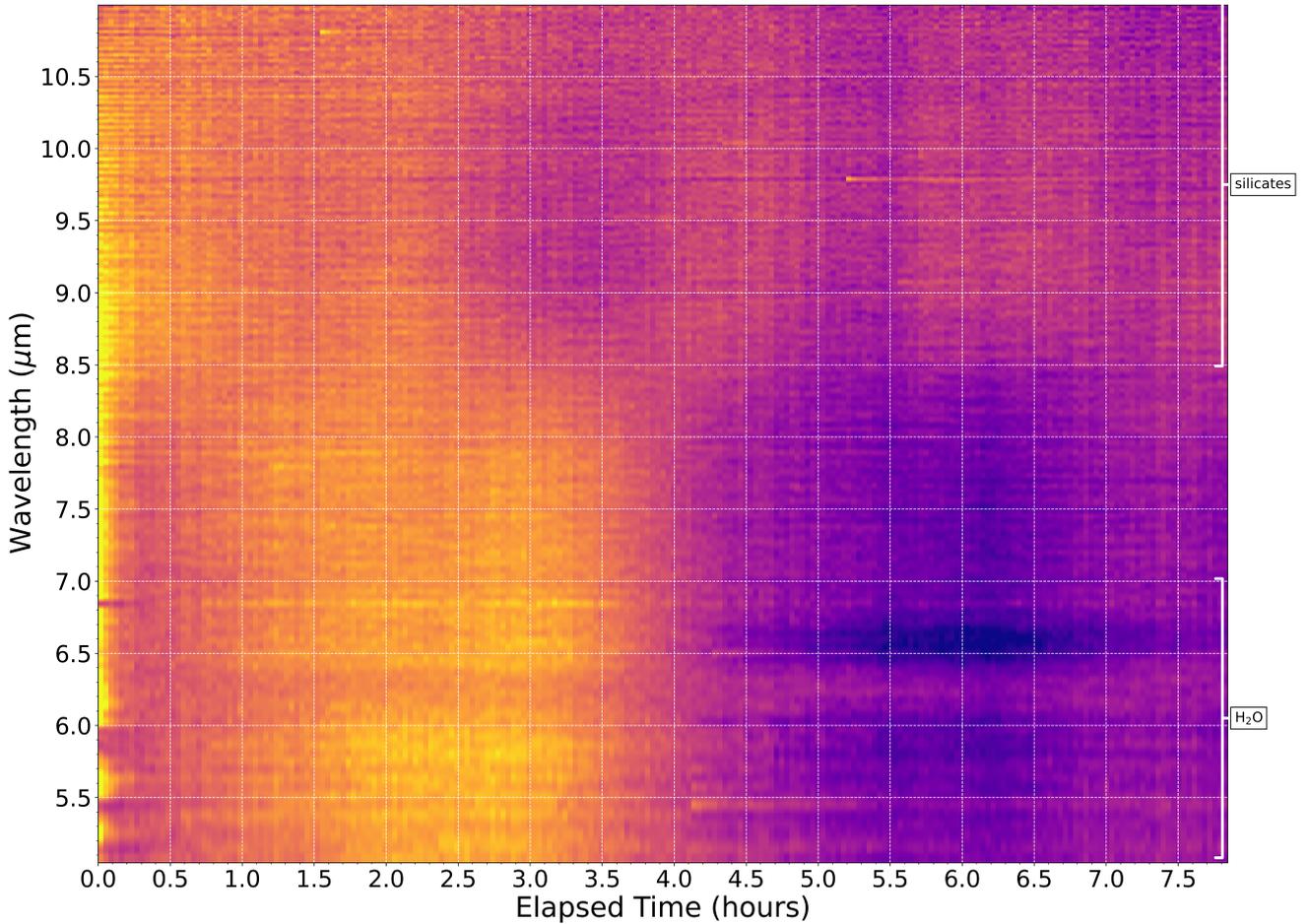}
    \caption{MIRI variability map for WISE 1049A, generated by producing median-normalized lightcurves for each wavelength bin in the NIRSpec spectra, then plotting as a function of time and wavelength.  The color-bar ranges from fractional fluxes of 0.98 to 1.02, in order to highlight wavelength-dependent changes in amplitude.  We label the wavelength ranges where water and small-grain silicate absorption features may be present. }
    \label{fig:MIRI_WISE1049A_map}
\end{figure*}

\begin{figure*}
	\includegraphics[width=\textwidth]{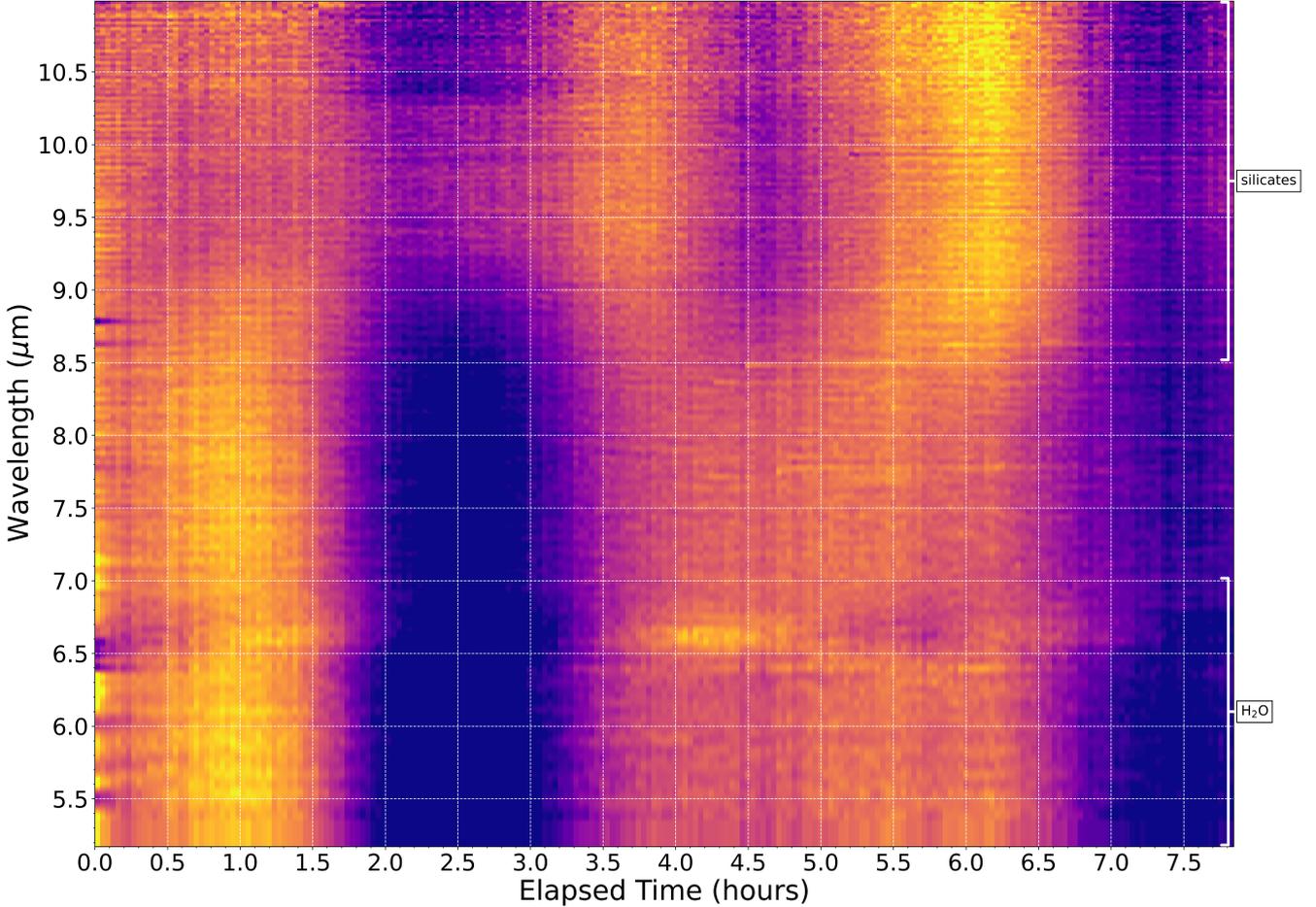}
    \caption{MIRI variability map for WISE 1049B, generated by producing median-normalized lightcurves for each wavelength bin in the NIRSpec spectra, then plotting as a function of time and wavelength.  The color-bar ranges from fractional fluxes of 0.98 to 1.02, in order to highlight wavelength-dependent changes in amplitude.  We label the wavelength ranges where water and small-grain silicate absorption features may be present.}
    \label{fig:MIRI_WISE1049B_map}
\end{figure*}

As we did for NIRSpec in Section \ref{sec:NSlc}, we construct variability maps for the MIRI data by producing median-normalized lightcurves for every intrinsic wavelength bin in the MIRI spectra and then rebinning onto a uniformly spaced wavelength grid of 0.01 $\mu$m per bin. 
The MIRI variability map for WISE 1049A is presented in Fig.~\ref{fig:MIRI_WISE1049A_map} and the MIRI variability map for WISE 1049B is presented in Fig.~\ref{fig:MIRI_WISE1049B_map}.  Both maps show some substructures between 6 -- 7 $\mu$m which may be connected to water absorption features at these wavelengths.  
For both WISE 1049A and B, there is an abrupt change in lightcurve properties around 8.5 $\mu$m, coincident with the onset of small grain silicate absorption at wavelengths $\geq$8.5 $\mu$m.  In Section~\ref{sec:tests}, we consider whether silicate absorption could be the driver of the change of variability properties seen in both objects at this wavelength.  There is also some substructure in the longer wavelength, double-peaked lightcurves for WISE 1049B -- the minima at these wavelengths appear shallower between 8.5 to 10 $\mu$m compared to wavelengths $\geq$10 $\mu$m.

\section{Analysis and Discussion}

\subsection{Period-by-period lightcurve evolution \label{sec:longtermlc}}

\begin{figure*}
	\includegraphics[width=\textwidth]{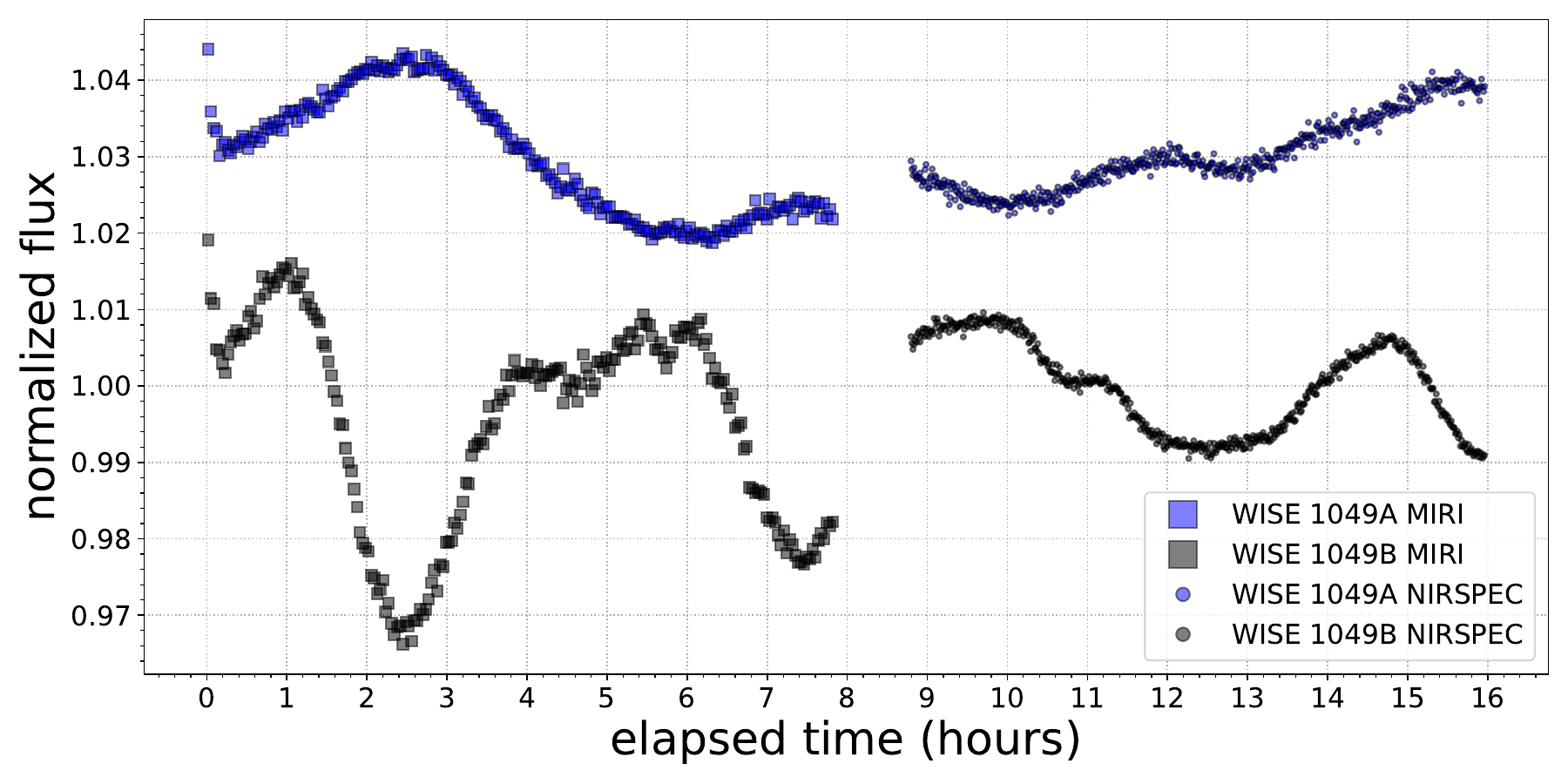}
    \caption{MIRI + NIRSpec lightcurves built from the median combination of individual channel lightcurve from 5--5.2 $\mu$m.  A constant has been added to the WISE 1049A MIRI and NIRSpec lightcurves to visually separate them on the plot from the WISE 1049B lightcurves.  The MIRI and NIRSpec lightcurves have both been normalized separately by their respective median values; there remains an unknown scaling factor between the two sets of observations.  These lightcurves cover more than 3 periods for B and likely $>$2 periods for A; period-to-period differences are evident in both binary components.}
    \label{fig:MIRINIRSpec_overlap_lightcurve}
\end{figure*}

In Fig.~\ref{fig:MIRINIRSpec_overlap_lightcurve}, we plot lightcurves spanning $\lambda$ between 5 to 5.2 $\mu$m, a region covered by both the MIRI and the NIRSpec observations.  While the NIRSpec spectrum is affected by 1/f noise at longer wavelengths, this affects only the absolute calibration (as the additional 1/f noise is not expected to vary significantly with time), hence, relative measurements such as lightcurves are still valid for NIRSpec at wavelengths between 5 and 5.5 $\mu$m.  However, given the issues calibrating absolute flux at the red edge of the NIRSpec wavelength range, we normalize the MIRI and NIRSpec lightcurves separately, hence, there is an unknown scaling factor between the two sets of observations.  The combined lightcurves cover nearly 17 hours in total, more than 3 periods for the B component, and likely $>$2 periods for A \citep[assuming~a~period~of~$
\sim$~7~hours,][]{Apai2021}.  The shapes of the lightcurves for both components differ significantly from period-to-period and the maximum change in flux between maxima and minima is larger for both components during the MIRI observations compared with the NIRSpec observation.  Similar significant changes over short timescales have been reported for WISE 1049B since the earliest monitoring campaigns for its variability \citep{Gillon2013, Smart2018, Apai2021}; WISE 1049A is comparatively less well-monitored, given its lower variability amplitude.  If the $\sim$7 hour period is correct, this is the first continuous, spatially-resolved, multi-period lightcurve for A.

\begin{figure*}
	\includegraphics[width=\textwidth]{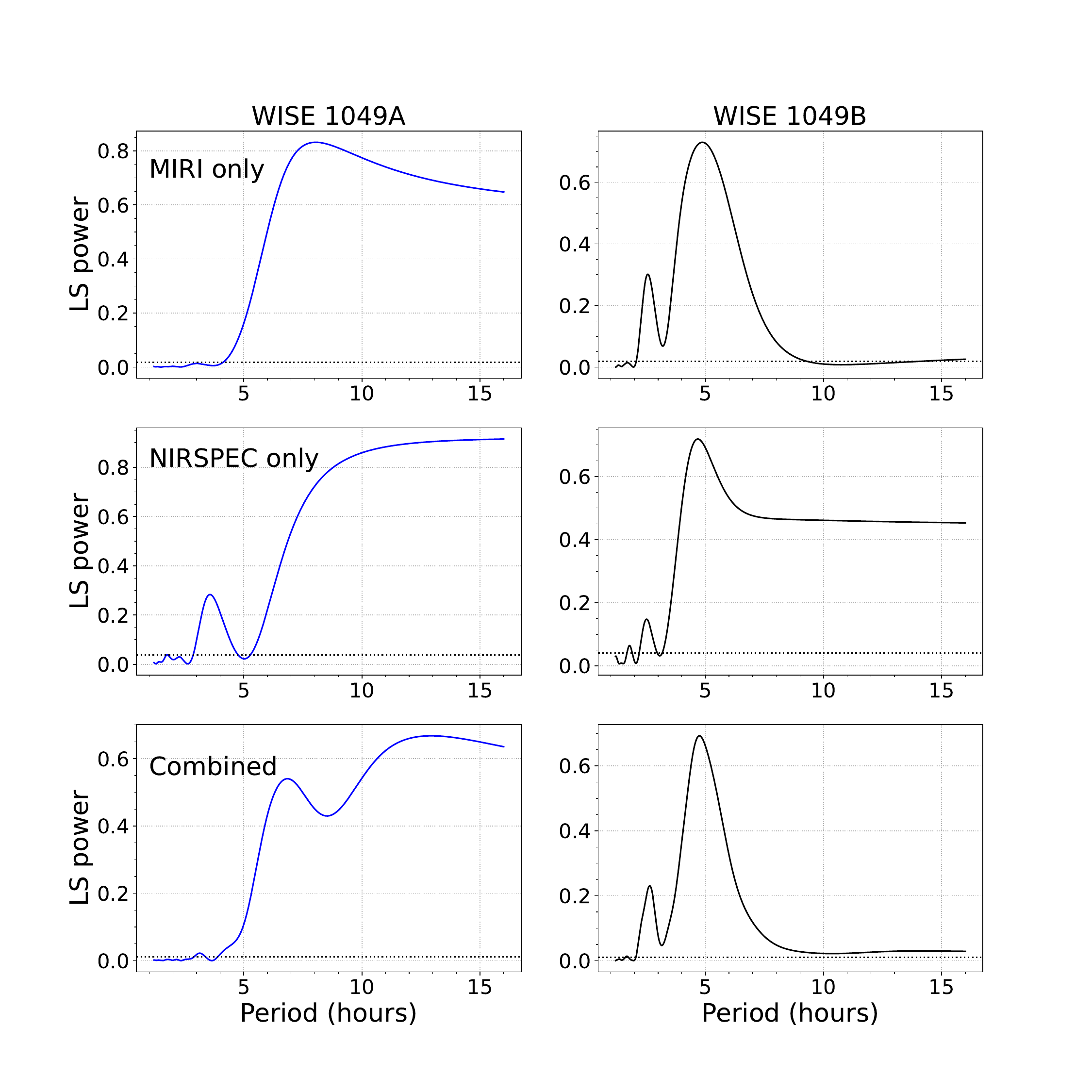}
    \caption{Lomb-Scargle periodograms for WISE 1049A (left column) and WISE 1049B (right column), considering periods up to 16 hours, and using the MIRI+NIRSpec lightcurves shown in Fig.~\ref{fig:MIRINIRSpec_overlap_lightcurve}.  The top row shows results using only the $\sim$8 hour MIRI observation, the middle row shows results using only the $\sim$7 hour NIRSpec observation, and the bottom row shows results using the combined lightcurve (and neglecting the unknown scaling factor between MIRI and NIRSpec lightcurves). The black dotted line shows the 99$\%$ false alarm probability power level; all periodogram peaks are clearly significant.  The peak power in the periodograms for WISE 1049A is consistent with the $\sim$7 hour period from \citet{Apai2021}, but with significant periodogram power remaining at longer periods as well.  The peak power in the periodograms for WISE 1049B is consistent with the $\sim$5 hour period for this object from the literature \citep{Gillon2013, Apai2021}.}
    \label{fig:MIRINIRSpec_overlap_periodogram}
\end{figure*}

With the caveat that our observations cover less than 3 rotation periods for either component and that both components of the binary have been known to show rapid evolution on short timescales, we used the Lomb-Scargle periodogram \citep{VanderPlas2018}, as implemented in \texttt{astropy}\footnote{https://docs.astropy.org/en/stable/timeseries/lombscargle.html}, to determine what periodicities may be present in our lightcurves.  Periodogram results are shown in Fig.~\ref{fig:MIRINIRSpec_overlap_periodogram}. We considered periods from 1.2 to 16 hours and calculated the 99$\%$ false alarm probability power using the built-in bootstrap method. A number of significant peaks that are well above the 99$\%$ false alarm probability power are found for each binary component.  The peak power in the periodograms for WISE 1049A are consistent with the $\sim$7 hour period from \citet{Apai2021}, but the periodograms still have significant power remaining at periods $>$ 10 hours.  Adopting a projected rotational velocity $v \sin i$ value of 17.6$\pm$0.1~km/s for WISE 1049A \citep{Crossfield2014}, and assuming a radius of $\sim$1 R$_{Jup}$ and an equator-on inclination (NB: the actual inclination may differ, but assuming an equator-on viewing angle provides the maximum possible period value) limits the  maximum period of WISE 1049A to $\leq$7 hours, suggesting that the periodogram power at periods $>$10 hours is a result of the relative shortness of these observations, covering only $\sim$2 periods for WISE 1049A. The most significant peak in power for WISE 1049B is found at the known period of $\sim$5 hours \citep{Gillon2013, Apai2021}, with a second significant peak at around 2.5 hours, indicating that some double-peaked structure may be present in the lightcurve.  

\subsection{Trends in "Amplitude" / Maximum Deviation as a function of wavelength\label{sec:maxdev}}

\begin{figure*}
 	\includegraphics[width=\textwidth]{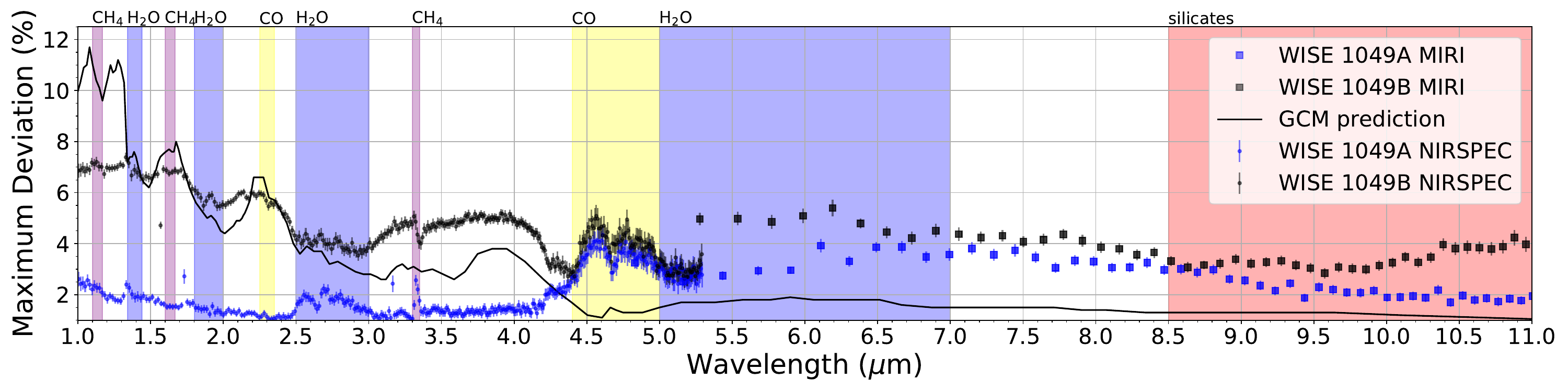}
    \caption{Wavelength vs. maximum deviation (percent variation between the highest peak and the deepest trough observed) for WISE 1049AB, derived from both MIRI and NIRSpec observations.  Shaded regions denote wavelengths of molecular and small-grain silicate absorption features. Uncertainties at each wavelength have been estimated using $\sigma_{pt}$ \citep{Radigan2014a}.  Predictions from a General Circulation Model (GCM) are shown as a black line and discussed in Sec.~\ref{sec:tests}}
    \label{fig:maxdeviation}
\end{figure*}

As the lightcurves for both WISE 1049A and B display non-sinusoidal changes between period-to-period, it is not fully appropriate to fit a sinusoidal amplitude or phase to our lightcurves.  Instead, we define the maximum deviation for each lightcurve as the percent variation between the highest peak and deepest trough observed over the course of each 7--8 hour MIRI or NIRSpec observation.  It is to be emphasized that this quantity will vary from period-to-period, and should not be taken as an inherent property of either binary component, especially considering the period-by-period changes discussed in Section~\ref{sec:longtermlc}.  Nonetheless, it is informative to investigate how maximum deviation varies as a function of wavelength.  Wavelength vs. maximum deviation is plotted in Fig.~\ref{fig:maxdeviation}. As with the lightcurves, uncertainties at each wavelength have been estimated using $\sigma_{pt}$, as defined in \citet{Radigan2014a}.

We suggest here interpretations for some of the features seen in both components' curves, however, we note that these interpretations should be considered preliminary, as future monitoring will be necessary to determine if any of these features persist over multi-period monitoring.  We find sharp changes in maximum deviation occurring over very narrow wavelength ranges for both components.  These are likely correlated with increased opacity in spectral absorption features, 
specifically, for both components around the 3.3 $\mu$m methane feature. Both display almost the same maximum deviation values from 4.5 - 5.0 $\mu$m, which may be related to the onset of the CO fundamental band at these wavelengths.  The maximum deviation decreases monotonically between 1 and 2.5 $\mu$m for B, but stays relatively more constant for A in the same wavelength region.  This agrees with previous results for WISE 1049B as well as other high-amplitude early T-dwarfs such as SIMP 0136 and 2M2139 \citep{Buenzli2015a, Apai2013}, where variability is found to peak around $J$ band and monotonically decrease across the $H$ and $K$ bands.  At longer wavelengths, the maximum deviation remains fairly constant for both components between 3.5 to 4 $\mu$m and between 5 to 6 $\mu$m.  We note that the maximum deviation at 5 $\mu$m for B measured with MIRI is quite a bit higher than that measured with NIRSpec, mirroring real changes in the morphology of the lightcurve from period-to-period, as discussed in Section~\ref{sec:longtermlc}.   The maximum deviation then decreases monotonically for A towards longer wavelengths, with B displaying increasing maximum deviation values at wavelengths $>$8.5 $\mu$m.

\subsection{How many different distinct lightcurve "behaviors" are there?\label{sec:clustering}}

The lightcurves for both WISE 1049A and B display complex, wavelength-dependent variations in behavior that are not possible to quantify using the standard definitions of phase and amplitude for a simple sinusoidal curve.  Thus, to quantify how many separate "classes" of lightcurves each of the components of the binary displays, we turned to machine-learning clustering algorithms.  We used K-means clustering, as implemented in the python 
\texttt{scikit-learn} package to delineate between different distinct classes of lightcurve behaviors.  

We started with the arrays of normalized single-wavelength-bin lightcurves displayed in Figures~\ref{fig:NIRSpec_WISE1049A_map}, \ref{fig:NIRSpec_WISE1049B_map}, \ref{fig:MIRI_WISE1049A_map}, and \ref{fig:MIRI_WISE1049B_map}.  The K-means algorithm considers the distance between different data points and centroids of a set number of $n_\mathrm{cluster}$ clusters.  For each instantiation of the algorithm, the K-means algorithm implemented in \texttt{scikit-learn} follows the following steps:

\begin{itemize}
\item Step 1: set number of centroids $n_\mathrm{cluster}$ to consider.  In the case of x, y positions, then the centroid would be a 2-d position; in the case of lightcurves, each centroid is an array with the same dimension as the input lightcurve.  
\item Step 2: choose $n_\mathrm{cluster}$ random centroids.  
\item Step 3: calculate the distance from each lightcurve as a function of wavelength from each centroid.  
\item Step 4: assign lightcurve for each wavelength to the closest centroid.  
\item Step 5: From the points assigned to each cluster centroid, calculate a new centroid value for each cluster.  
\end{itemize}

Using the new centroids, the algorithm then repeats steps 3 to 5 until the centroid positions no longer change.  For that instantiation, the quality of the clusters is determined by calculating the sum of the squared error ($SSE$), essentially the sum of all the distances between individual cluster points and their respective centroids.  This algorithm is non-deterministic, so the exact clusters chosen and $SSE$ will vary from instantiation to instantiation.  To choose the best set of clusters for each potential $n_{cluster}$ value, we then run 10 instantiations with different initial random centroid positions, then choose the instantiation with the lowest value of $SSE$.  

We must also decide how many clusters are appropriate to use for a given dataset.  As more clusters / centroids are added, the distance between each cluster member and centroid will decrease, eventually reaching the pathological case of $n_{cluster}$ = number of datapoints, and zero distance between each data point and its respective cluster centroid.  To find the appropriate number of clusters that balances the errors and numbers of clusters, we used the elbow point method.  In plotting $n_{clusters}$ vs $SSE$, the curve will first rapidly drop and then reach an "elbow", after which the curve decreases much less rapidly.  The $n_{cluster}$ value where this elbow occurs provides a reasonable trade-off between number of clusters and the $SSE$ for each k-means run.  To select the best elbow point and $n_{cluster}$ value, we ran 10 instantiations for $n_{cluster}$ running from 1 to 11, plotted $n_{cluster}$ vs. $SSE$, then used \texttt{KneeLocator} from \texttt{kneed}\footnote{https://pypi.org/project/kneed/} to select the elbow point and hence best $n_{cluster}$ value.

\begin{figure*}
 	\includegraphics[width=\textwidth]{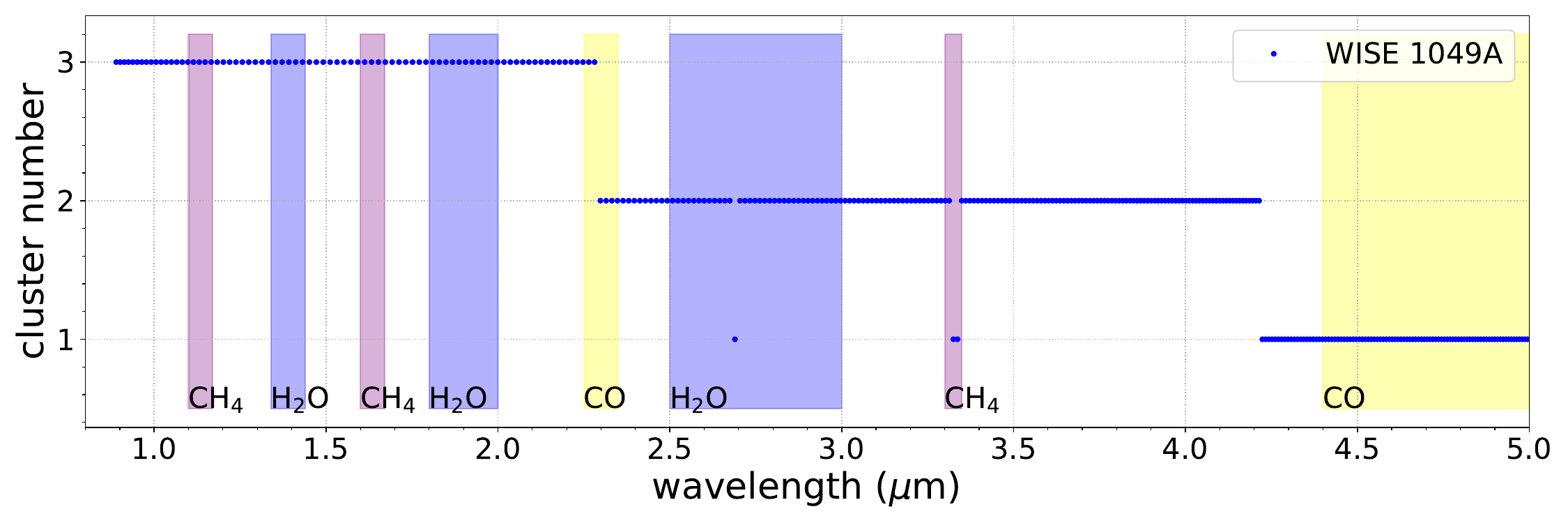}
        \includegraphics[width=\textwidth]{figures/fig18b.pdf}  
    \caption{Assigned cluster as a function of wavelength for WISE 1049A (top) and WISE 1049B (bottom) for NIRSpec lightcurves.  Shaded regions denote wavelengths of molecular absorption features.  Major transitions between lightcurve clusters occur around 2.3 $\mu$m and 4.2 $\mu$m in both binary components, with shifts between clusters also occurring in discrete wavelength ranges associated with molecular absorption features.  For both WISE 1049A and B, the lightcurves for the bluest wavelengths predominantly fall into one cluster, while the lightcurves for intermediate wavelengths form a second cluster, with a final cluster comprised mostly of the lightcurves from the reddest wavelengths.}
    \label{fig:NIRSpec_clustering_breaks}
\end{figure*}

\begin{figure*}
 	\includegraphics[width=0.45\textwidth]{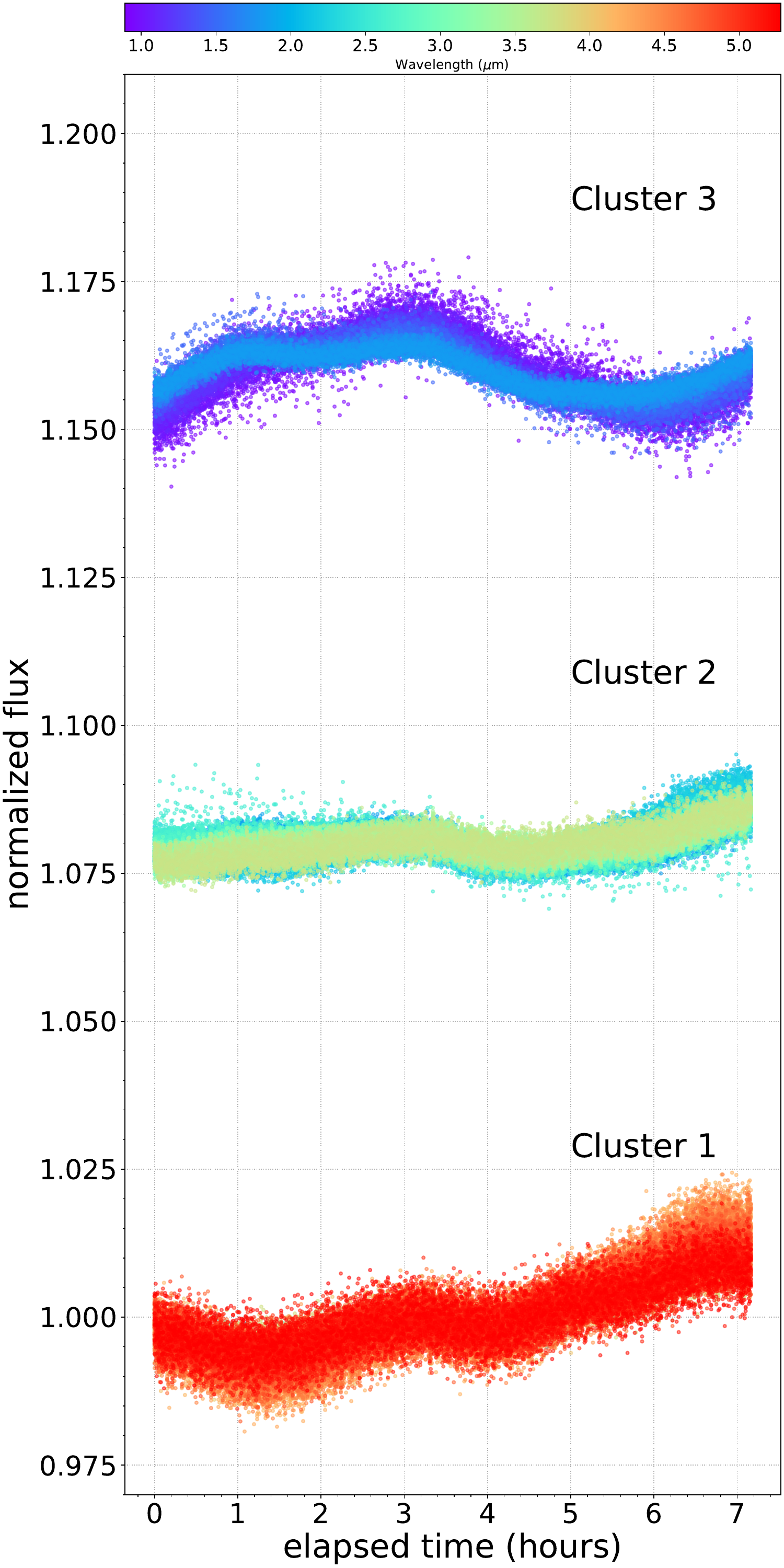} 
	\includegraphics[width=0.45\textwidth]{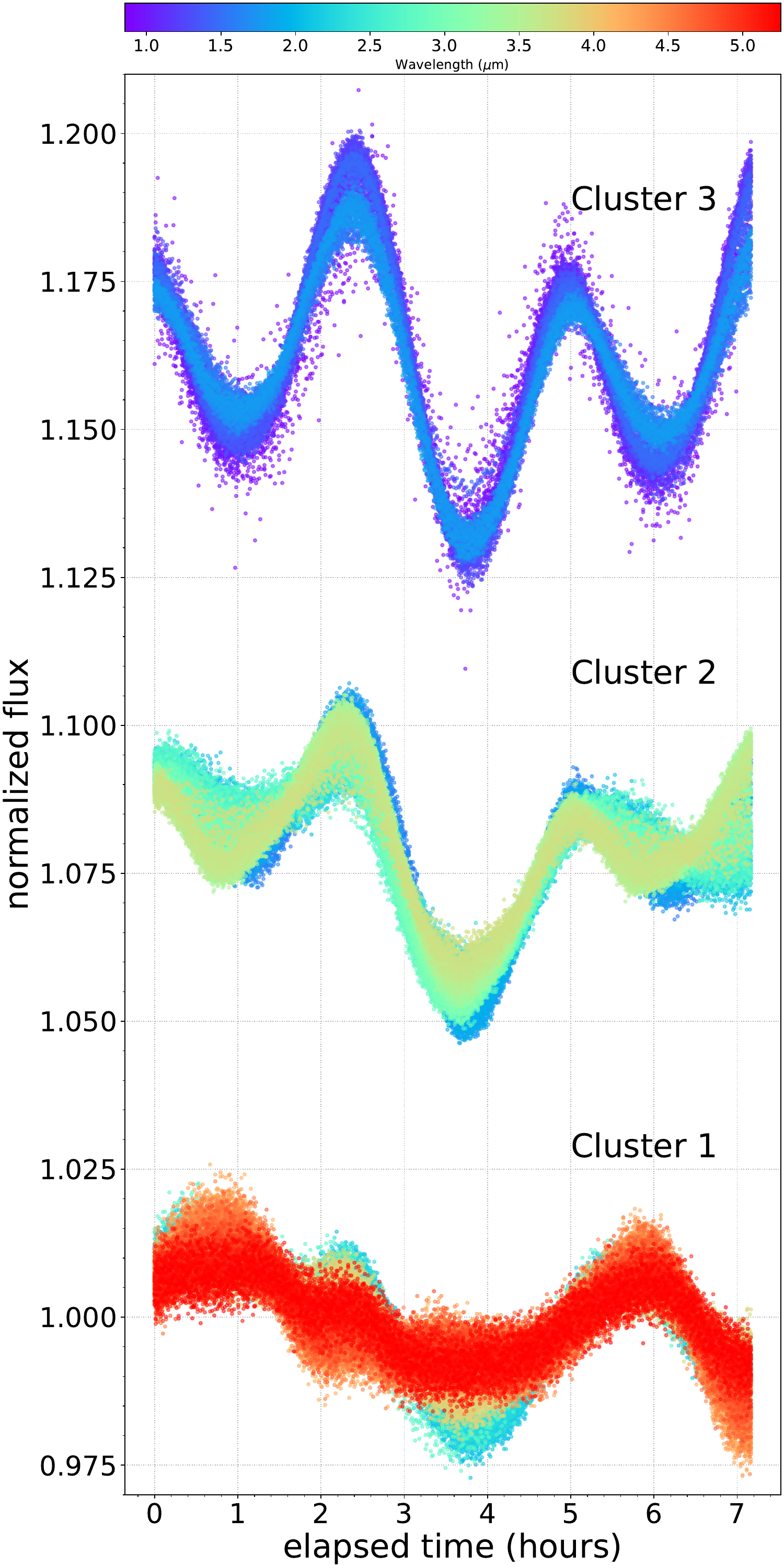} 
    \caption{NIRSpec single-channel lightcurves shown as a function of assigned cluster for WISE 1049A (left) and WISE 1049B (right).  Lightcurves have been binned to a cadence of 45 s and the wavelength for each lightcurve corresponds to the plot marker color.}
    \label{fig:NIRSpec_clustering_clusters}
\end{figure*}

\begin{figure*}
 	\includegraphics[width=\textwidth]{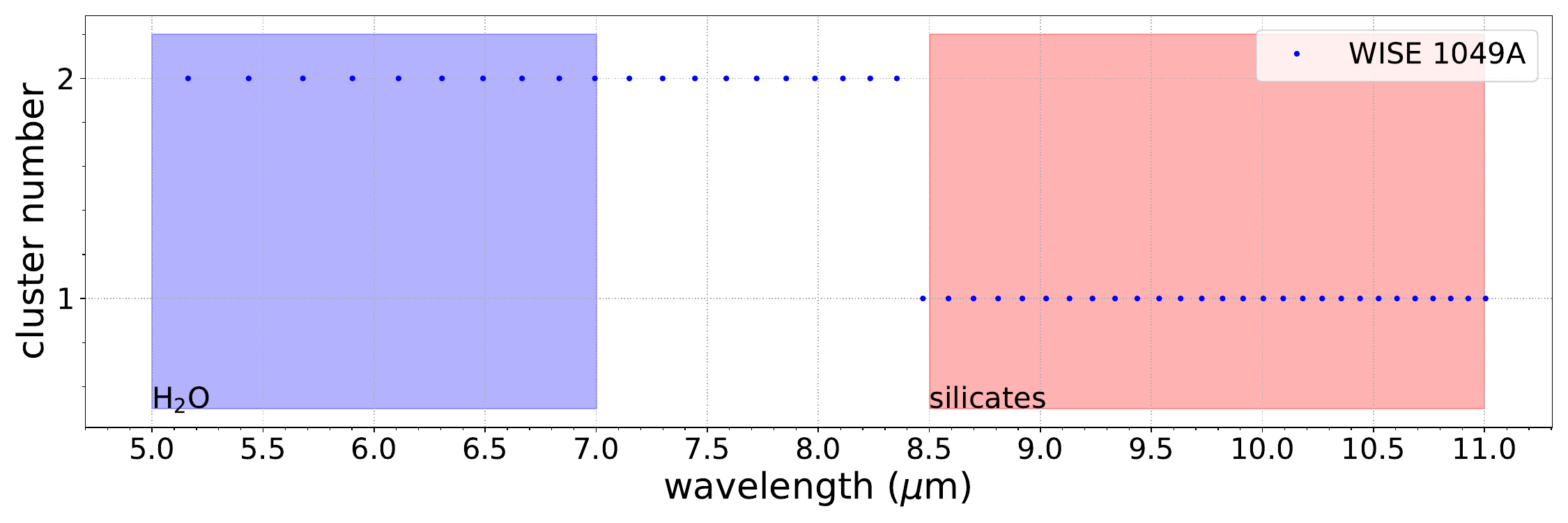}
 	\includegraphics[width=\textwidth]{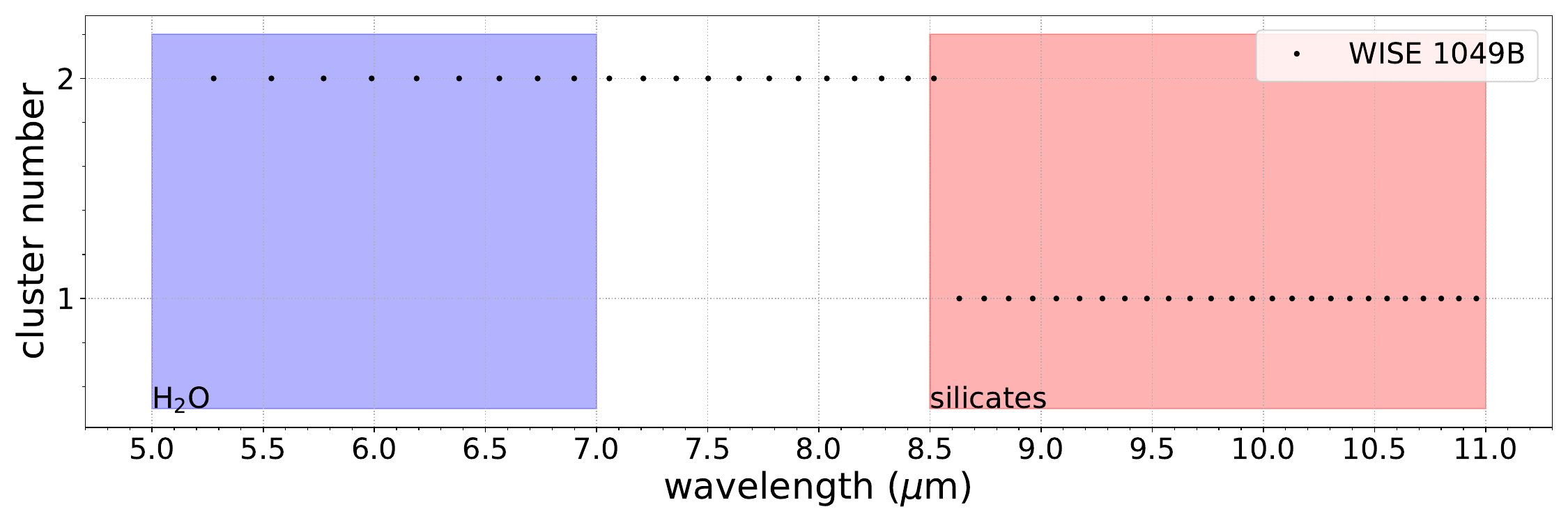}  
    \caption{Assigned cluster as a function of wavelength for WISE 1049A (top) and WISE 1049B (bottom) for MIRI lightcurves.  Shaded regions denote water absorption and the onset of small-grain silicate absorption.  A major transition between lightcurve clusters occurs around 8.5 $\mu$m.}
    \label{fig:MIRI_clustering_breaks}
\end{figure*}

\begin{figure*}
  	\includegraphics[width=0.45\textwidth]{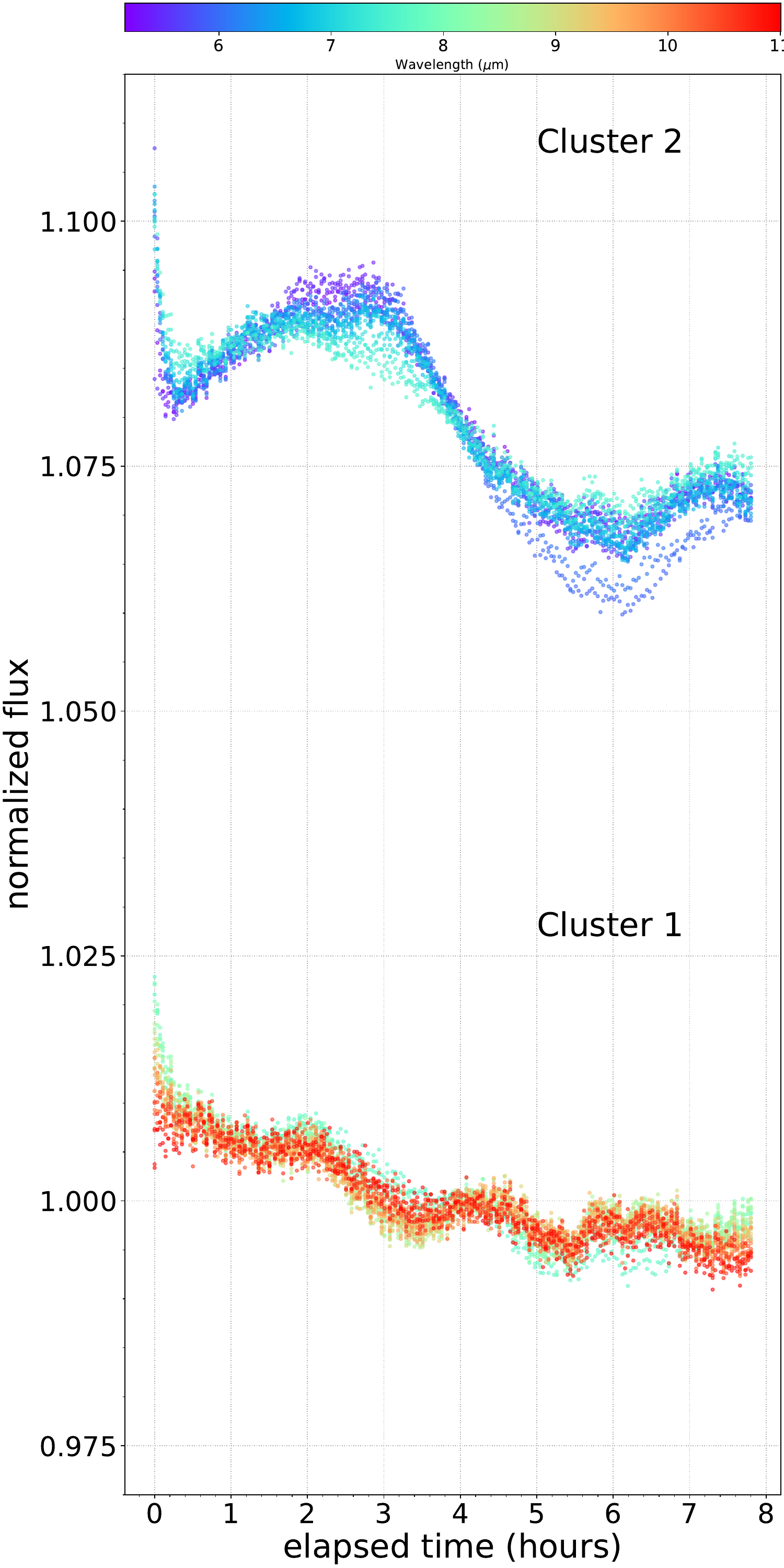} 
	\includegraphics[width=0.45\textwidth]{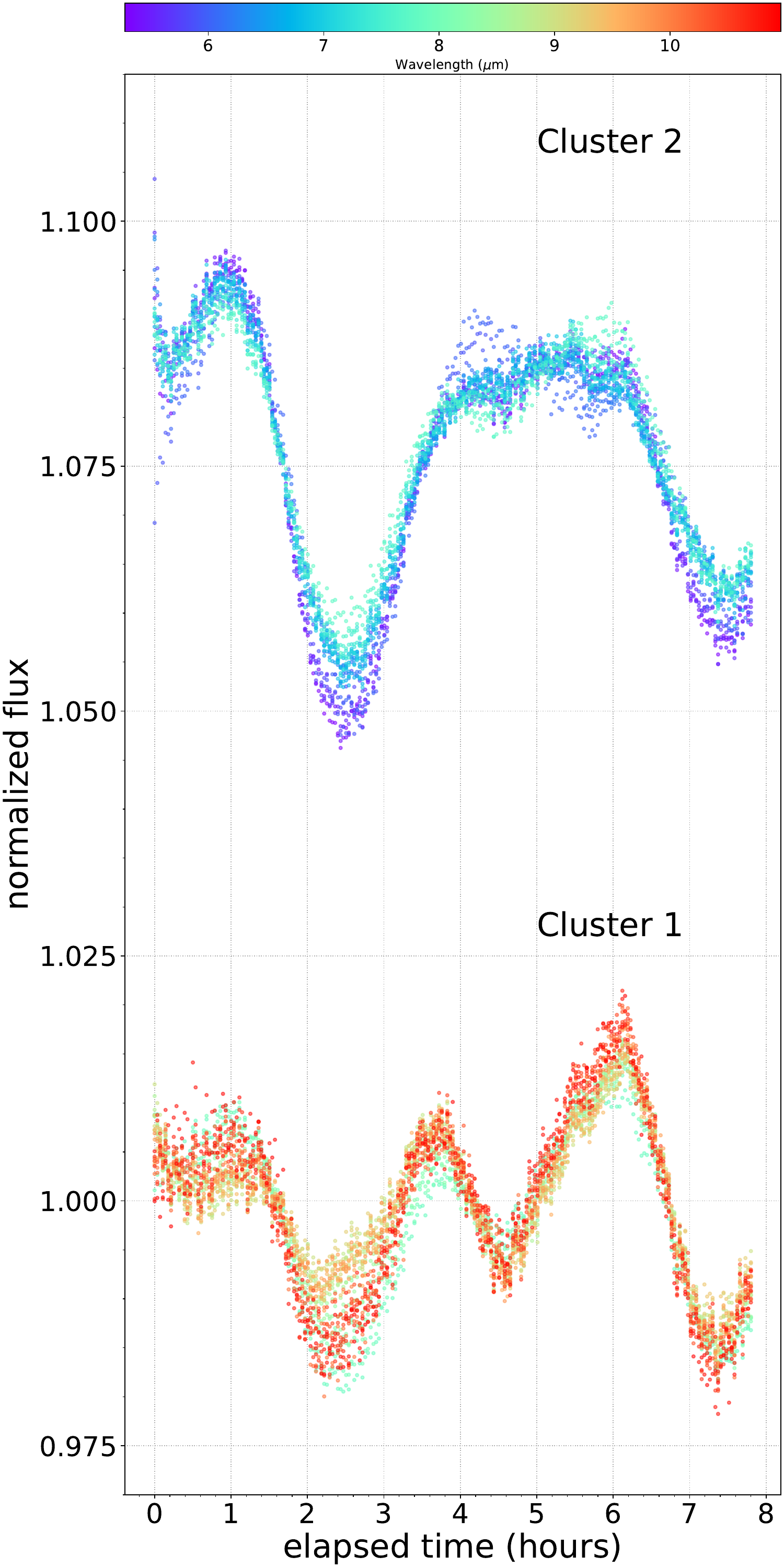} 
    \caption{MIRI single-channel lightcurves shown as a function of assigned cluster for WISE 1049A (left) and WISE 1049B (right).  Lightcurves have been binned to a cadence of 129 s and the wavelength for each lightcurve corresponds to the plot marker color.}
    \label{fig:MIRI_clustering_clusters}
\end{figure*}

Running this algorithm on both NIRSpec and MIRI lightcurves produces a best number of clusters of 3 clusters for NIRSpec and 2 clusters for MIRI.  The cluster assignments as a function of wavelength for both binary components for NIRSpec are shown in Fig.~\ref{fig:NIRSpec_clustering_breaks} and the resulting clusters of lightcurves are plotted in Fig.~\ref{fig:NIRSpec_clustering_clusters}.  
Cluster assignments as a function of wavelength for MIRI are shown in Fig.~\ref{fig:MIRI_clustering_breaks} and the resulting clustered lightcurves are shown in Fig.~\ref{fig:MIRI_clustering_clusters}.

We find three main transitions in behavior / cluster for both components of the binary: 1) change in behavior at 2.3 $\mu$m, corresponding to a CO absorption bandhead, 2) change in behavior at 4.2 $\mu$m, slightly bluewards to the onset of the CO fundamental band at wavelengths $>$4.4 $\mu$m, 3) change in behavior at 8.3--8.5 $\mu$m, potentially corresponding to silicate absorption.  These transition points are visually evident in the variability maps shown in Figures~\ref{fig:NIRSpec_WISE1049A_map}, \ref{fig:NIRSpec_WISE1049B_map}, \ref{fig:MIRI_WISE1049A_map}, and \ref{fig:MIRI_WISE1049B_map} as well.
For WISE 1049B, the lightcurve behaviors for the shortest ($<$2.3 $\mu$m) and longest ($>$8.5 $\mu$m) look qualitatively similar, displaying "double-peaked" behavior (as noted in Section~\ref{sec:mirilc}), while lightcurves between 4.2 and 8.5 $\mu$m for WISE 1049B appear to be "single-peaked". Interestingly, \citet{Vos2023} found from spectral retrieval studies of two planetary-mass T2 type objects that the wavelengths where WISE 1049B shows double-peaked behavior ($<$2.3 $
\mu$m and $>$8.5 $\mu$m) correspond to the wavelengths most affected by high-altitude silicate slab clouds in their retrievals, see their Fig.~5. WISE1049B likely has a similar atmospheric structure to the objects studied in \citet{Vos2023}, which may suggest that silicate clouds play a role in driving variability at these wavelengths.  For both WISE 1049A and B, there is a distinctive and sharp change in behavior right around 3.3 $\mu$m, corresponding to a methane feature clearly seen in both components.  Similarly, a sharp change seen around 2.6-2.7 $\mu$m in both components may stem from water absorption at these wavelengths, but appears more complicated in WISE 1049B than A.  WISE 1049B also exhibits shifts in behavior around water absorption features at $\sim$1.8 $\mu$m and $\sim$1.9 $\mu$m.


\subsection{The effect of molecular opacity on lightcurve behavior}

From the clustering analysis in the previous section, we find that specific changes in lightcurve behavior seem to be correlated with molecular absorption features.  Here we consider that behavior for each of the three molecules that display absorption features in the NIRSpec spectra of WISE 1049AB, specifically, methane, water, and carbon monoxide.

\begin{figure*}
        \includegraphics[width=0.48\textwidth]{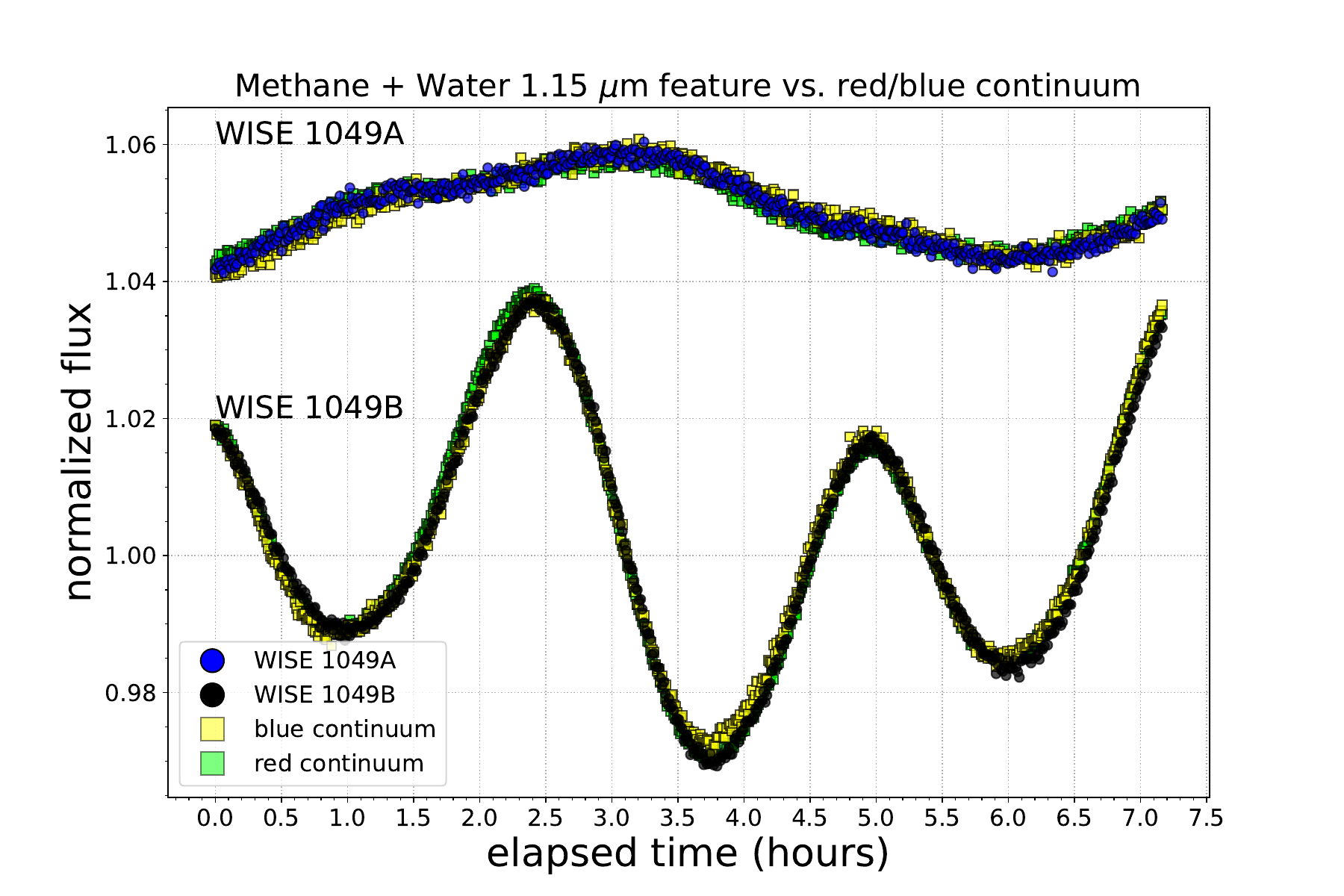} 
  	\includegraphics[width=0.48\textwidth]{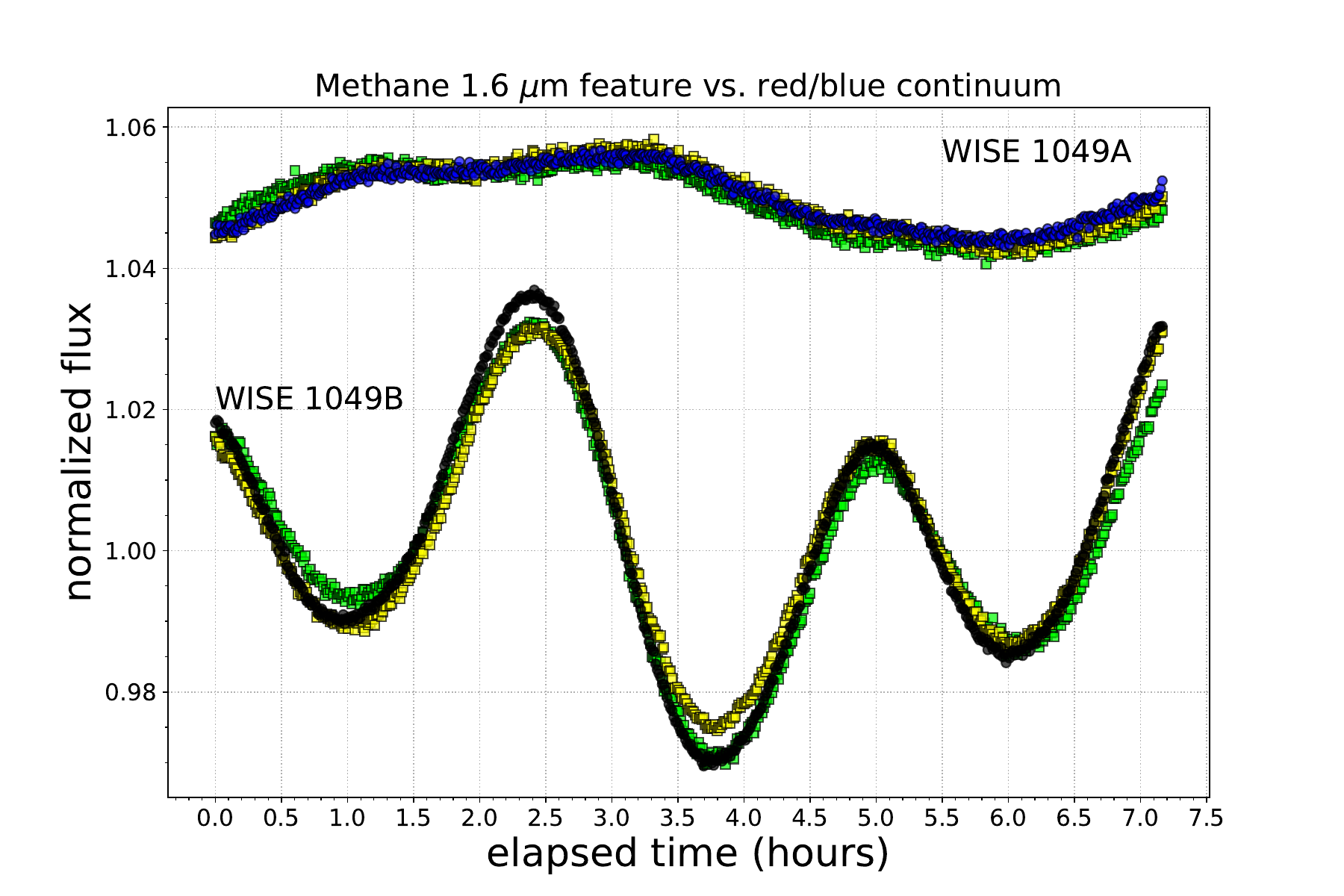} 
	\includegraphics[width=0.48\textwidth]{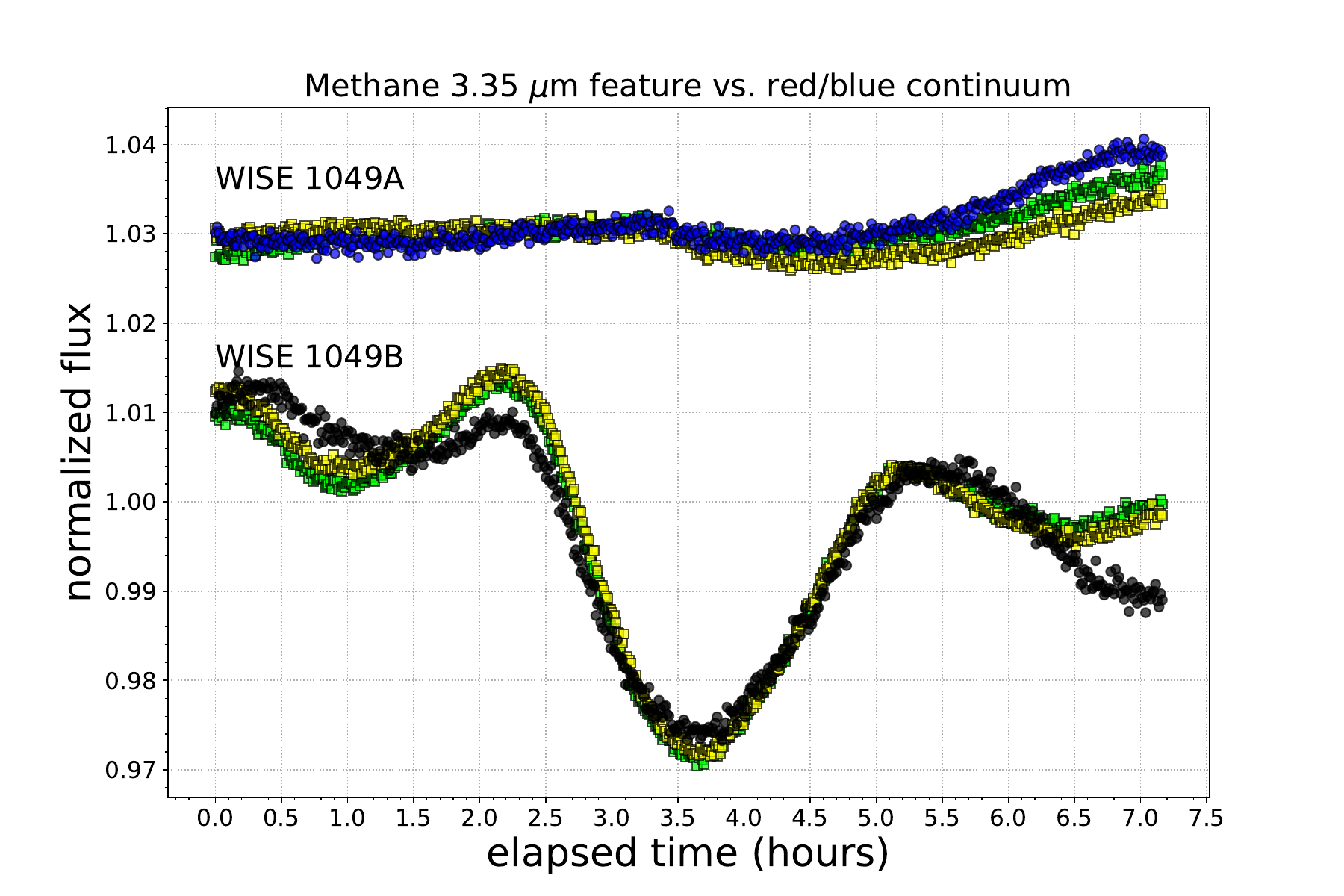}
    \caption{Narrowband lightcurves inside and outside of methane absorption features.  For the 1.15 $\mu$m feature, wavelengths between 1.1 and 1.17 $\mu$m were taken as being on the absorption feature, with a blue continuum lightcurve taken at wavelengths 1.0 to 1.07 $\mu$m and a red continuum lightcurve taken at wavelengths 1.23 to 1.3 $\mu$m.  For the 1.6 $\mu$m feature, wavelengths between 1.6 and 1.67 $\mu$m were taken as being on the absorption feature, with a blue continuum lightcurve taken at wavelengths 1.5 to 1.57 $\mu$m and a red continuum lightcurve taken at wavelengths 1.73 to 1.8 $\mu$m. For the 3.3 $\mu$m feature, wavelengths between 3.3 and 3.35 $\mu$m were taken as being on the absorption feature, with a blue continuum lightcurve taken at wavelengths 3.2 to 3.3 $\mu$m and a red continuum lightcurve taken at wavelengths 3.4 to 3.5 $\mu$m.  Only the lightcurve for the 3.3 $\mu$m feature shows significant differences in behavior between the absorption feature and continuum lightcurves.}  
    \label{fig:methane_lightcurves}
\end{figure*}

The NIRSpec spectra cover methane absorption features at 1.15 $\mu$m (NB: there is some degeneracy with water absorption opacity in this feature), 1.6 $\mu$m, and 3.3 $\mu$m.  Both 1.15 $\mu$m and 3.3 $\mu$m absorption features are detected in the spectra of WISE 1049A and B (Fig.~\ref{fig:NIRSpecspectra}), but the 1.6 $\mu$m feature is not clearly detected in either binary component.  In Fig.~\ref{fig:methane_lightcurves}, we plot narrowband lightcurves inside and outside of these 3 methane features, taking the median lightcurve inside each absorption feature as well as in continuum regions placed both blue-ward and red-ward of the feature.  As reflected by the clustering analysis, only the 3.3 $\mu$m feature shows significant differences in lightcurve behavior inside and outside of this feature.  The shapes of the lightcurves for both A and B in this case differ sufficiently within the absorption feature compared to outside the feature that the clustering algorithm assigns the lightcurves within the absorption feature to a different lightcurve cluster compared to the continuum lightcurves.  This likely traces real differences in pressure probed inside and outside of this spectral feature, where at the highest opacities in the core of this feature, the lightcurves probe a higher altitude, lower pressure region of the atmosphere compared to the continuum lightcurves.

\begin{figure*}
        \includegraphics[width=0.48\textwidth]{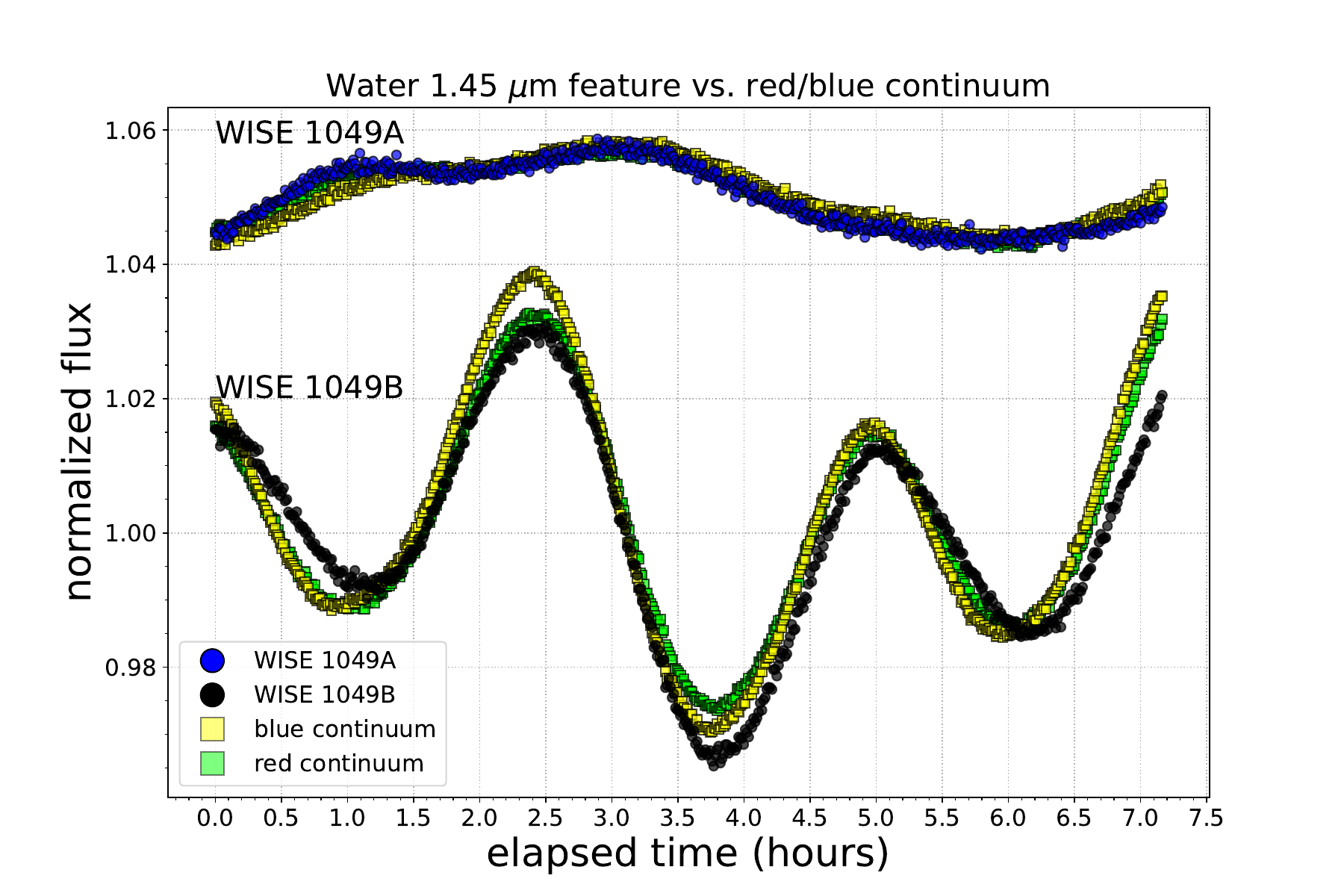}  
  	\includegraphics[width=0.48\textwidth]{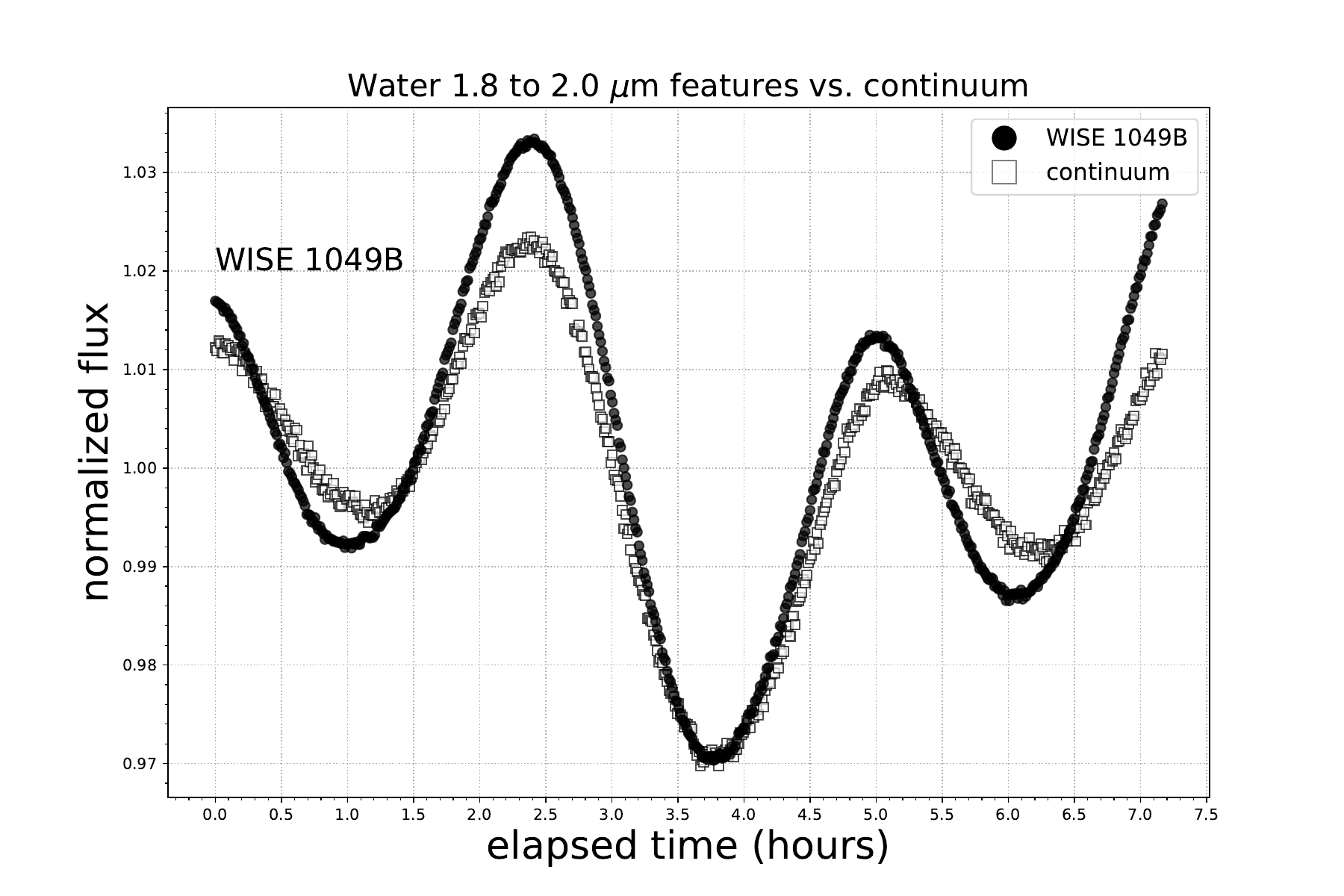} 
	\includegraphics[width=0.48\textwidth]{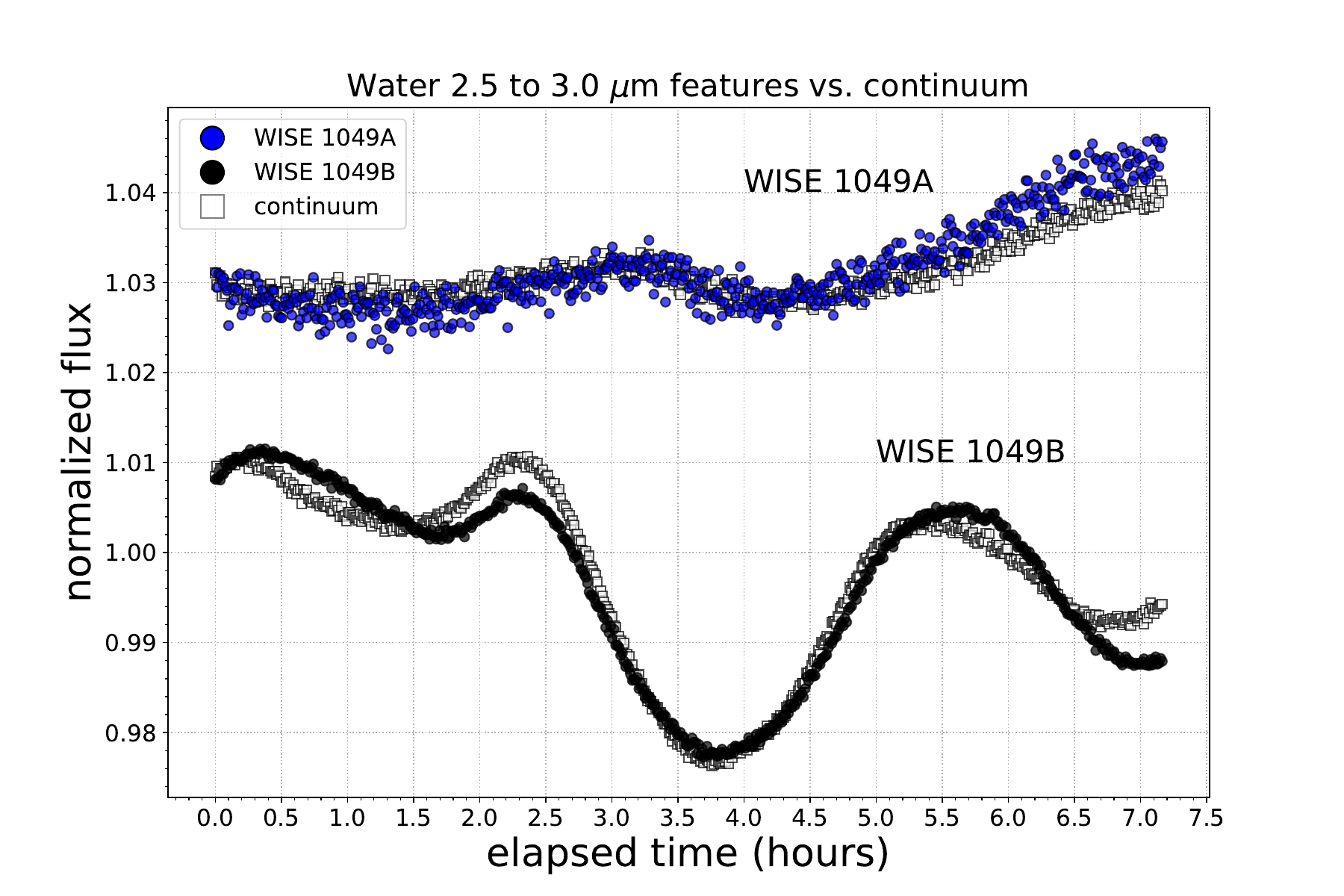} 
   \caption{Narrowband lightcurves inside and outside of water absorption features.  For the 1.45 $\mu$m feature, wavelengths between 1.34 and 1.44 $\mu$m were taken as being on the absorption feature, with a blue continuum lightcurve taken at wavelengths 1.2 to 1.3 $\mu$m and a red continuum lightcurve taken at wavelengths 1.5 to 1.5 $\mu$m. For the water features between 1.8 and 2.0 $\mu$m in the spectrum of WISE 1049B, lightcurves at wavelengths identified by the clustering algorithm as being on an absorption feature were median combined to produce a composite lightcurve from both absorption features.  Lightcurves at wavelengths between 1.6 and 2.0 $\mu$m that the algorithm did not consider part of the feature were median-combined to produce the continuum lightcurve.  A similar procedure was followed between 2.5 and 3.0 $\mu$m for both WISE 1049A and B to produce lightcurves within and outside of water absorption opacity respectively.}  
    \label{fig:water_lightcurves}
\end{figure*}

Water absorption features are present in the NIRSpec spectra of WISE 1049AB at 1.4 $\mu$m, between 1.8 and 2 $\mu$m, and between 2.5 and 3 $\mu$m.  The clustering analysis reveals shifts in assigned cluster as a function of wavelength in WISE 1049B co-located with the known water absorption features between 1.8 and 2 $\mu$m and between 2.5 and 3 $\mu$m, but does not find a similar shift for the 1.4 $\mu$m feature.  For WISE 1049A, the clustering algorithm only pinpoints a shift in lightcurve cluster around 2.7 $\mu$m.  As with methane, we consider the narrowband lightcurves constructed inside and outside of these absorption features.  In Fig.~\ref{fig:water_lightcurves}, we plot these lightcurves for the three regions of molecular opacity described above. For the 1.4 $\mu$m and 1.8--2.0 $\mu$m features, the lightcurves for WISE 1049B appear to decrease in amplitude inside the water feature compared to outside, and display minima / maxima that occur at slightly later times compared to the continua lightcurves, with no variation in lightcurve behavior seen in WISE 1049A between wavelengths inside and outside these absorption features.  In contrast, for the 2.5--3.0 $\mu$m water features, both WISE 1049A and WISE 1049B exhibit significant changes in lightcurve shape, transitioning between different lightcurve clusters inside and outside of the absorption features in this wavelength range. 

NIRSpec covers both a CO bandhead at $\sim$2.3 $\mu$m and the CO fundamental band at 4.4 $\mu$m -- 5.0 $\mu$m \citep{Miles2023}.  Both WISE 1049A and B show a transition between different clusters of lightcurves (as identified by the K-means clustering algorithm) either at or near the location of these bandheads.  However, given that these bandheads cover fairly wide spectral ranges and have significant substructure, it is more difficult to assign appropriate in-feature and continuum wavelength regions.  Thus, we did not attempt to plot lightcurves inside and outside of these features, as we did for the water and methane features.  


\subsection{Updated estimate of bolometric luminosity and inferred physical properties from evolutionary models\label{sec:Lbol}}

To estimate the bolometric luminosities of both WISE 1049A and B, we first constructed "white-light" lightcurves across both NIRSpec (0.5--5 $\mu$m) and MIRI (5--11 $\mu$m) bandpasses.  Starting with each NIRSpec and MIRI spectrum in F$_{\lambda}$ units, we summed across each respective bandpass, taking into account the changing widths of spectral bins due to the varying resolution as a function of wavelength.  We converted to luminosity in units of W, adopting a parallax of 501.557$\pm$0.082 mas to WISE 1049AB \citep{Lazorenko2018}. We estimate the minimum luminosity across the full JWST bandpass, L$_{\rm minJWST}$, as the minimum luminosity across the full NIRSpec bandpass, L$_{\rm minNIRSpec}$, plus the minimum luminosity across the full MIRI bandpass, L$_{\rm minMIRI}$.  Similarly, we estimate the maximum luminosity across the full JWST bandpass, L$_{\rm maxJWST}$, as the maximum luminosity across the full NIRSpec bandpass, L$_{\rm maxNIRSpec}$, plus the maximum luminosity across the full MIRI bandpass, L$_{\rm maxMIRI}$.  To account for flux received from wavelengths not covered by JWST, we attempted to fit a number of model grids to the full JWST median spectra of both components using \texttt{species} \citep{Stolker2020}, but we were unable to find satisfactory fits over such a wide bandwidth.  However, as brown dwarfs output the vast majority of their light in the infrared, the luminosity measured by JWST, L$_{\rm JWST}$, should be very close to the bolometric luminosity, with $<$10$\%$ of total light emitted outside the wavelength range covered by JWST \citep{Miles2023}.  Thus, the full bolometric luminosity will be largely insensitive to the selection of model to fill in the remaining wavelengths.  To determine how much of the full luminosity of WISE1049AB is covered in the JWST bandpass and calculate a "bolometric correction", we calculated L$_{\rm JWST}$ / L$_{\rm bol}$ for model atmospheres drawn from the ExoRem model grids \citep{Charnay2018}, with T$_\mathrm{eff}$ ranging from 1100 to 1900 K, and a range of different log(g) and metallicity values.  For all model atmospheres considered, L$_{\rm JWST}$ / L$_{\rm bol}$ $\geq$0.9, and in most cases, L$_{\rm JWST}$ / L$_{\rm bol}$ ranged from 0.95 to 0.97.  We found similar results for other model grids -- in all cases, the JWST bandpass covers $\geq$90$\%$ of the emitted light from both binary components.  We estimate a conservative range of potential values of bolometric luminosity, taking into account the intrinsic variability of each component, ranging from $\frac{L_{\rm minJWST}}{0.97}$ to $\frac{L_{\rm maxJWST}}{0.9}$.  This yields a range in bolometric luminosity from log(L$_{\rm bol}$ / L$_{\odot}$) = $-$4.63 to $-$4.59 for WISE 1049A and from log(L$_{\rm bol}$ / L$_{\odot}$) = $-$4.71 to $-$4.66 for WISE 1049B.  These values agree well with the bolometric luminosities reported in \citet{Faherty2014} of log(L$_{\rm bol}$ / L$_{\odot}$) =  $-$4.67$\pm$0.04 for WISE 1049A and log(L$_{\rm bol}$ / L$_{\odot}$) = $-$4.71$\pm$0.1 for WISE 1049B.

\begin{figure*}
  	\includegraphics[width=\textwidth]{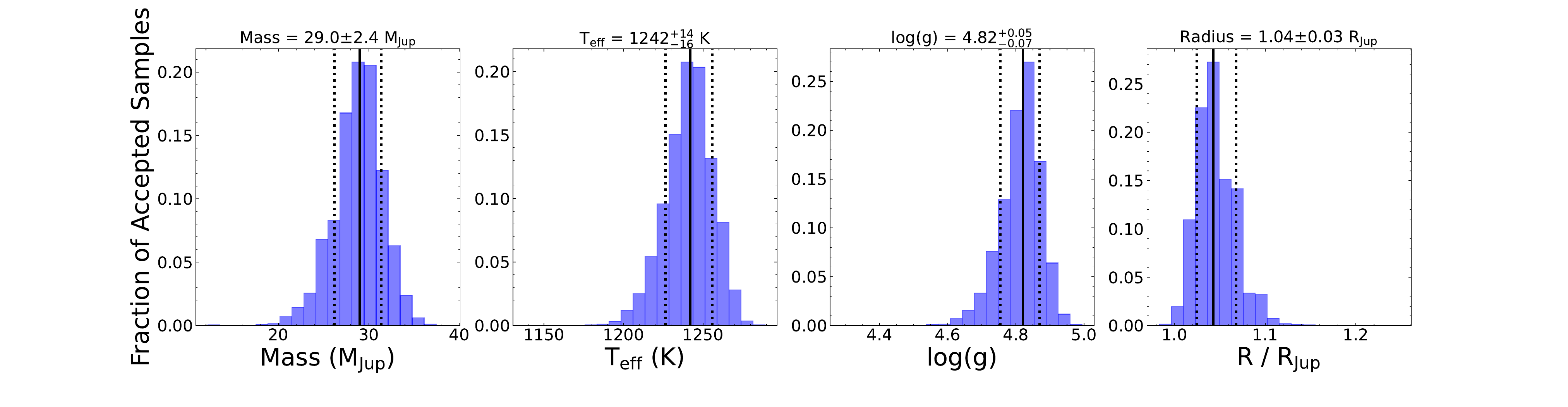} 
        \includegraphics[width=\textwidth]{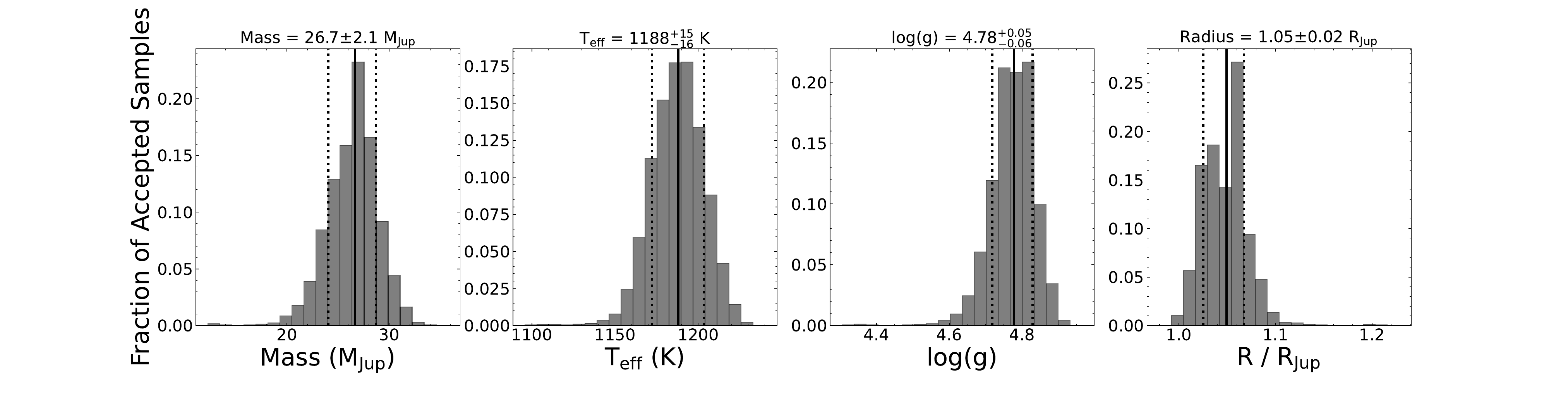}
    \caption{Physical properties estimated using the hybrid model grid from \citet{Saumon2008}, which includes cloud opacity for L-type objects but models T-type objects as clear atmospheres. Top row: results for WISE 1049A, shown as blue histograms, bottom row: results for WISE 1049B, shown as black histograms.  The vertical dotted lines show the 67\% confidence interval region for each parameter.}
    \label{fig:SMhybrid_histograms}
\end{figure*}

\begin{figure*}
  	\includegraphics[width=\textwidth]{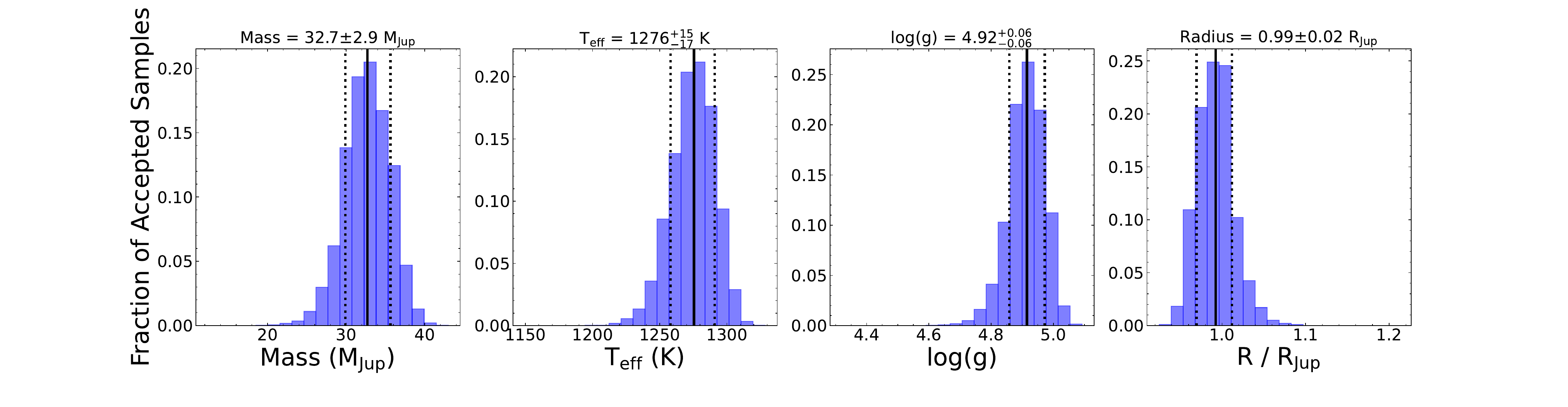} 
        \includegraphics[width=\textwidth]{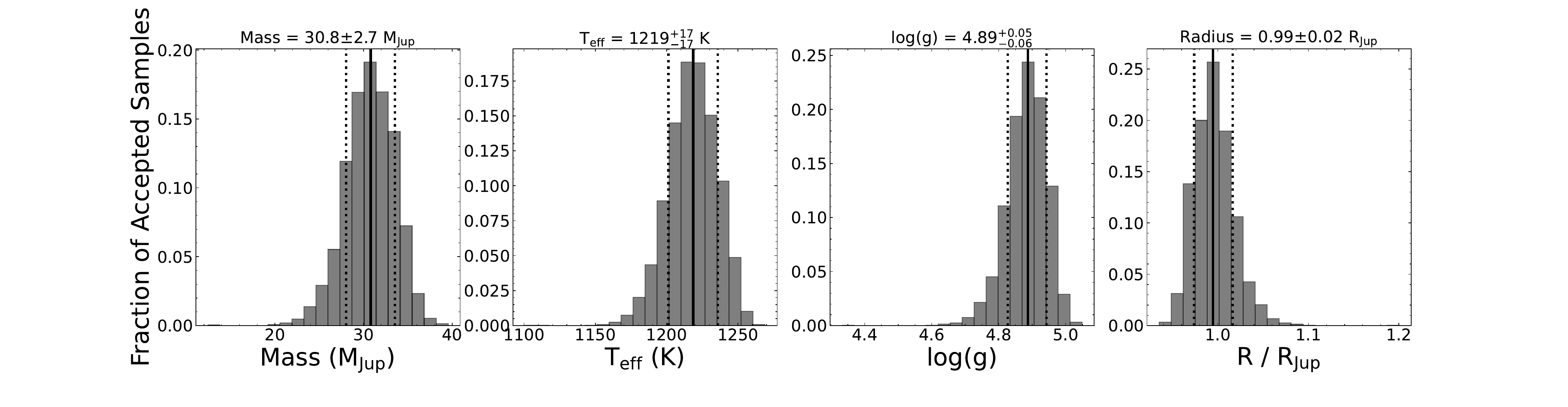}
    \caption{Physical properties estimated using the cloud-free Sonora-Bobcat grid from \citet{Marley2021}.}
    \label{fig:SonoraBobcat_histograms}
\end{figure*}

Using the JWST-derived bolometric luminosity ranges for WISE 1049A and B, with an age of 510$\pm$9.5 Myr from membership in the Oceanus moving group \citep{Gagne2023} and parallax measurements from \citet{Lazorenko2018}, we estimate effective temperature, T$_\mathrm{eff}$, and surface gravity, log(g), using the method described in \citet{Carter2023} with both the hybrid solar-metallicity evolutionary models (which include clouds for L-type objects) from \citet{Saumon2008} and the cloudless Sonora-Bobcat evolutionary models \citep{Marley2021}.  We draw n=$10^6$ samples from a Gaussian distribution centered around the 510 Myr age estimate, with a Gaussian FWHM of 9.5 Myr, and a uniform distribution of masses ranging from 1 to 50 M$_{\rm Jup}$, then interpolate the respective grid model at that age, mass point to find the corresponding luminosity, effective temperature, and radius.  We retain samples with interpolated bolometric luminosities falling within the JWST-derived bolometric luminosity ranges for WISE 1049A and B.  Histograms of the retained mass, T$_\mathrm{eff}$, radius, and log(g) samples are plotted in Fig.~\ref{fig:SMhybrid_histograms} for the \citet{Saumon2008} models and Fig.~\ref{fig:SonoraBobcat_histograms} for the Sonora-Bobcat evolutionary models \citep{Marley2021} respectively.  The presence of cloud opacity does not strongly affect the bulk physical properties estimated from evolutionary models.  Using either model grid, we find very similar parameters for both components -- T$_\mathrm{eff}$ between 1150 and 1300 K and log(g) between 4.7 and 5, which are in general agreement with values in the literature \citep{Faherty2014, Filippazzo2015}.

\subsection{Do different lightcurve clusters correspond to different atmospheric pressures / depths? \label{sec:pressure}}

\begin{figure*}
  	\includegraphics[width=\textwidth]{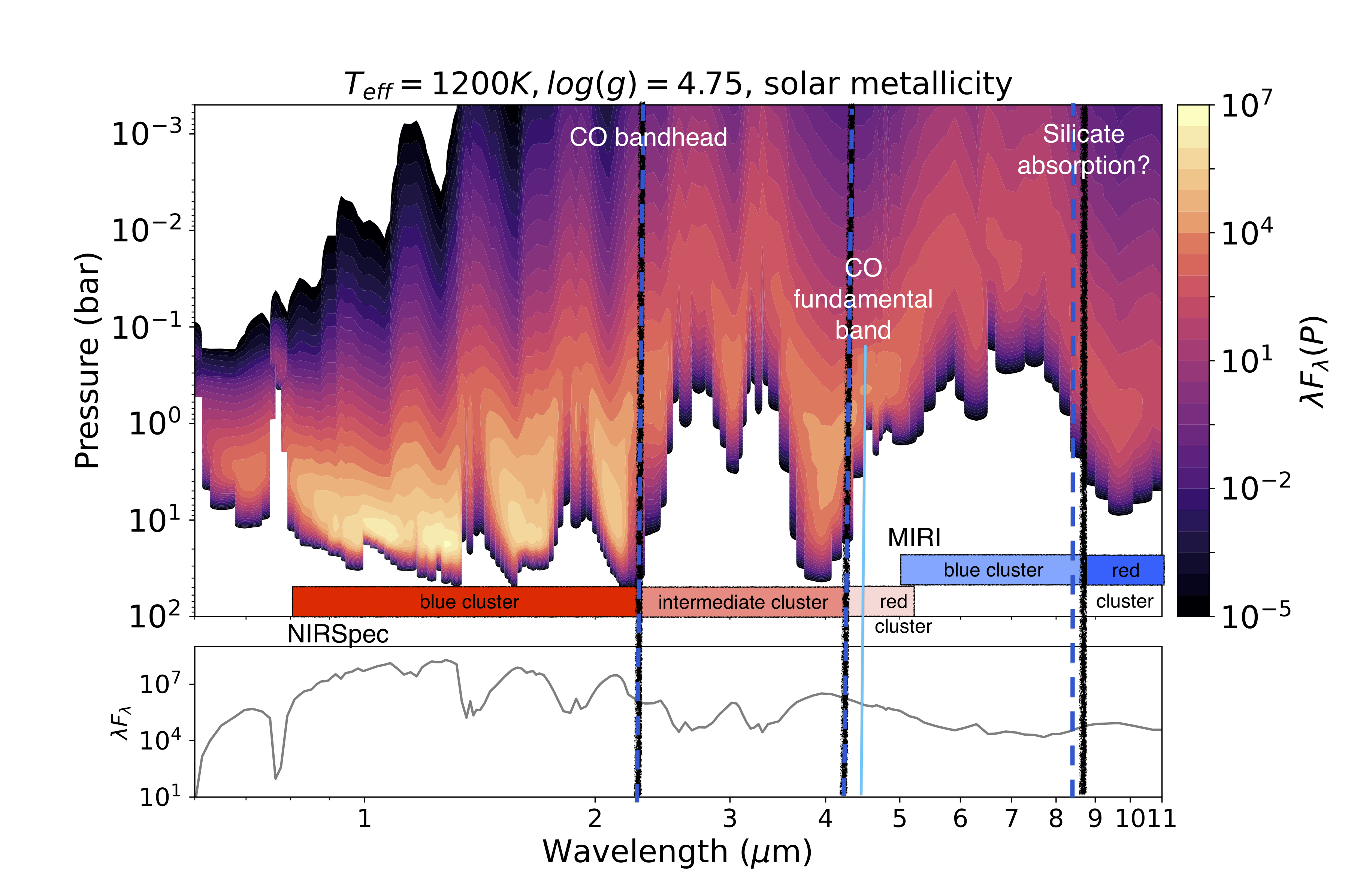} 
    
    \caption{Contribution plot computed from a Sonora-Bobcat cloud-free model for a model atmosphere with T$_\mathrm{eff}$ and log(g) similar to both components of WISE 1049AB.  Light is emitted at a range of pressures for each wavelength, but tends to be emitted at higher pressures / greater depths for wavelengths $<$ 2.3 $\mu$m, with the average depth that light is emitted at a given wavelength strongly influenced by opacity sources at that wavelength.  The three wavelengths at which lightcurves transition between different clusters are shown as dashed blue vertical lines for WISE 1049A and solid black vertical lines for B; the first two of these occur near CO opacity features and all three are located at regions on the plot where the average pressure at which light is emitted at a given wavelength changes significantly.  The dominant lightcurve cluster as a function of wavelength is shown as rectangular boxes for both the NIRSpec and MIRI bandpasses.}  
    \label{fig:contribution_plot}
\end{figure*}

We plot the contribution function and the resulting spectrum for a Sonora-Bobcat cloudless model in chemical equilibrium, with T$_\mathrm{eff}$=1200 K, log(g)=4.75, and solar metallicity in Fig.~\ref{fig:contribution_plot}.  Following the procedure described in \cite{McCarthy2024}, we calculated the flux contribution for each pressure and wavelength for an atmosphere similar to that of both components of WISE 1049AB. We used the publicly available Sonora Bobcat correlated k-coefficients \citep{SonoraKcoefs} and PT profile \citep{SonoraBobcat}, and assuming an atmosphere in local thermal equilibrium, we solved the radiative transfer equation along the PT profile, resulting in the flux per pressure per wavelength. From the discussion in Section~\ref{sec:Lbol}, this model atmosphere is a reasonable match to the properties of both components of WISE 1049AB and should provide an indicative guide to the impact of TOA structure at different pressure levels on the resulting lightcurve as a function of wavelength.  

We found significant global changes in lightcurve behavior at three wavelengths in both WISE 1049A and B, using the clustering analysis discussed in Section~\ref{sec:clustering}: 1) at the onset of the 2.3 $\mu$m CO bandhead, 2) at 4.2 $\mu$m, slightly bluewards of the onset of the CO fundamental band at 4.4 $\mu$m, and 3) around 8.5 $\mu$m, potentially connected with the onset of silicate absorption.  These changes in lightcurve behavior are also connected to changes in the average pressure probed at each wavelength, as seen in Fig.~\ref{fig:contribution_plot}.  We interpret the lightcurves observed with both NIRSpec and MIRI as likely stemming from 3 discrete pressure levels -- a deep pressure level that drives the double-peaked variability seen at wavelengths $<$2.3 $\mu$m and $>$8.5 $\mu$m, an intermediate pressure level that shapes lightcurve morphology between 2.3 and 4.2 $\mu$m, and a higher-altitude pressure level that produces the single-peaked and plateaued lightcurve behavior between 4.2 and 8.5 $\mu$m.    

\subsection{Tests of variability driving mechanisms\label{sec:tests}}

The JWST variability monitoring observations described here enable access to key molecular features across the near-IR with NIRSpec \citep{Faherty2014, Miles2023} as well as to the silicate feature at $\lambda$ $>$ 8.5 $\mu$m with MIRI \citep{Suarez2022}.  This broad wavelength coverage enables us to test specific variability predictions for different driving mechanisms for variability.

Considering the case of variability driven by inhomogeneous coverage due to clouds composed of small grain silicates, \citet{Luna2021} modeled a high patchy cloud layer composed of small grain silicates over a deeper, uniform cloud layer.  This two-layer cloud model was used to recreate the $>$8.5 $\mu$m silicate feature observed in field brown dwarfs with \textit{Spitzer} \citep{Suarez2022} and then to simulate variability measured by \textit{Spitzer} in field brown dwarfs \citep{Metchev2015a}. Based on this model, \cite{Luna2021} predict enhanced variability $>$9 $\mu$m relative to 6--7 $\mu$m. Considering the maximum deviation as a function of wavelength (Fig.~\ref{fig:maxdeviation}), the maximum deviation appears to decrease monotonically for WISE 1049A, leading to reduced variability $>$9 $\mu$m compared to 6--7 $\mu$m.  For WISE 1049B, the maximum deviation at wavelengths $>$9 $\mu$m is comparable or slightly lower than at 6--7 $\mu$m.  This rules out the particular scenario modeled of a high-altitude patchy cloud specifically composed of small grain silicates ($<$1 $\mu$m grain size), but other scenarios with larger silicate grains deeper in the atmosphere may still drive variability without producing a clear silicate absorption feature at $>$8.5 $\mu$m.  As WISE 1049A does display small-grain silicate absorption at $>$8.5 $\mu$m, there is likely a layer of high-altitude silicate clouds producing this feature, but since maximum deviation is not enhanced at these wavelengths, this suggests that this high altitude silicate layer may not be patchy.

The atmospheric general circulation models (GCMs) of brown dwarfs of \cite{Tan2021a} and \cite{Tan2021b} suggested that the surface patchiness of brown dwarfs can be driven by the coupling between the atmospheric winds and cloud radiative feedback, naturally producing rotational lightcurves and their irregular time evolution. Inheriting from the GCM described in \cite{showman2009}, we update the brown dwarf GCM to use a non-grey radiative transfer to calculate more realistic heating/cooling rates and thermal structures and assume equilibrium chemistry with a solar atmospheric composition. The GCM includes the formation, evaporation and sedimentation of MgSiO$_3$ and Fe clouds, and their feedbacks generate the circulation (Tan et al. in prep). We applied the relevant parameters of WISE 1049B (gravity of $588~{\rm ms^{-2}}$, radius of 1.05 $R_{\rm J}$, rotation period of 5.28 hours) to conduct preliminary modeling of the spectroscopic variability of WISE 1049B.
After the GCM reached equilibrium, we computed spectroscopic light curves by post-processing the 3D, time-dependent temperature and cloud outputs of the GCM with PICASO \citep{Batalha_2019,Mukherjee2023}. The GCM and PICASO have the same opacity source and method of computing radiative transfer which ensures accurate spectroscopic calculations. 

We computed spectroscopic light curves over 11 rotations of the WISE 1049B model assuming an equator-on viewing angle. The maximum deviation of normalized light curves was derived the same way as for the observed lightcurves (Sec.~\ref{sec:maxdev}). Fig.~\ref{fig:maxdeviation} shows the comparison of the observed and modeled maximum deviations as a function of wavelength. We find that at relatively short wavelengths of $\lesssim 4.2~\mu{\rm m}$, the simulated maximum deviations qualitatively reproduce the observed trends that higher amplitudes occur at molecular opacity windows, in particular, at $1-1.2$, 1.6, and $2.2~\mu{\rm m}$ regions. This is because fluxes at these wavelengths can escape from deeper and hotter layers and can scan higher brightness contrasts between cloudy and relatively cloud-free areas as the object rotates. This comparison indicates that cloud patchiness associated with the somewhat deep silicate and iron clouds is partly responsible in generating the observed spectroscopic variability. 

Meanwhile,  disagreements are prominent. First, the simulated deviations at longer wavelengths of $\gtrsim 4.3~\mu{\rm m}$ are significantly lower than the observed deviations by a factor of more than 2. Second, at shorter wavelengths, the observed deviations are much more muted over wavelength than the simulated ones. Both of these deviations suggest that the upper atmosphere is more active than predicted by the GCM. The current GCM includes only silicate and Fe clouds that are too deep to be sensed by longer wavelengths. Additional upper cloud layers of perhaps KCl or Na$_2$S \citep{Morley2012} might help with two aspects: 1) directly increasing the inhomogeneous cloud opacity  at shallow pressures and thus increasing the variability amplitudes at longer wavelengths; and 2) increasing the overall column cloud opacity, helping to mute the variability variation across shorter wavelengths. This would be tested in more dedicated GCM studies in the future. 

\citet{Tremblin2020} model the expected variability as driven by non-equilibrium chemistry producing vertical mixing and hot-spots.  They predict little variability at 10 $\mu$m, but enhanced variability in methane and CO absorption features due to potential composition variations as well as anti-correlated variability (180$^{\circ}$ phase shift) between the 1.1--1.5 $\mu$m and the 3--5 $\mu$m bandpasses.  Like in the previous cases considered, the observed lightcurves display complex behaviors that do not definitively follow the predictions from this model.  We still find significant variability in both binary components at 10 $\mu$m.  We do not find clear "phase shifts" between the 1.1--1.5 $\mu$m and 3--5 $\mu$m bandpasses, but we do find that lightcurves in these bandpasses fall into two quite different clusters of lightcurve behaviors.  We do not find enhanced variability in methane and CO absorption features relative to the continuum, but we do find differences in behaviors of the lightcurves inside and outside of these features.  

\subsection{Comparison with previous studies}

\begin{figure*}
  	\includegraphics[width=\textwidth]{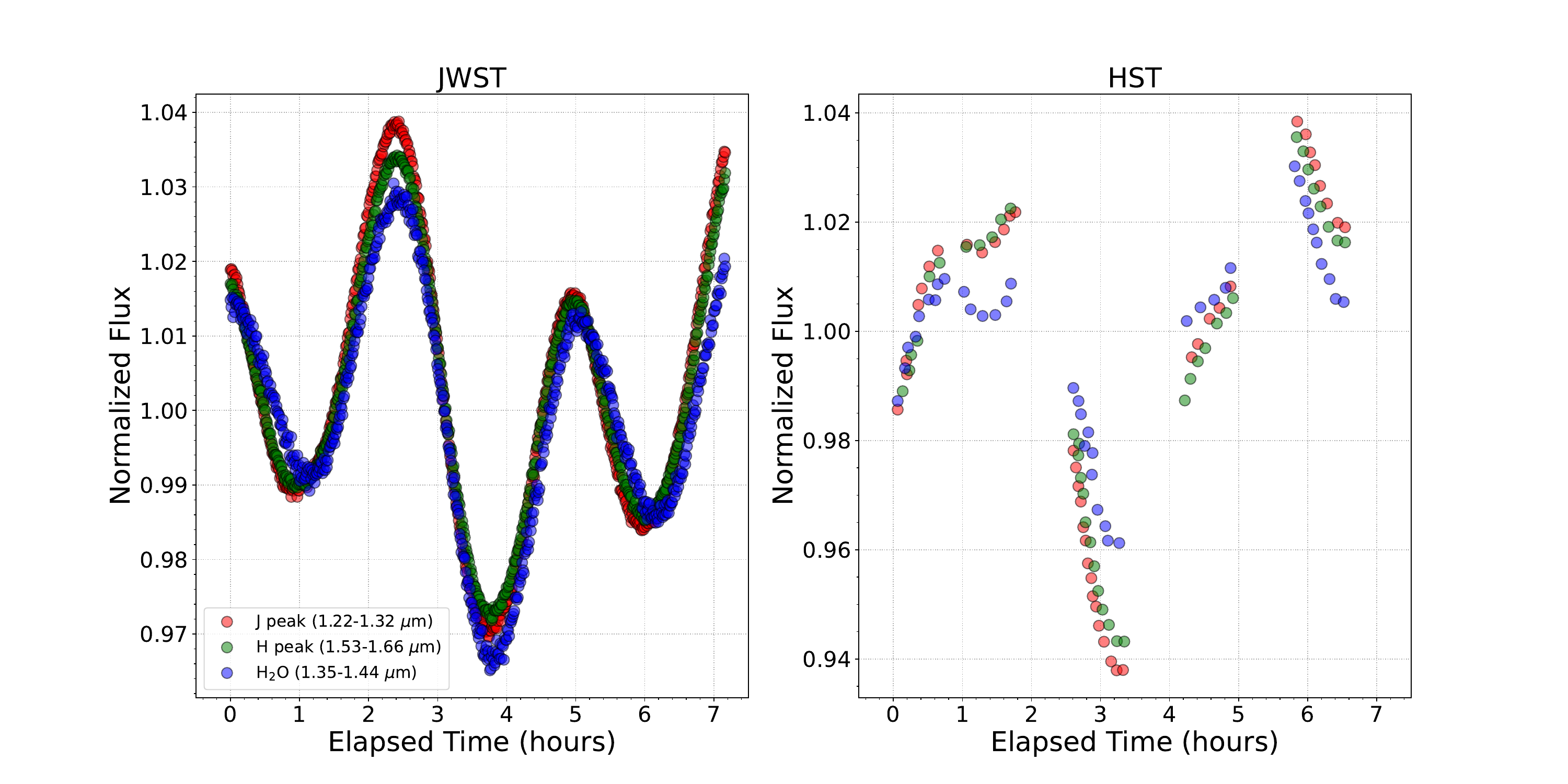}    
    \caption{Comparison of JWST NIRSpec lightcurves (left) with HST lightcurves from \citet{Buenzli2015a} (right), in the same three spectral regions as Fig.~5 from \citet{Buenzli2015a}: J peak (1.22 - 1.32 $\mu$m), which captures the peak of J band emission, H peak (1.53 - 1.66 $\mu$m), which captures the peak of H band emission, and H$_{2}$O (1.35 - 1.44 $\mu$m), situated on a water absorption feature.  In contrast with the HST lightcurves, the shapes of the JWST lightcurves are similar for each spectral region, but do show offsets in the timing of successive maxima and minima between the water feature lightcurve and the J and H band lightcurves.}
    \label{fig:JWST_vs_HST_comparison}
\end{figure*}

Given the proximity and brightness of WISE 1049AB, numerous studies have probed its variability at both infrared and optical wavelengths \citep{Gillon2013, Biller2013a, Burgasser2014, Mancini2015, Street2015, Buenzli2015, Buenzli2015a, Kellogg2017, Heinze2021, Apai2021}.  However, the number of studies that have robustly resolved both components in the infrared is considerably more limited.  A number of studies look only at the combined unresolved lightcurve of both components \citep{Gillon2013, Street2015, Apai2021} or make the assumption that WISE 1049A is non-variable and then use the A component to detrend lightcurves for the more variable B component \citep{Kellogg2017, Heinze2021}. The precursor dataset which provides the best match and hence most robust comparison to the JWST data described in this paper is the 1.1--1.7 $\mu$m HST WFC3 G141 grism dataset presented in \citet{Buenzli2015a}.  In Fig.~\ref{fig:JWST_vs_HST_comparison}, we present our JWST NIRSpec lightcurves side-by-side with the HST WFC3 lightcurves from \citet{Buenzli2015a}.  We construct lightcurves in the same three spectral regions as shown in Fig.~5 of \citet{Buenzli2015a}, specifically: J peak (1.22 -- 1.32 $\mu$m), which captures the peak of J band emission, H peak (1.53 -- 1.66 $\mu$m), which captures the peak of H band emission, and H$_{2}$O (1.35 -- 1.44 $\mu$m), situated on a water absorption feature.  The HST lightcurves showed a reduced maximum deviation between successive maxima and minima in the water absorption feature relative to J and H peak, as well as a somewhat different lightcurve shape in the water feature compared to the two continuum spectral regions.  In comparison, the shape of all three lightcurves are quite similar in the JWST data, but with timing offsets between successive maxima and minima, with extrema in the water feature occurring slightly later than extrema in J and H peak. 

\section{Conclusions}

We report results from 8 hours of MIRI LRS observations followed by 7 hours of NIRSpec BOTS PRISM mode variability monitoring observations of the benchmark brown dwarf binary WISE 1049AB.  This is the first variability monitoring study at wavelengths $>$5 $\mu$m for either component of the binary as well as the first study to simultaneously cover the full bandpass between 1-5 $\mu$m.  We report the following key conclusions:

\begin{enumerate}
\item We obtained very high S/N ratio lightcurves for WISE 1049AB with both MIRI and NIRSpec, yielding sensitivity to variations as low as $\sim$0.2$\%$ with  MIRI (125 s cadence, 0.5 $\mu$m wavelength bins) and as low as $\sim$0.1$\%$ with NIRSpec (45 s cadence, 0.2 $\mu$m wavelength bins). JWST's sensitivity to extremely small variations across a wide wavelength range will be transformative for understanding the atmospheres of brown dwarfs via variability monitoring.
\item The MIRI LRS traces were separated by $\sim$2 pixels and required PSF-fitting techniques to retrieve the individual traces for both components.  In this paper, we demonstrate multiple methods for deblending the traces using PSF-fitting techniques, all of which produce consistent results.  Due to the increased PSF FWHM at longer wavelengths, we were only able to extract absolute photometry up to $\lambda$=10 $\mu$m and relative photometry up to $\lambda$=11 $\mu$m. 
\item We find several water, methane, and CO absorption features across the NIRSpec bandpass for both components of the binary.  The MIRI spectra display water and methane absorption, with a potential detection in WISE 1049A of the plateau feature attributed to small-grain ($\lesssim$1 $\mu$m) silicates seen in other mid-to-late-L brown dwarfs \citep{Cushing2006, Suarez2022}.
\item Both components are significantly variable across the full NIRSpec+MIRI bandpass ($>$2$\%$ variation over a given observation), but, as has been found in previous studies, WISE 1049B is still considerably more variable than WISE 1049A at most wavelengths.
\item We find complex, wavelength-dependent trends in lightcurve shape for both components of the binary.  These trends cannot be modeled as sinusoidal variations, or easily described in terms of a single "amplitude" or "phase".  In particular, for WISE 1049B, lightcurves vary between a double-peaked shape (time between successive minima or maxima of half of the known period) at wavelengths $<$4.2 $\mu$m and wavelengths $>$8.5 $\mu$m, and a single-peaked shape (time between successive minima or maxima on order of the known period) at wavelengths intermediate between 4.2 and 8.5 $\mu$m. 
\item Considering the MIRI + NIRSpec lightcurves as a single time series over the wavelengths where the spectra overlap (5-5.2 $\mu$m), we find clear period-to-period variation, with higher maximum deviation (difference between highest peak and deepest trough in a given observation) during the MIRI observation compared to NIRSpec.  A periodogram analysis retrieves the $\sim$5 hour period previously measured for WISE 1049B \citep{Gillon2013, Apai2021} and is consistent with the reported $\sim$7 hour period for WISE 1049A \citep{Apai2021}.
\item From the NIRSpec and MIRI spectra, we determine updated bolometric luminosity ranges of log(L$_{\rm bol}$ / L$_{\odot}$) = $-$4.63 to $-$4.59 for WISE 1049A and from log(L$_{\rm bol}$ / L$_{\odot}$) = $-$4.71 to $-$4.66 for WISE 1049B. Adopting an age of 510$\pm$9.5 Myr from membership in the Oceanus moving group \citep{Gagne2023} and parallax measurements from \citet{Lazorenko2018}, we estimate effective temperatures, T$_\mathrm{eff}$, between 1150 and 1300 K and surface gravities, log(g), between 4.7 and 5 for both components, in general agreement with values in the literature \citep{Faherty2014, Filippazzo2015}.  
\item Using the K-means algorithm to sort lightcurves into distinct clusters based on the dominant trends identified within each cluster of lightcurves, we find 3 distinct lightcurve clusters for both WISE 1049A and B across the NIRSpec bandpass, and 2 distinct lightcurve clusters for each across the MIRI bandpass.
We find three main transitions in behavior / cluster for both components of the binary: 1) change in behavior at 2.3 $\mu$m coincident with a CO absorption bandhead, 2) change in behavior at 4.2 $\mu$m, close to the onset of the
CO fundamental band at $\lambda >$ 4.4 $\mu$m, 3) change in behavior at 8.3-8.5 $\mu$m, potentially corresponding to silicate absorption.  For WISE 1049B, the lightcurve behaviors for the shortest ($<$2.3 $\mu$m) and longest ($>$8.5 $\mu$m) look qualitatively similar, displaying "double-peaked" behavior (as noted in Section~\ref{sec:mirilc}), while lightcurves between 4.2 and 8.5 $\mu$m for WISE 1049B appear to be "single-peaked".  For both WISE 1049A and B, there is a distinctive and sharp change in behavior right around 3.3 $\mu$m, corresponding to a methane feature clearly seen in both components.  Similarly, a sharp change seen around 2.6-2.7 $\mu$m in both components may stem from water absorption at these wavelengths, but appears more complicated in WISE 1049B than A.  WISE 1049B also exhibits shifts in behavior around water absorption features at $\sim$1.8 and $\sim$1.9 $\mu$m.
\item We argue that the distinct changes in lightcurve behavior found at 2.3 $\mu$m, 4.2 $\mu$m, and 8.5 $\mu$m are connected to changes in the average pressure probed at each wavelength, as seen in Fig.~\ref{fig:contribution_plot}.  We interpret the lightcurves observed with both NIRSpec and MIRI as likely stemming from 3 discrete pressure levels -- a deep pressure level driving the double-peaked variability seen at wavelengths $<$2.3 $\mu$m and $>$8.5 $\mu$m, an intermediate pressure level shaping the lightcurve morphology between 2.3 and 4.2 $\mu$m, and a higher-altitude pressure level producing single-peaked and plateaued lightcurve behavior at wavelengths between 4.2 and 8.5 $\mu$m.  
\item Our observations test predictions for variability driven by inhomogeneous, small-grain silicate clouds \citep{Luna2021}, general circulation models driven by cloud radiative feedback \citep{Tan2021b}, and driven by hotspots produced by non-equilibrium chemistry \citep{Tremblin2020}.  However, in all cases, our lightcurves display considerable complexity that cannot easily be described by these of predictions.
\end{enumerate}

These observations demonstrate the transformational power of JWST to reveal the complex vertical structure of brown dwarf atmospheres.  While WISE 1049AB are the two brightest brown dwarfs known, dozens of others are amenable to similar studies with JWST.  JWST also enables similar studies of young, giant exoplanets, the lower surface gravity, lower-mass cousins of brown dwarfs.  This is the first such study, but will not be the last -- in the next few observing cycles, JWST will transform our understanding of both brown dwarf and young, giant exoplanet atmospheres.

\section*{Acknowledgements}

B.A.B. and B.J.S. acknowledge funding by the UK Science and Technology Facilities Council (STFC) grant no. ST/V000594/1. J. M. V. acknowledges support from a Royal Society - Science Foundation Ireland University Research Fellowship (URF$\backslash$1$\backslash$221932). A.M.M. acknowledges support from the National Science Foundation Graduate Research Fellowship under Grant No. DGE-1840990.  B.A.B. thanks Sarah Kendrew and Greg Sloan for providing a MIRI PSF built from commissioning data and for advice on how to implement PSF-fitting for MIRI LRS data.

\section*{Data Availability}

All raw and pipeline-processed data presented in this article are available via the MAST archive. After publication, we will provide the custom-reduced data products described in this article on Zenodo at DOI 10.5281/zenodo.12531991.



\bibliographystyle{mnras}
\bibliography{Full} 








\bsp	
\label{lastpage}
\end{document}